\documentclass[10pt,letterpaper]{article}

\usepackage{natbib}
\usepackage{hyperref}

\usepackage{booktabs}

\usepackage[latin1]{inputenc}							
\usepackage{amsmath}									
\usepackage{empheq}

\usepackage{graphicx}

\usepackage{xcolor}
\definecolor{bl}{rgb}{0.0,0.2,0.6}

\usepackage{sectsty}
\usepackage[compact]{titlesec} 
\allsectionsfont{\color{bl}\scshape\selectfont}

\makeatletter							
\def\printtitle{%						
    {\color{bl} \centering \huge \sc \textbf{\@title}\par}}		
\makeatother							

\title{Weighted skewness and kurtosis\\unbiased by sample size}

\makeatletter							
\def\printauthor{%					
    {\centering \@author}}				
\makeatother							

\author{%
\vspace{8pt}
\large Lorenzo Rimoldini \\
\vspace{5pt}
\small Observatoire astronomique de l'Universit\'e de Gen\`eve, chemin des Maillettes 51, CH-1290 Versoix, Switzerland \\
\small ISDC Data Centre for Astrophysics, Universit\'e de Gen\`eve, chemin d'Ecogia 16, CH-1290 Versoix, Switzerland \\
\vspace{3pt}
\small \texttt{lorenzo@rimoldini.info}\\
	\vspace{12pt}
Draft version: April 28, 2013\\
	\vspace{8pt}
	}

\usepackage{fancyhdr}
\usepackage{lastpage}	
\usepackage{abstract}			
\abslabeldelim{\quad}						% 
\begin{document}

\printtitle 

\printauthor

\begin{abstract}
Central moments and cumulants are often employed to characterize the distribution of data. The skewness and kurtosis are particularly useful for the detection of outliers, the assessment of departures from normally distributed data, automated classification techniques and other applications. Robust definitions of higher order moments are more stable but might miss characteristic features of the data, as in the case of astronomical time series with rare events like stellar bursts or eclipses from binary systems. Weighting can help identify reliable measurements from uncertain or spurious outliers, so unbiased estimates of the weighted skewness and kurtosis moments and cumulants, corrected for sample-size biases, are provided under the assumption of independent data. The comparison of biased and unbiased weighted estimators is illustrated with simulations as a function of sample size, employing different data distributions and weighting schemes.
\end{abstract}

\section{Introduction}
Descriptive statistics provide essential tools to quantify the main features of data and typically consist of simple quantities which can be computed efficiently and easily included in the analysis of large data volumes.
The ability to summarize essential information in a few parameters has found widespread interdisciplinary applications.
Central moments 
characterize the shape of the distribution of measurements around the mean value for most distributions occurring in practice. 
The familiar moments of  variance, skewness and kurtosis give indications on the dispersion, asymmetry and peakedness or weight of the tails of the distribution, respectively.

Moments are usually computed on random variables. 
% common
Herein, their application is extended to data generated from deterministic functions  
and randomized by the uneven sampling of a finite number of measurements and by their uncertainties, 
% common
whereas the corresponding `population' statistics are defined in the 
limit of an infinite regular sampling with no random or systematic errors.
% common
This scenario is common in astronomical time series, 
where measurements are typically non-regular due to observational constraints, 
% common
they are unavoidably affected by noise and sometimes also not very numerous: 
all of these aspects introduce some level of randomness in the characterization of the underlying signal of a star. 

While the effects of noise and sampling on time series are studied in \citet{RimoldiniIntrinsic,RimoldiniWeighted}, 
% common
this work addresses the bias, precision and accuracy of weighted estimators in the case of small sample sizes. 
% common
Bias is defined as the difference between expectation and 
population values and thus expresses a systematic deviation from the true value.
% common
Precision is described by the dispersion of measurements, 
while accuracy is related to the distance of an estimator from the true value and thus combines the bias and precision concepts 
% common
(e.g., accuracy can be measured by  the mean square error, defined by the sum of bias and uncertainty in quadrature).

Higher moments such as skewness and kurtosis have received particular attention for the detection of outliers and of departures from normally distributed data \citep{Dagostino}.
The underlying concepts can be expressed in many alternative ways, some of which \citep[e.g.,][]{Moors,Hosking,Groeneveld,Bowley} avoid sample means or non-linear transformations and favour more robust results.
While  insensitivity to a few botched measurements is a clear advantage, sometimes outliers are part of the targeted signal, as in the case of eclipses occurring in the light curves of binary star systems.
Since weighting can help distinguish meaningful outliers (e.g., the data corresponding to eclipsed phases) from spurious measurements,  the present work focusses on the conventional definitions of skewness and kurtosis in terms of central moments and cumulants, and provides sample-size corrected (`unbiased') estimates of the weighted formulations.

Sample moments do not provide unbiased estimates of population moments.
As the sample size decreases, the uncertainty of the sample mean around the population value increases and 
 higher order central moments can become  biased as a result.
Statistical estimators which remain unbiased as a function of sample size are relevant to those applications which aim at the characterization of the population  which the measured sample represents. This approach is particularly important for the interpretation and comparison of data in a  broader context  than the sole description of a sample.

Weighting can
quantify the relevance of  measurements (e.g., by inverse-squared uncertainties), enhance targeted features of the data 
depending on the objectives of the analysis, and have different implications on the precision and accuracy of estimators:
\begin{itemize}
\item[(i)] They might decrease, because weights assign more importance to some data at the expense of other ones, 
% common
effectively reducing the sample size as results depend mostly on fewer `relevant' measurements. This case is apparent in Sec.~\ref{sec:simulation}, for example, when weighting by  the inverse-squared uncertainties at high signal-to-noise ($S/N$) ratios. 
\item[(ii)] They might increase, when weights reduce greater dispersions and biases than the ones caused by an effectively smaller sample.
For example, weighting by inverse-squared uncertainties was shown to improve both precision and accuracy at low $S/N$ levels \citep{RimoldiniIntrinsic}. 
\end{itemize}
Weighting might exploit correlations in the data to improve precision \citep{RimoldiniWeighted}. 
% common
Since correlated data do not satisfy the assumptions of the expressions derived herein, their application might return biased results. 
% common
However, small biases could be justified if  improvements in precision are significant and, 
depending on the extent of the application, larger biases could be mitigated with mixed weighting schemes, such as the one described in Sec.~\ref{sec:simulation}.

Similarly to the pros and cons of weighting, unbiased estimators are expected to be more accurate but less precise than the biased counterparts, since they take into account the uncertainty of the sample mean. Thus, they are favoured when biases from small sample sizes are larger than the dispersion of unbiased statistics.
A compromise solution (weighted or unweighted, biased or unbiased) should balance biases against the  dispersion of weighted or unbiased estimators (which might depend on the statistics and the data), improve the overall accuracy and be applied uniformly to all data.

Unbiased expressions are derived for the weighted skewness and kurtosis (central moments and cumulants) in the case of independent measurements. 
The results are  
illustrated with simulated data and the dependence of unbiased weighted estimators on sample size is shown for two weighting schemes: the common inverse-squared uncertainties and interpolation-based weights as described in \citet{RimoldiniWeighted}. 
The latter  demonstrated a significant improvement in the precision of weighted moments and cumulants for data sets with at least a few tens of measurements.

This paper is organized as follows. 
The notation employed throughout is defined in Sec.~\ref{sec:notation}.
Sample (biased) weighted moments and cumulants are recalled in Sec.~\ref{sec:biased}.
Sample-size unbiased weighted and unweighted moments and cumulants are presented in Sec.~\ref{sec:unbiased}.
Biased and unbiased estimators are compared with simulated signals as a function of sample size in Sec.~\ref{sec:simulation}, including weighted and unweighted estimators and two different signal shapes. 
Conclusions are drawn in Sec.~\ref{sec:concl}, followed by detailed derivations of the sample-size unbiased weighted estimators in App.~\ref{app:derivations}.

\newpage

\section{Notation}
\label{sec:notation}
% common
For a set of $n$ measurements $\mathbf{x}=(x_1,x_2,...,x_n)$, the following quantities are defined.
\begin{itemize}
\item[(i)]
% common
Population central moments $\mu_r\!=\!E[(\mathbf{x}-\mu)^r]$ with mean $\mu=E(\mathbf{x})$, 
where $E(.)$ denotes expectation, and cumulants $\kappa_2=\mu_2$, $\kappa_3=\mu_3$, $\kappa_4=\mu_4-3\mu_2^2$  \citep[e.g.,][]{Kendall}.\footnote{Cumulants, first derived by \citet{Thiele}, have also been named `cumulative moment functions' \citep{Fisher} and `semi-invariants' by other authors \citep[e.g.,][]{Cramer,Dressel}.}
\item[(ii)]
% common
The sum of the $p$-th power of weights is defined as $V_p=\sum_{i=1}^n w_i^p$.
\item[(iii)]
% common
Sample central moments
$m_r=\sum_{i=1}^n w_i (x_i-\bar{x})^r/V_1$ and 
corresponding cumulants $k_r$.
\item[(iv)]
% common
Sample-size unbiased estimates of central 
moments  $M_i$ and cumulants $K_i$, i.e., $E(M_i)=\mu_i$ and $E(K_i)=\kappa_i$.
\item[(v)]
% common
The standardized skewness and kurtosis are 
defined as $g_1=k_3/k_2^{3/2}$,~ $g_2=k_4/k_2^2$,~ $G_1=K_3/K_2^{3/2}$, and
$G_2=K_4/K_2^2$, with population values $\gamma_1=\kappa_3/\kappa_2^{3/2}$ and $\gamma_2=\kappa_4/\kappa_2^2$. 
$G_1$ and $G_2$ satisfy consistency (for $n\rightarrow\infty$) but are not unbiased in general \citep[e.g., see][for exceptions]{Heijmans}.
\item[(vi)]
No systematics or other instrumental errors are considered herein and uncertainties are often referred to as errors.
\item[(vii)]
Statistics weighted by the inverse-squared uncertainties are called `error-weighted' for brevity and interpolation-based weights computed in phase \citep{RimoldiniWeighted} are named `phase weights'.
\end{itemize}

\section{Sample moments and cumulants}
\label{sec:biased}
The sample weighted central moments, such as the variance $m_2$, skewness $m_3$,  kurtosis $m_4$ and the respective cumulants, are defined in terms of the weighted mean $\bar{x}$ as follows:
\begin{align}
&\bar{x} =  \frac{1}{V_1}\sum_{i=1}^{n}w_i x_i \\
&m_2 = \frac{1}{V_1}\sum_{i=1}^{n}w_i (x_i-\bar{x})^2 = k_2 \\
&m_3 =  \frac{1}{V_1}\sum_{i=1}^{n}w_i (x_i-\bar{x})^3 = k_3 \\
&m_4  =   \frac{1}{V_1}\sum_{i=1}^{n}w_i (x_i-\bar{x})^4 \\
&k_4  =  m_4-3\,m_2^2. 
\end{align}
The unweighted forms can be obtained by substituting $w_i=1$ (for all $i$) and $V_1=n$ in the above equations.

\newpage

\section{Sample-size unbiased moments and cumulants}
\label{sec:unbiased}
The sample-size bias corrected weighted central moments, such as the variance $M_2$, skewness $M_3$,  kurtosis $M_4$ and the respective cumulants
are derived assuming independent measurements and weights, as described in full detail in App.~\ref{app:derivations}. They are defined in terms of  sample estimators as follows:
\begin{align}
M_2 =\,& \frac{V_1^2}{V_1^2-V_2}\,m_2 = K_2 \\
M_3 =\,&   \frac{V_1^3}{V_1^3-3V_1V_2+2V_3}\,m_3 = K_3 \label{eq:startResult}\\
M_4  =\,&   \frac{V_1^2(V_1^4-3V_1^2V_2+2V_1V_3+3V_2^2-3V_4)}{(V_1^2-V_2)(V_1^4-6V_1^2V_2+8V_1V_3+3V_2^2-6V_4)}\,m_4 \,+\notag\\
\,&  -\frac{3V_1^2(2V_1^2V_2-2V_1V_3-3V_2^2+3V_4)}{(V_1^2-V_2)(V_1^4-6V_1^2V_2+8V_1V_3+3V_2^2-6V_4)}\,m_2^2 \\
K_4  =\,&   \frac{V_1^2(V_1^4-4V_1V_3+3V_2^2)}{(V_1^2-V_2)(V_1^4-6V_1^2V_2+8V_1V_3+3V_2^2-6V_4)}\,m_4 \,+\notag\\
\,&  -\frac{3V_1^2(V_1^4-2V_1^2V_2+4V_1V_3-3V_2^2)}{(V_1^2-V_2)(V_1^4-6V_1^2V_2+8V_1V_3+3V_2^2-6V_4)}\,m_2^2\,. \label{eq:endResult}
\end{align}

The corresponding unweighted forms can be achieved by direct substitution $V_p=n$ for all $p$, 
 leading to the known relations \citep[e.g., see][]{Cramer}:
\begin{align}
M_2 =\,& \frac{n}{n-1}\,m_2 = K_2 \\
M_3 =\,&   \frac{n^2}{(n-1)(n-2)}\,m_3 = K_3 \\
M_4  =\,&   \frac{n(n^2-2n+3)}{(n-1)(n-2)(n-3)}\,m_4 -\frac{3n(2n-3)}{(n-1)(n-2)(n-3)}\,m_2^2 \\
K_4  =\,&   \frac{n^2(n+1)}{(n-1)(n-2)(n-3)}\,m_4 -\frac{3n^2}{(n-2)(n-3)}\,m_2^2.
\end{align}

\section{Estimators as a function of sample size}
\label{sec:simulation}
The effect of different weighting schemes on sample and population estimators is illustrated as a function of sample size with simulated data, through which
biased and unbiased estimators are compared for specific periodic signals, sampling and error laws.
The  values of the population moments of the continuous simulated periodic `true' signal $\xi(\phi)$ are computed averaging in phase $\phi$ as follows:
\begin{align}
\mu_r=\frac{1}{2\pi}\int_0^{2\pi}\left[\xi(\phi)-\mu \right]^r {\mathrm d}\phi, ~~~~~~\mbox{where}~~~\mu=\frac{1}{2\pi}\int_0^{2\pi}\xi(\phi)\,{\mathrm d}\phi.
\end{align}

 \subsection{Simulation}
Simulated signals are described by a simple sinusoidal function to the first and the fourth powers.
The signal $\xi(\phi)$ with variance $\mu_2$ and signal-to-noise ratio $S/N$ (estimated by the ratio of the standard deviation $\sqrt{\mu_2}$ and the root of the mean of squared measurement uncertainties $\epsilon_i$) is  sampled from $n=10$ to 1000 times at phases $\phi_i$ randomly drawn from a uniform distribution:
\begin{empheq}[left=\empheqlbrace]{align}
&\xi(\phi)=\xi_o+A\sin^{\alpha}\phi \label{eq:simuStart}\\
&x_i\sim {\cal N}(\xi_i,\epsilon_i^2) ~~~~~~~~~\mbox{for~~}\xi_i=\xi(\phi_i) \mbox {~~and~~}\phi_i\sim {\cal U}(0,2\pi) \\
&\epsilon_i^2=\left(1+\rho_i \right)\,\mu_2 \,/\, (S/N)^2 ~~~~~~~~~~~\mbox{for~~}\rho_i\sim{\cal U}(-0.8,0.8), \label{eq:simuCoreEnd}
\end{empheq}
where $\alpha=1,4$ and the $i$-th measurement $x_i$ is drawn from a normal distribution ${\cal N}(\xi_i,\epsilon_i^2)$ of mean $\xi_i$ and variance $\epsilon_i^2$.
The latter is defined in terms of a variable $\rho_i$  randomly drawn from a uniform distribution ${\cal U}(-0.8,0.8)$ so that measurement uncertainties vary by up to a factor of 3 for a given $\mu_2$ and $S/N$ ratio.
Simulations were repeated $10^4$ times for each sample size. 

\begin{figure}
\centering
\includegraphics[width=8cm]{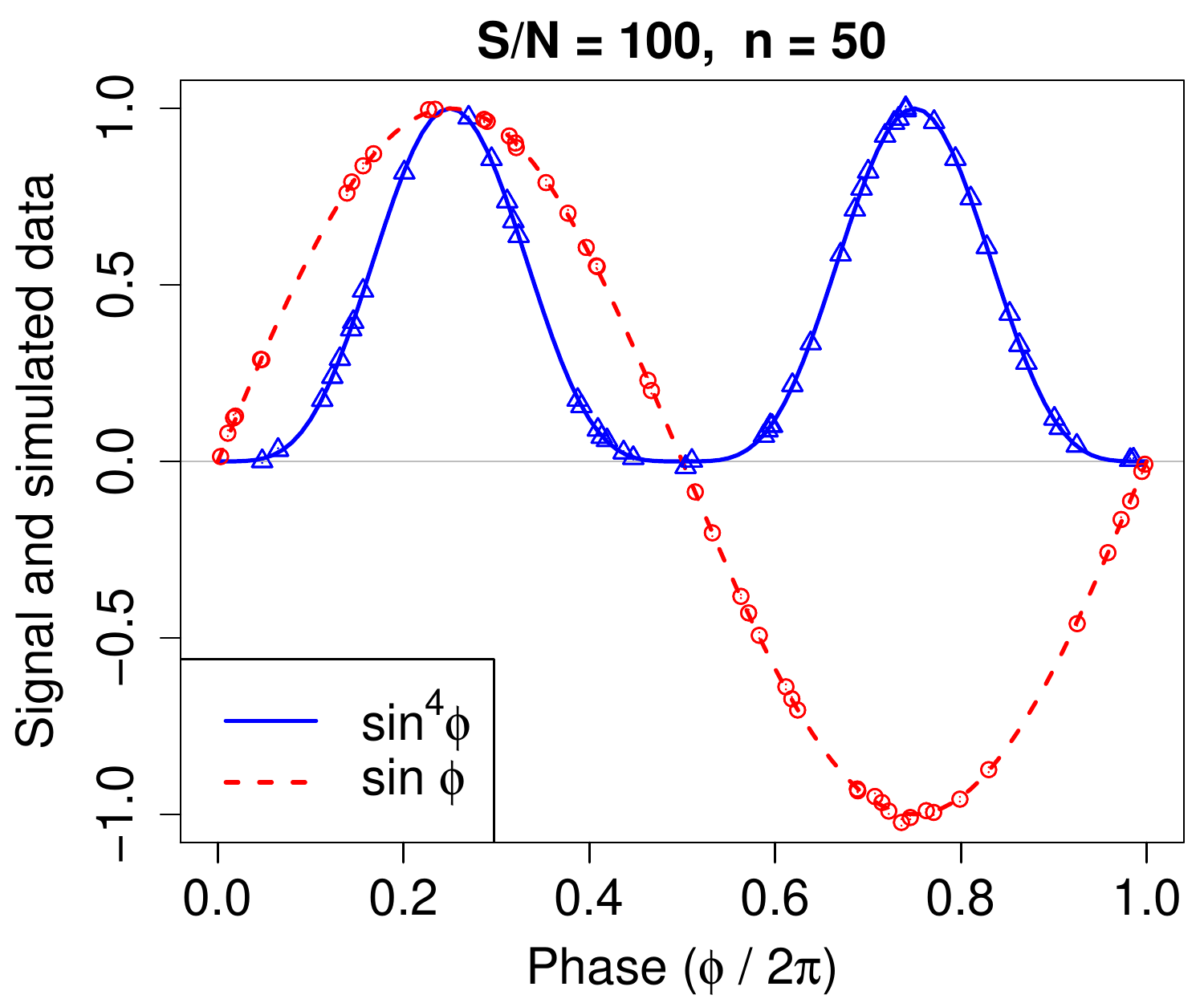} \\
\caption{Simulated signals of the forms of $\sin^4 \phi$ (solid blue curve) and $\sin\phi$ (dashed red curve) are irregularly sampled by 50 measurements (denoted by blue triangles and red circles, respectively) with $S/N=100$.
}
\label{fig:simu}
\end{figure}

The dependence of estimators on noise and the corresponding unbiased expressions were presented in \citet{RimoldiniIntrinsic}. Herein, the $S/N$ ratio is set to 100 so that noise biases are negligible with respect to the ones resulting from small sample sizes. 
Sample signals and simulated data are illustrated in Fig.~\ref{fig:simu} for $n=50$.
The reference population values of the mean, variance, skewness and kurtosis of the simulated signals are listed in Table~\ref{tab:pop}.

\begin{table}
 \centering
  \caption{Population values of the estimators illustrated in Figs~\ref{fig:M1}--\ref{fig:K4_100wb}.}
\label{tab:pop}
  \begin{tabular}{@{}clll@{}}
  \toprule
 Population&Equivalent&\multicolumn{2}{c}{\!\!\!\!\!\!\!\!\!\!\!\!\!\!Population values for  $\xi(\phi)=\xi_o+A\sin^{\alpha}\phi$}\\
Statistics&Expression&$\alpha=1$&$\alpha=4$\\
\midrule
$\mu$&--&$\xi_o$&$\xi_o+3 A /8$\\
$\mu_2$&$\kappa_2$&$A^2/2$&$17 A^2/128$\\
$\mu_3$&$\kappa_3$&$0$&$3A^3/128$\\
$\mu_4$&$\kappa_4+3\kappa_2^2$&$3A^4/8$&$963 A^4/32768$\\
$\kappa_4$&$\mu_4-3\mu_2^2$&$-3A^4/8$&$-771 A^4 / 32768$\\
$\gamma_1$&$\kappa_3/\kappa_2^{3/2}$,~$\mu_3/\mu_2^{3/2}$&$0$&$\sqrt{1152/4913}\approx 0.484$\\
$\gamma_2$&$\kappa_4/\kappa_2^2$,~$\mu_4/\mu_2^2-3$&$-3/2$&$-771/578\approx -1.334$\\
\bottomrule 
\end{tabular} 
\end{table}
\newpage
Error weights are defined by $w_i=1/\epsilon_i^2$, while phase weights  follow \citet{RimoldiniWeighted}, assuming phase-sorted data:
\begin{empheq}[left=\empheqlbrace]{align}
&w_i= h(n|a,b)\,\frac{w'_i}{\sum_{j=1}^n w'_j}+ \left[1-h(n|a,b)\right]/n  ~~~~~~~\forall i\in(1,n)\label{eq:phaseGap}\\
&w'_i=\phi_{i+1}-\phi_{i-1}~~~~~~~~~~~~~~~~~~\forall i\in(2,n-1) \label{eq:w_gap_ia}\\
&w'_1=\phi_{2}-\phi_{n}+2\pi \label{eq:w_gap_1a}\\
&w'_n=\phi_{1}-\phi_{n-1}+2\pi \label{eq:w_gap_na}\\
&h(n|a,b)=\frac{1}{1+e^{-(n-a)/b}}~~~~~~~~\mbox{for}~~a,b>0. \label{eq:simuEnd}
\end{empheq}

Estimators derived herein assume a single weighting scheme and combinations of estimators 
% common
(like the variance and the mean in the standardized skewness and kurtosis) are expected to apply the same weights to terms associated with the same measurements.
% common
The function $h(n|a,b)$ constitutes just an example to achieve a mixed weighting scheme: 
% common
tuning parameters $a,b$ offer the possibility to control the transition from unweighted to phase-weighted estimators (in the limits of small and large $n$, respectively) and thus
% common
reach a compromise solution between precision and accuracy  for all values of $n$,  according to the specific estimators,  signals, sampling, errors,  sample sizes and their distributions in the data. 

\subsection{Results}
\begin{figure}
\begin{center}
~~~~~~~~{\bf\fbox{\parbox{0.15\textwidth}{\centering Mean \\ $(\sin\phi,~\sin^4 \phi)$}}}\\
\end{center}
\begin{minipage}{0.5\columnwidth}
\centering
~~~~~~Unweighted \\
\includegraphics[width=\columnwidth]{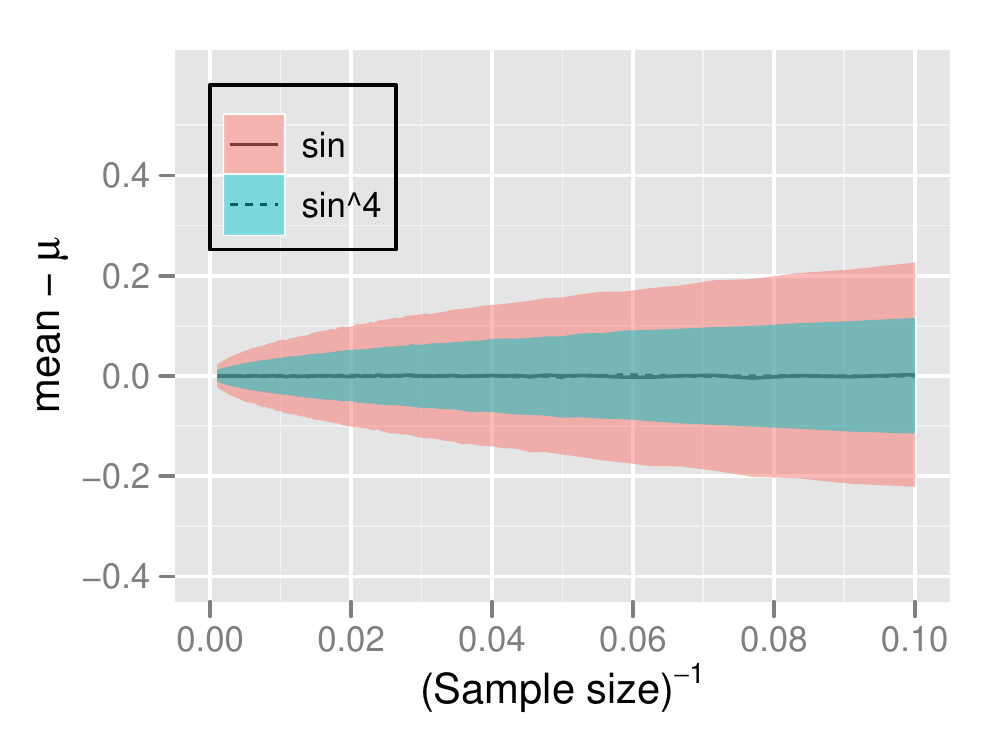} \\
~~~~~~~~Phase Weighted ($a,b\rightarrow 0$)\\
\includegraphics[width=\columnwidth]{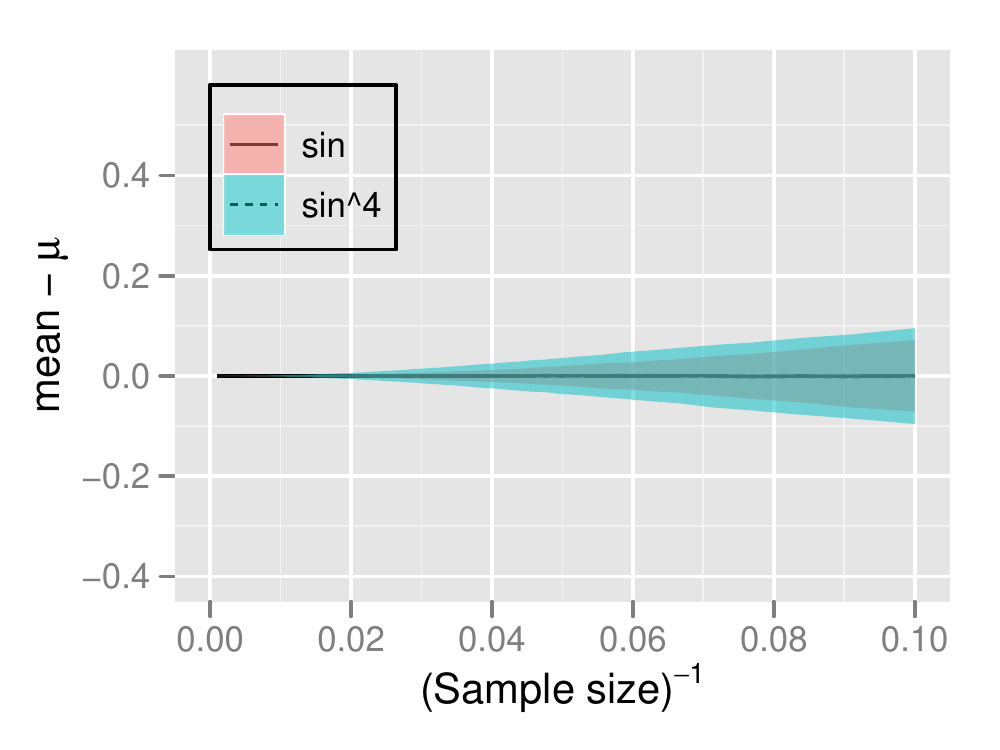} \\
\end{minipage}
\begin{minipage}{0.5\columnwidth}
\centering
~~~~~~~~Error Weighted\\
\includegraphics[width=\columnwidth]{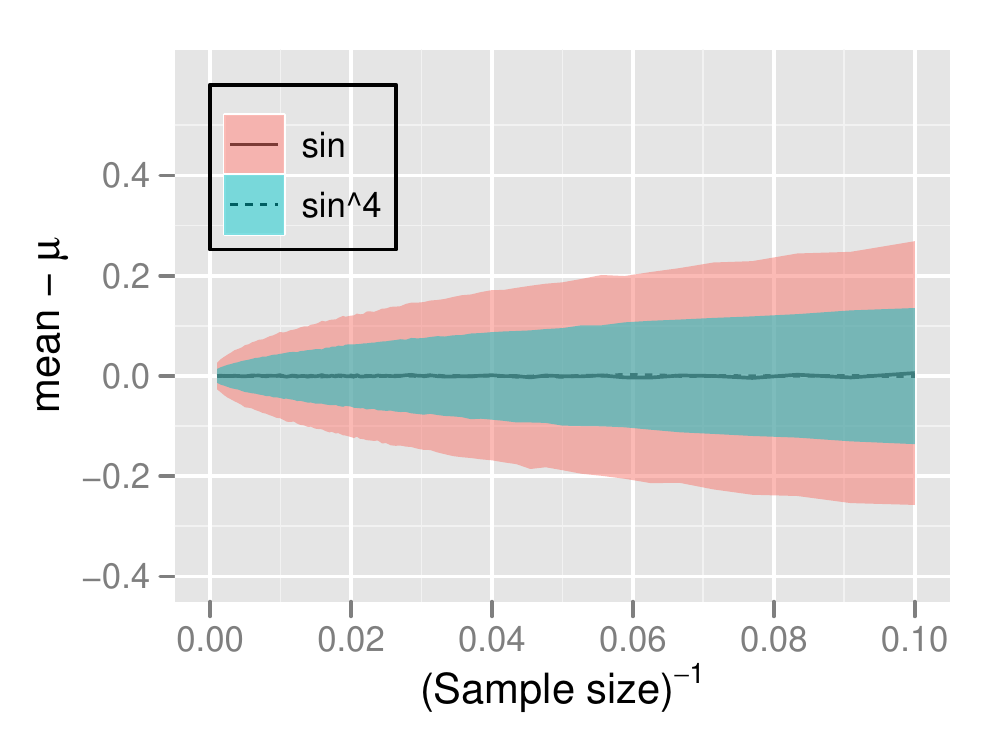}\\
~~~~~~~~Phase Weighted ($a=25,b=6$)\\
\includegraphics[width=\columnwidth]{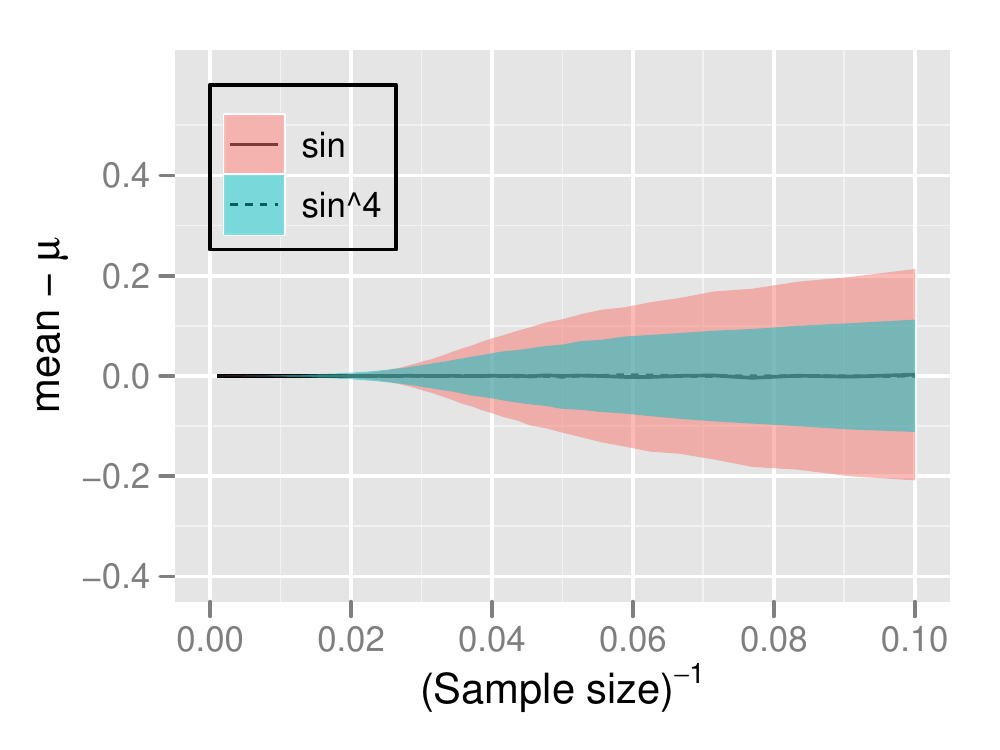} 
\end{minipage}
\caption{Sample mean of $\sin\phi$ and $\sin^4\phi$ for $n\in(10,1000)$ and $S/N=100$: unweighted on the top-left hand side, weighted by the inverse of squared measurement errors on the top-right hand side, weighted by phase gaps, as defined by Eq.~(\ref{eq:phaseGap}), with different parameter values, as specified above the lower panels. Shaded areas encompass one standard deviation from the average of the distribution of the mean employing simulations defined by Eqs~(\ref{eq:simuStart})--(\ref{eq:simuCoreEnd}).}
\label{fig:M1}
\end{figure}

Figure~\ref{fig:M1} illustrates the sample mean in the various scenarios considered in the simulations at $S/N=100$: unweighted and with different weighting schemes (error-weighted and phase-weighted). While accuracy is the same in all cases, the best precision of the mean is achieved employing phase weights, with no need to limit interpolation within large gaps for small sample sizes. 

\begin{figure}
\begin{center}
~~~~~~~~{\bf\fbox{\parbox{0.15\textwidth}{\centering Variance \\ $(\sin \phi)$}}}\\
\end{center}
\begin{minipage}{0.5\columnwidth}
\centering
~~~~~~Unweighted \\
\includegraphics[width=\columnwidth]{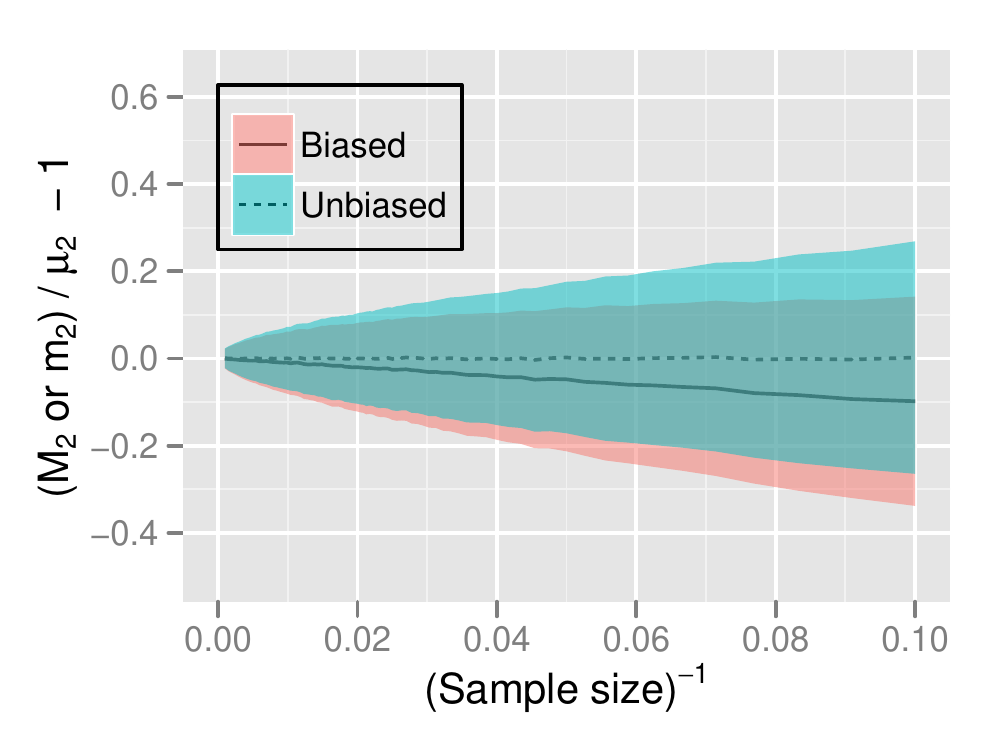} \\
~~~~~~~~Phase Weighted ($a,b\rightarrow 0$)\\
\includegraphics[width=\columnwidth]{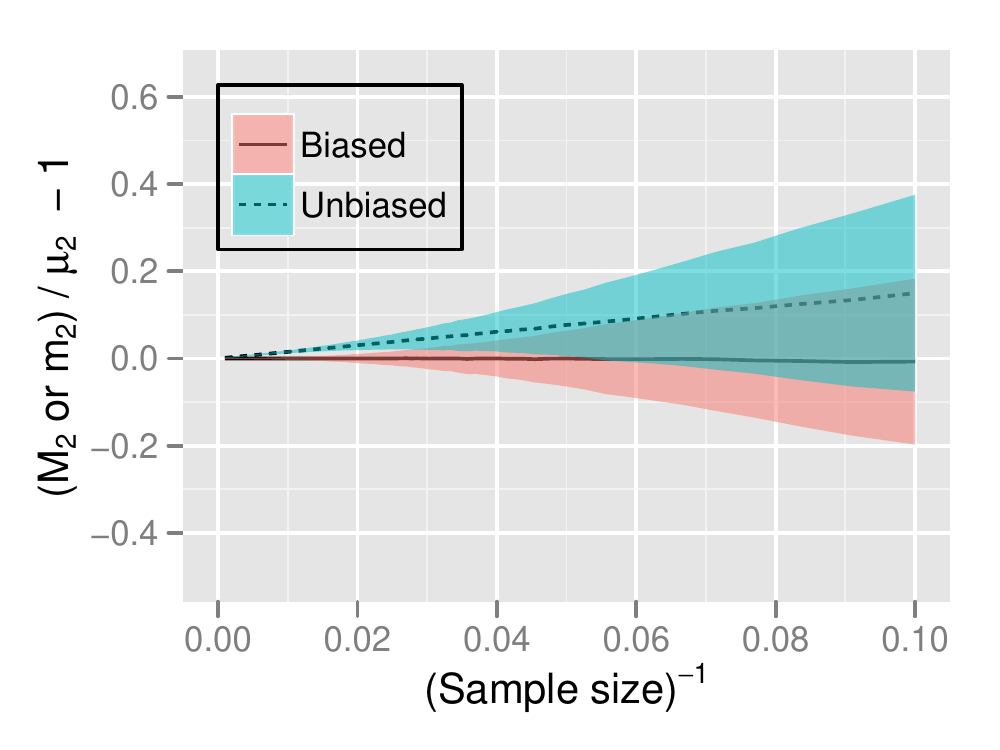} \\
\end{minipage}
\begin{minipage}{0.5\columnwidth}
\centering
~~~~~~~~Error Weighted\\
\includegraphics[width=\columnwidth]{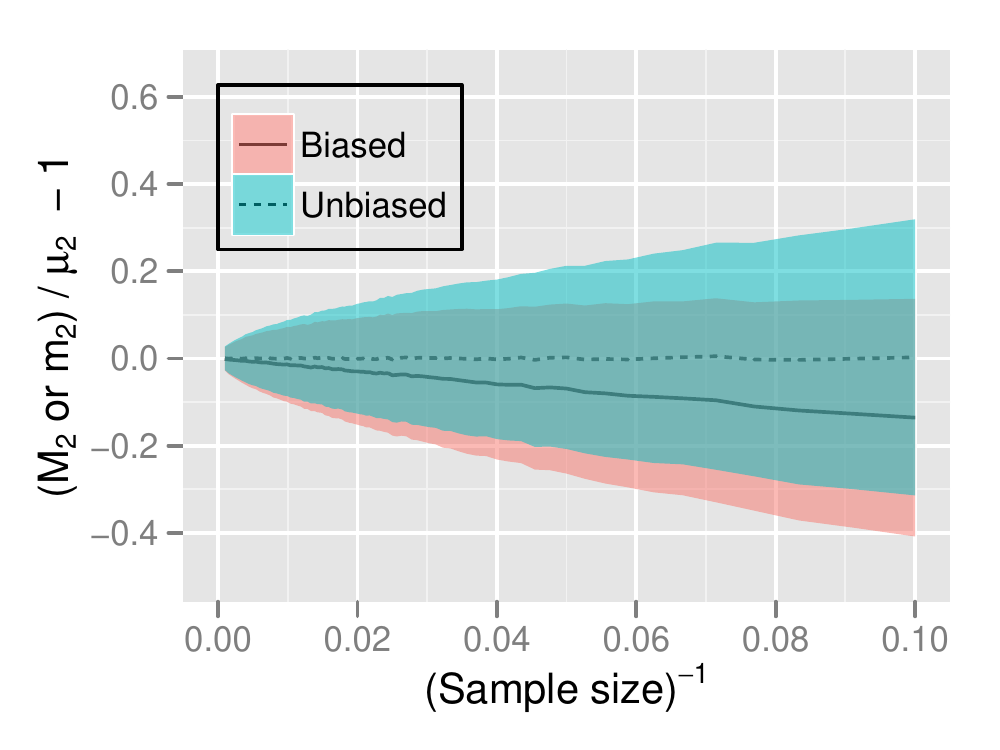}\\
~~~~~~~~Phase Weighted ($a=25,b=6$)\\
\includegraphics[width=\columnwidth]{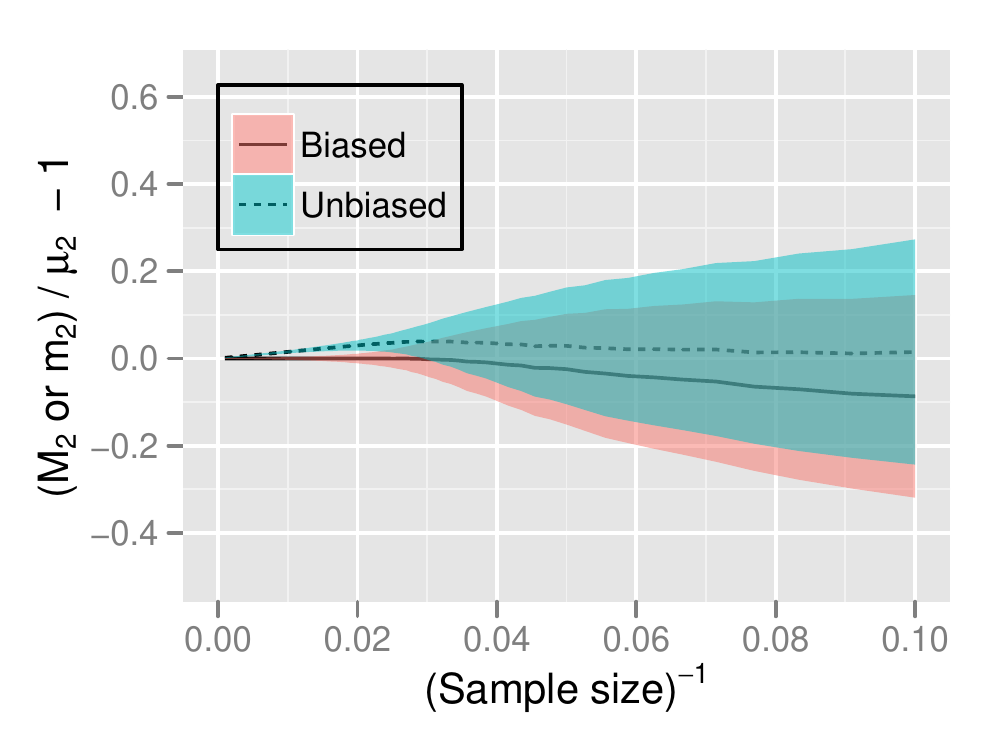} 
\end{minipage}
\caption{Sample ({\it `biased'}\,) variance $m_2$ of $\sin\phi$ versus its population ({\it `unbiased'}\,) estimate $M_2$ for $n\in(10,1000)$ and $S/N=100$: unweighted on the top-left hand side, weighted by the inverse of squared measurement errors on the top-right hand side, weighted by phase gaps, as defined by Eq.~(\ref{eq:phaseGap}), with different parameter values, as specified above the lower panels. 
The correlations introduced by phase weights are expected to bias the otherwise `unbiased' variance.
Shaded areas encompass one standard deviation from the mean of the distribution of the variance employing simulations defined by Eqs~(\ref{eq:simuStart})--(\ref{eq:simuCoreEnd}).}
\label{fig:M2}
\end{figure}

\begin{figure}
\begin{center}
~~~~~~~~{\bf\fbox{\parbox{0.15\textwidth}{\centering Variance \\ $(\sin^4 \phi)$}}}\\
\end{center}
\begin{minipage}{0.5\columnwidth}
\centering
~~~~~~Unweighted \\
\includegraphics[width=\columnwidth]{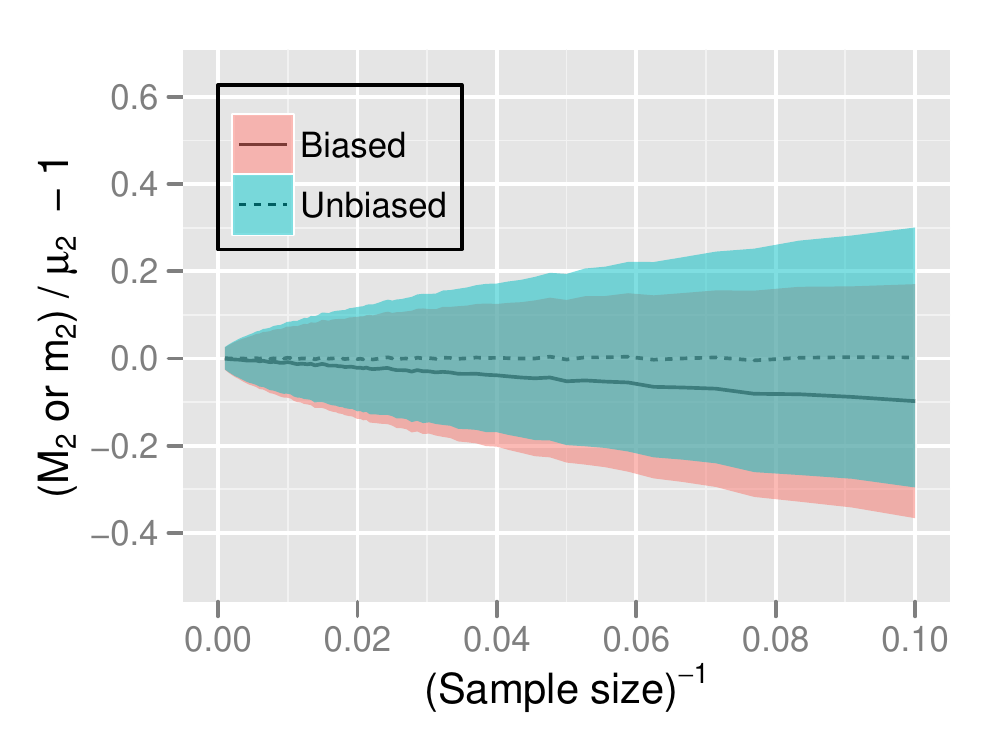} \\
~~~~~~~~Phase Weighted ($a,b\rightarrow 0$)\\
\includegraphics[width=\columnwidth]{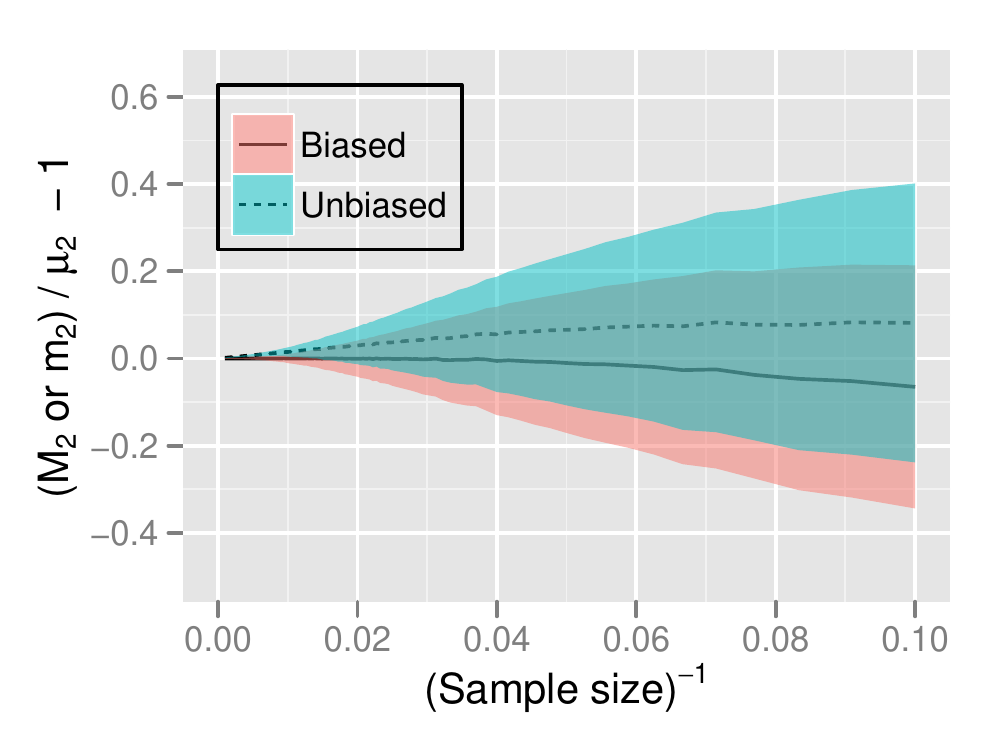} \\
\end{minipage}
\begin{minipage}{0.5\columnwidth}
\centering
~~~~~~~~Error Weighted\\
\includegraphics[width=\columnwidth]{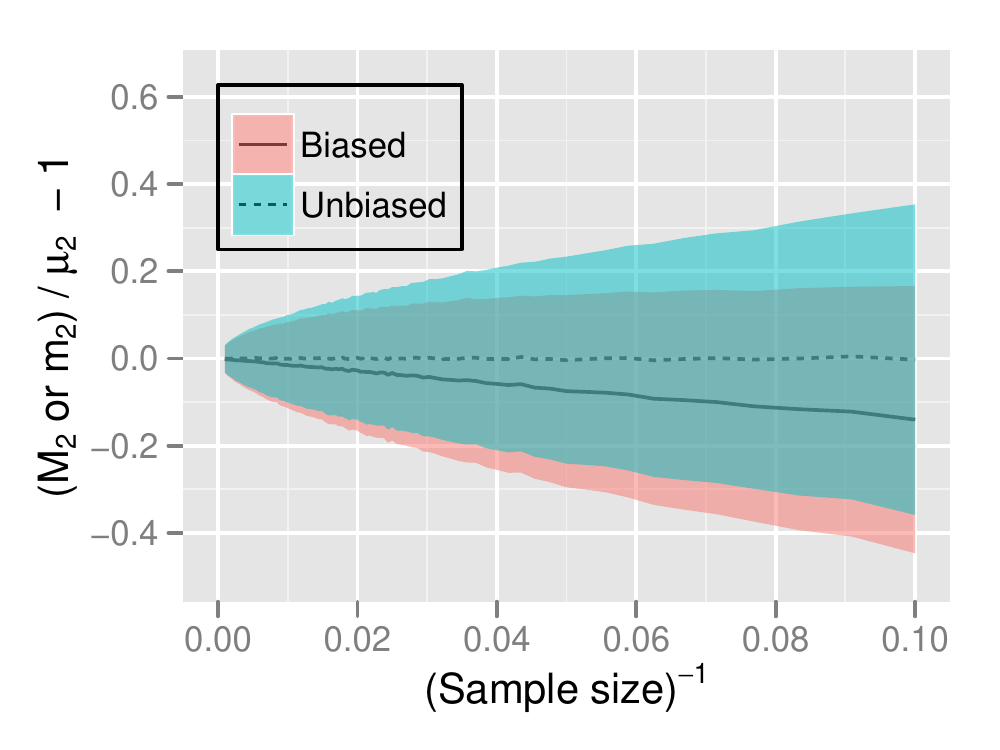}\\
~~~~~~~~Phase Weighted ($a=25,b=6$)\\
\includegraphics[width=\columnwidth]{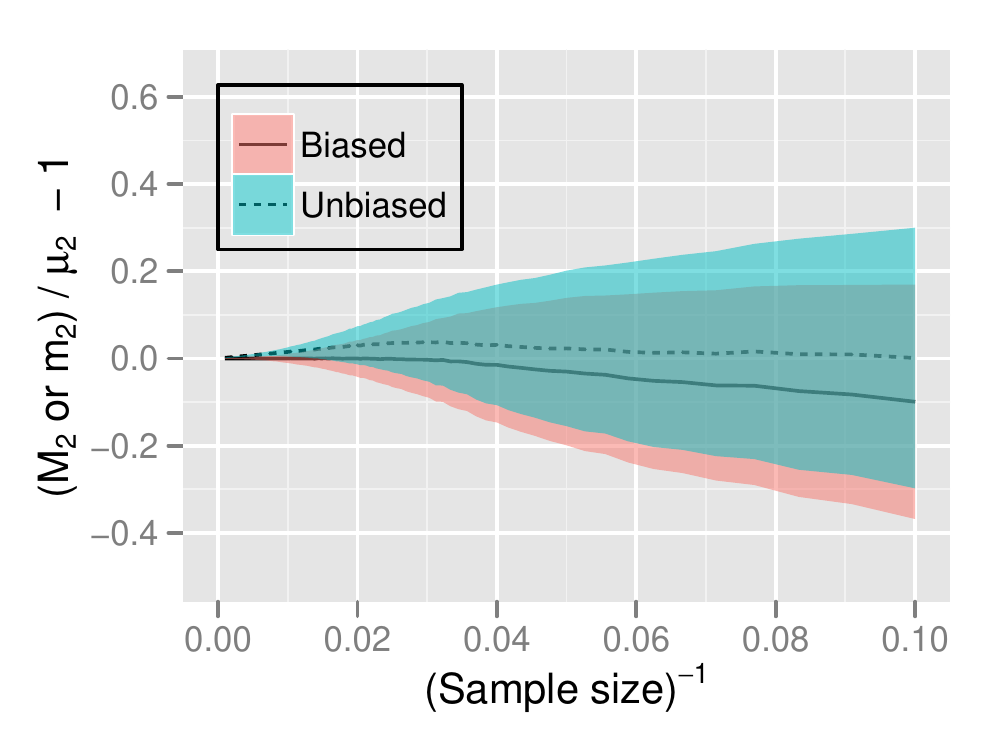} 
\end{minipage}
\caption{Sample ({\it `biased'}\,) variance $m_2$ of $\sin^4\phi$ versus its population ({\it `unbiased'}\,) estimate $M_2$ for $n\in(10,1000)$ and $S/N=100$: unweighted on the top-left hand side, weighted by the inverse of squared measurement errors on the top-right hand side, weighted by phase gaps, as defined by Eq.~(\ref{eq:phaseGap}), with different parameter values, as specified above the lower panels. 
The correlations introduced by phase weights are expected to bias the otherwise `unbiased' variance.
Shaded areas encompass one standard deviation from the mean of the distribution of the variance employing simulations defined by Eqs~(\ref{eq:simuStart})--(\ref{eq:simuCoreEnd}).}
\label{fig:M2b}
\end{figure}

\begin{figure}
\begin{center}
~~~~~~~~{\bf\fbox{\parbox{0.15\textwidth}{\centering Skewness \\ $(\sin \phi)$}}}\\
\end{center}
\begin{minipage}{0.5\columnwidth}
\centering
~~~~~~Unweighted \\
\includegraphics[width=\columnwidth]{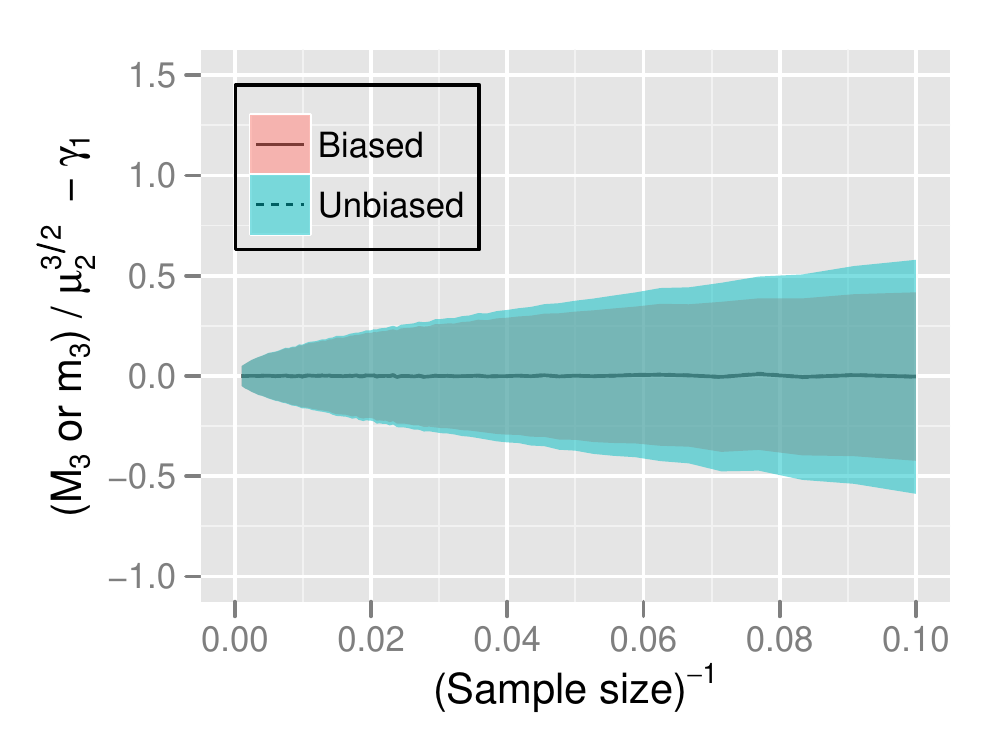}\\
~~~~~~~~Error Weighted \\
\includegraphics[width=\columnwidth]{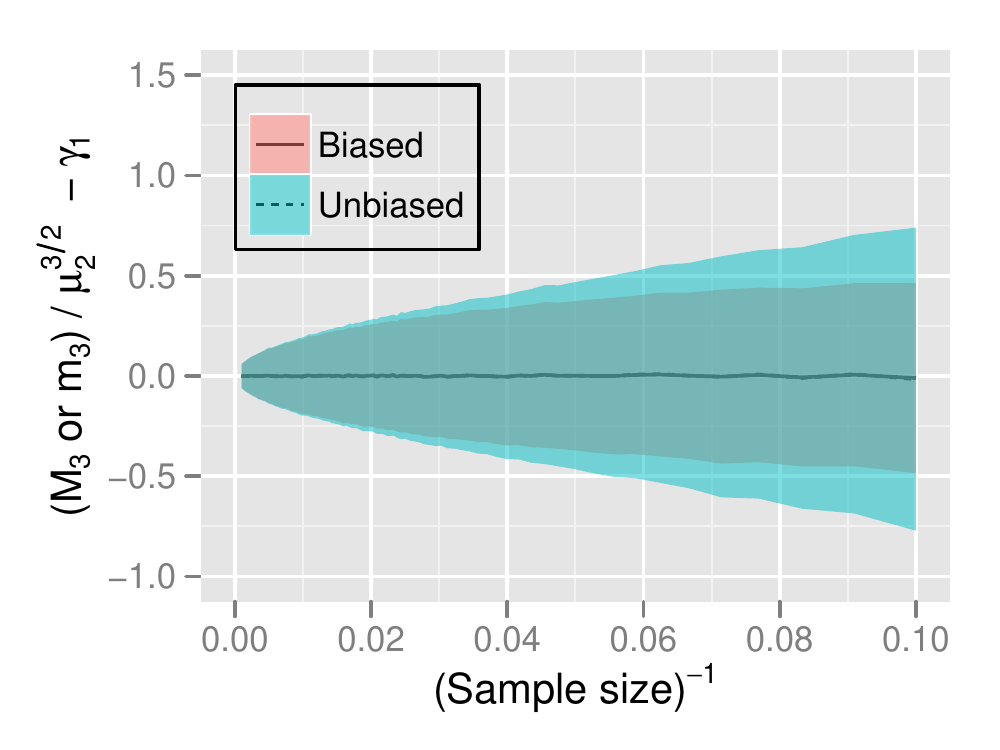}
\end{minipage}
\begin{minipage}{0.5\columnwidth}
\centering
~~~~~~Unweighted  \\
\includegraphics[width=\columnwidth]{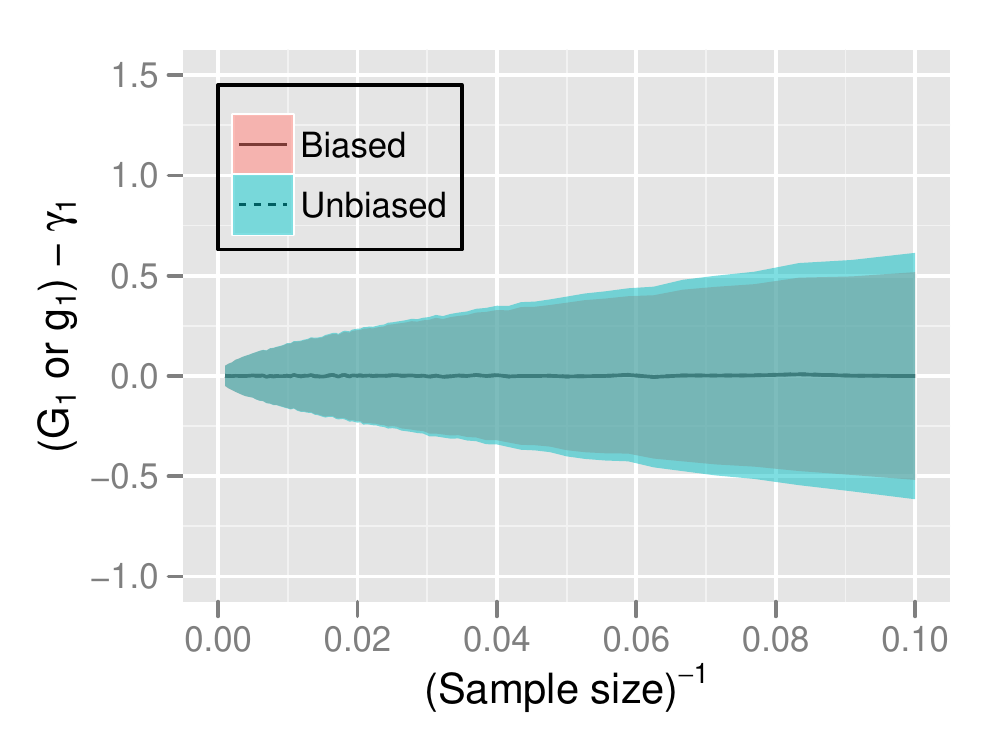}\\
~~~~~~~~Error Weighted \\
\includegraphics[width=\columnwidth]{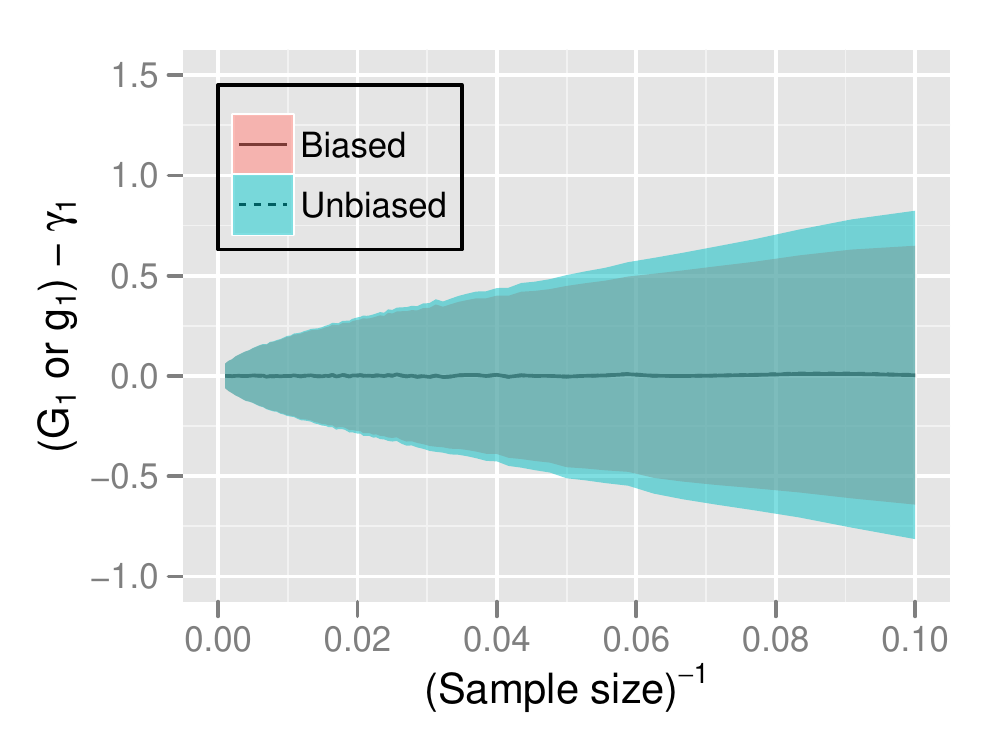}
\end{minipage}
\caption{Sample ({\it `biased'}\,) skewness $m_3$ of $\sin\phi$ versus its population ({\it `unbiased'}\,) estimate $M_3$ for $n\in(10,1000)$ and $S/N=100$: unweighted in the upper panels and weighted by the inverse of squared measurement errors in the lower panels. 
Shaded areas encompass one standard deviation from the mean of the distribution of the skewness employing simulations defined by Eqs~(\ref{eq:simuStart})--(\ref{eq:simuCoreEnd}). }
\label{fig:M3_100}
\end{figure}

\begin{figure}
\begin{center}
~~~~~~~~{\bf\fbox{\parbox{0.15\textwidth}{\centering Skewness \\ $(\sin \phi)$}}}\\
\end{center}
\begin{minipage}{0.5\columnwidth}
\centering
~~~~~~~~Phase Weighted ($a,b\rightarrow 0$)\\
\includegraphics[width=\columnwidth]{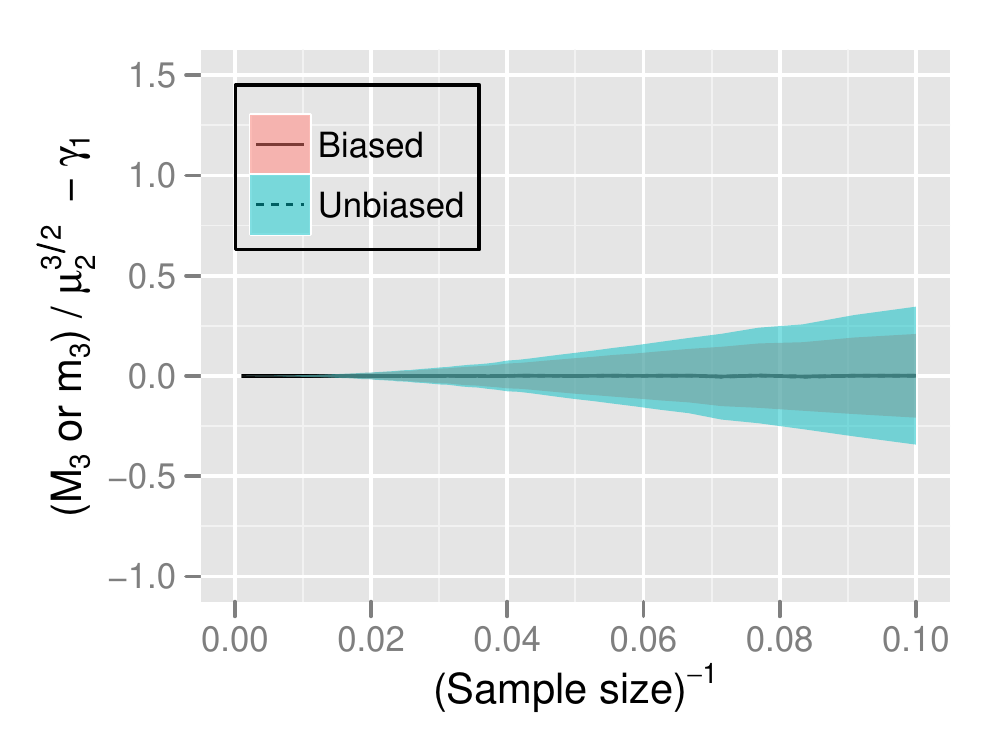}\\
~~~~~~~~Phase Weighted ($a=25,b=6$)\\
\includegraphics[width=\columnwidth]{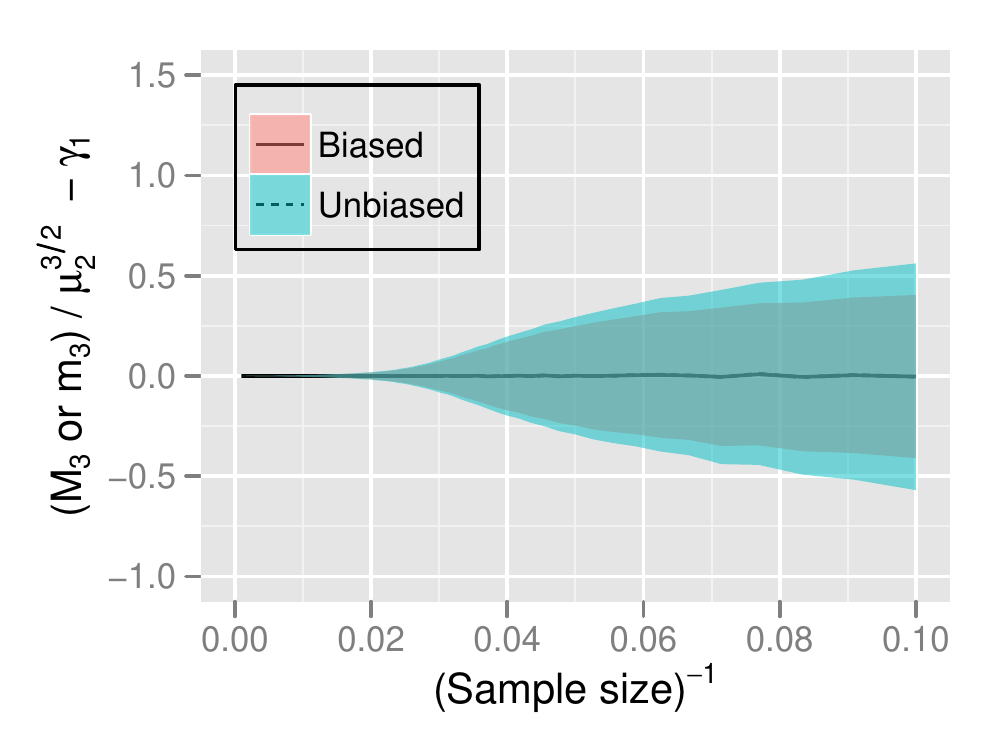}
\end{minipage}
\begin{minipage}{0.5\columnwidth}
\centering
~~~~~~~~Phase Weighted ($a,b\rightarrow 0$)\\
\includegraphics[width=\columnwidth]{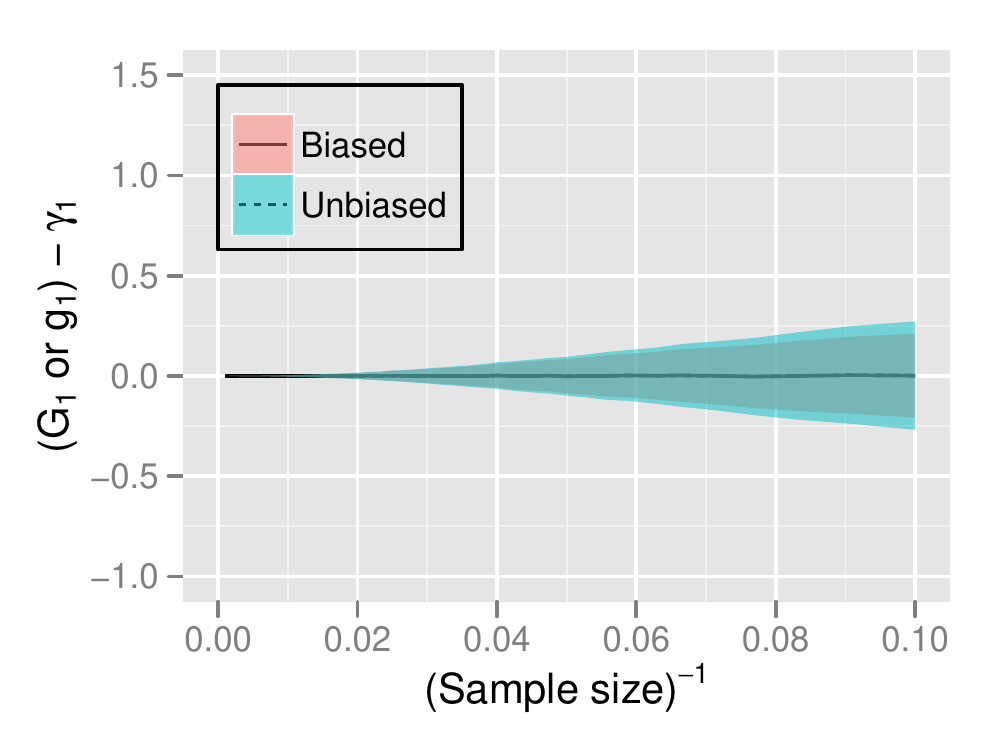}\\
~~~~~~~~Phase Weighted ($a=25,b=6$)\\
\includegraphics[width=\columnwidth]{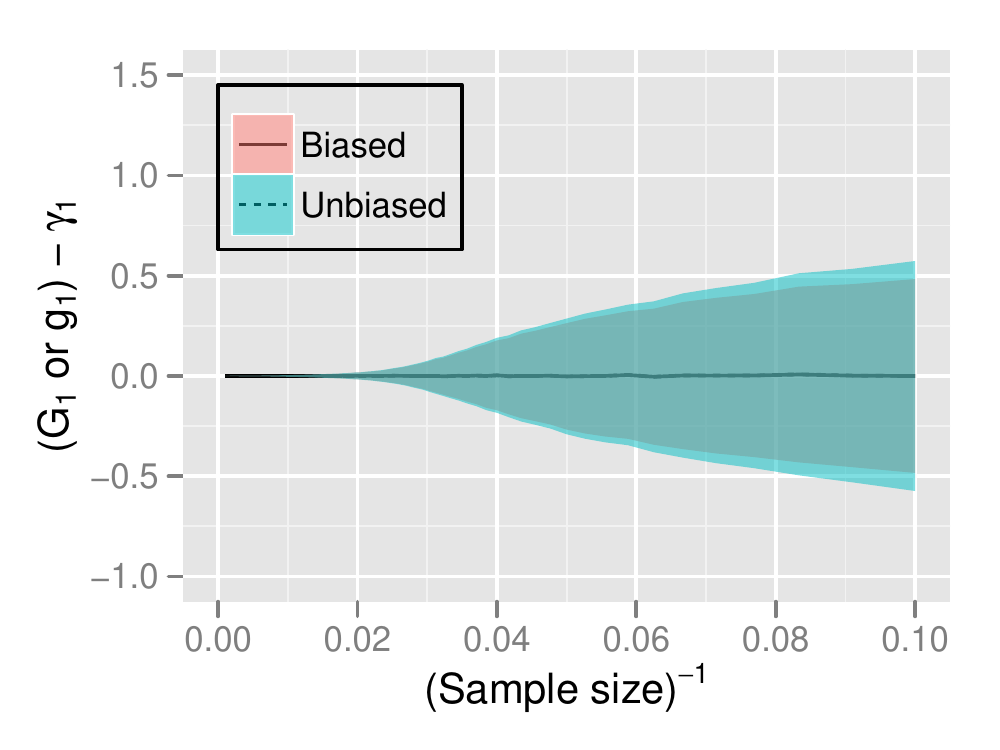}
\end{minipage}
\caption{Sample ({\it `biased'}\,) skewness $m_3$ of $\sin\phi$ versus its population ({\it `unbiased'}\,) estimate $M_3$ for $n\in(10,1000)$ and $S/N=100$, weighted by phase gaps, as defined by Eq.~(\ref{eq:phaseGap}), with different parameter values, as specified above each panel. 
 Shaded areas encompass one standard deviation from the mean of the distribution of the skewness employing simulations defined by Eqs~(\ref{eq:simuStart})--(\ref{eq:simuCoreEnd}). }
\label{fig:M3_100w}
\end{figure}

\begin{figure}
\begin{center}
~~~~~~~~{\bf\fbox{\parbox{0.15\textwidth}{\centering Skewness \\ $(\sin^4 \phi)$}}}\\
\end{center}
\begin{minipage}{0.5\columnwidth}
\centering
~~~~~~Unweighted \\
\includegraphics[width=\columnwidth]{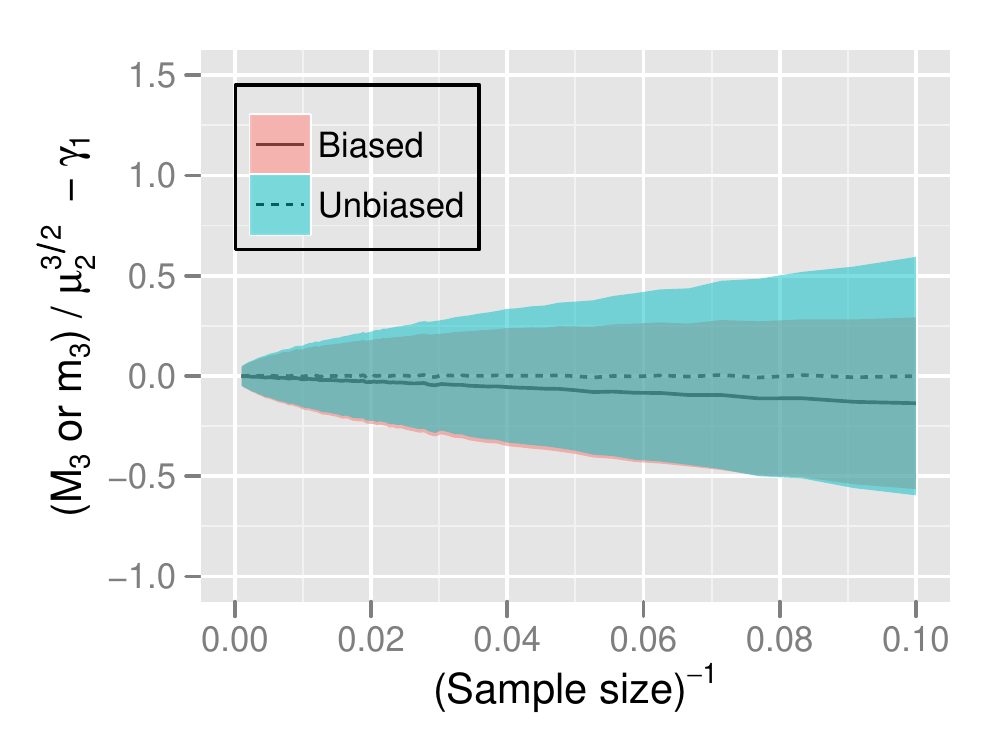}\\
~~~~~~~~Error Weighted \\
\includegraphics[width=\columnwidth]{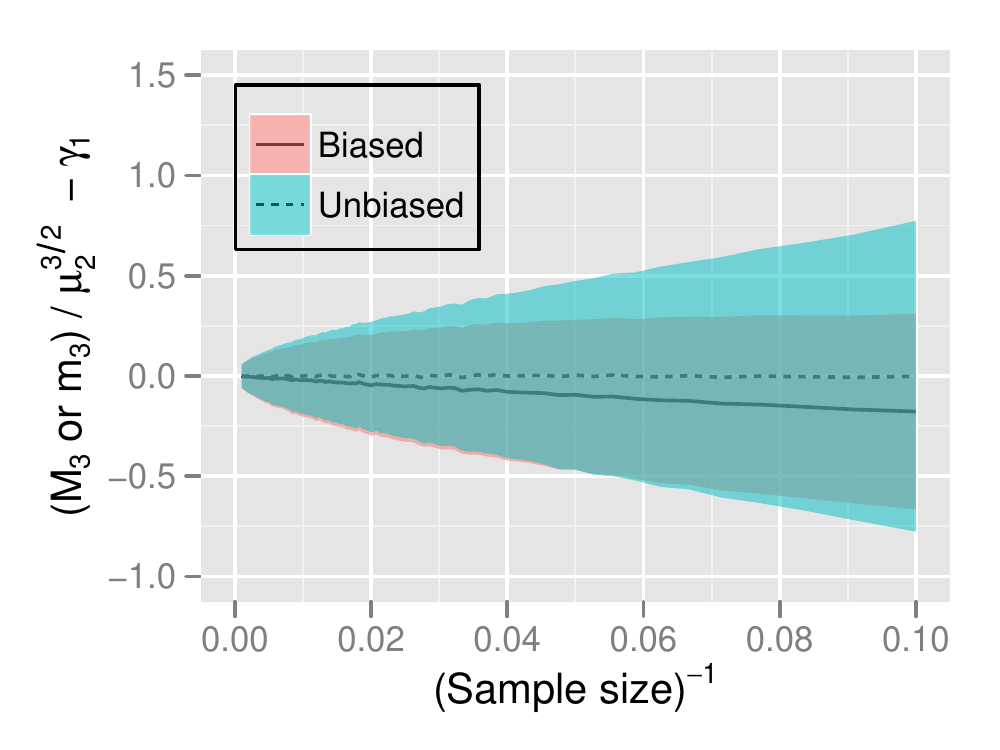}
\end{minipage}
\begin{minipage}{0.5\columnwidth}
\centering
~~~~~~Unweighted  \\
\includegraphics[width=\columnwidth]{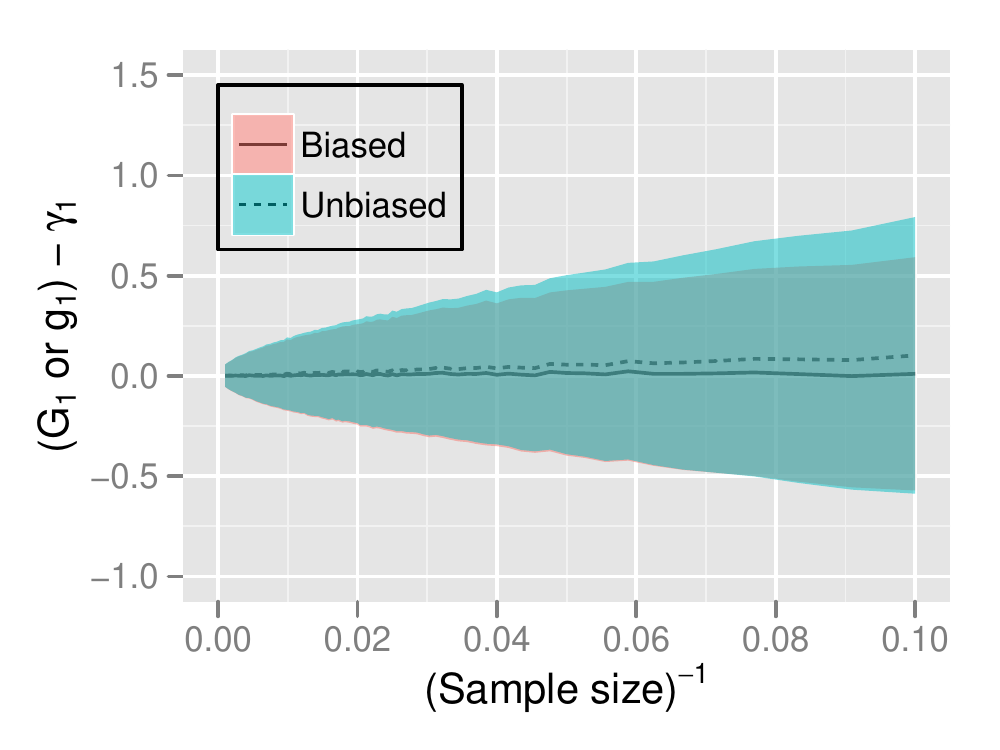}\\
~~~~~~~~Error Weighted \\
\includegraphics[width=\columnwidth]{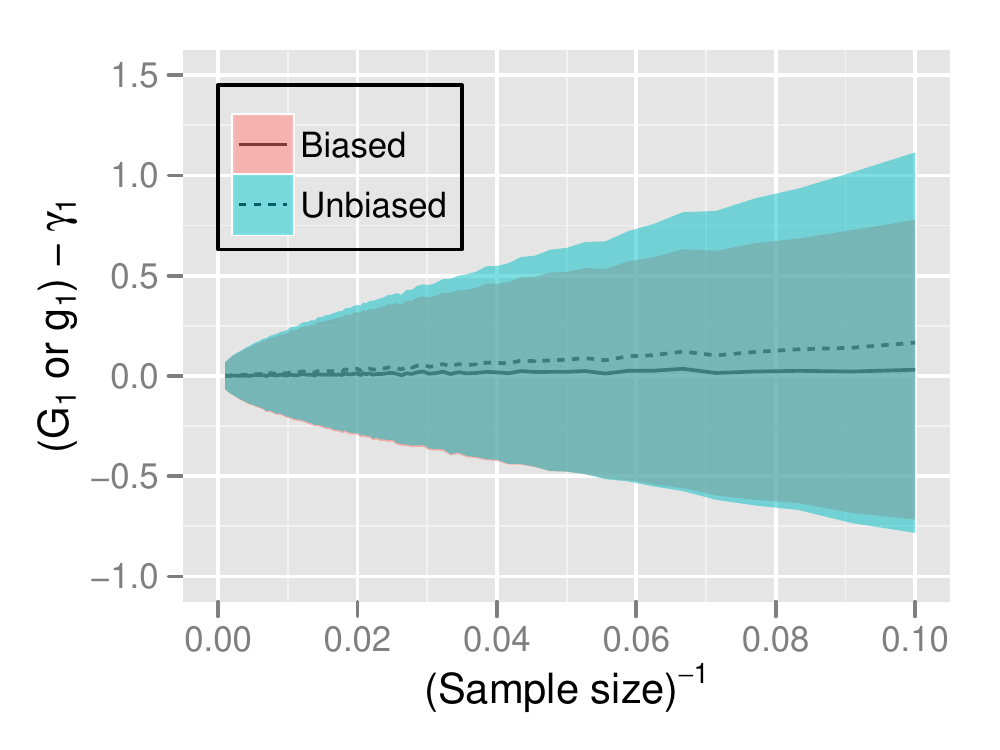}
\end{minipage}
\caption{Sample ({\it `biased'}\,) skewness $m_3$ of $\sin^4\phi$ versus its population ({\it `unbiased'}\,) estimate $M_3$ for $n\in(10,1000)$ and $S/N=100$: unweighted in the upper panels and weighted by the inverse of squared measurement errors in the lower panels. 
Estimators labeled as `unbiased' but involving ratios or powers of unbiased estimators are not expected to remain unbiased.
Shaded areas encompass one standard deviation from the mean of the distribution of the skewness employing simulations defined by Eqs~(\ref{eq:simuStart})--(\ref{eq:simuCoreEnd}). }
\label{fig:M3_100b}
\end{figure}

\begin{figure}
\begin{center}
~~~~~~~~{\bf\fbox{\parbox{0.15\textwidth}{\centering Skewness \\ $(\sin^4 \phi)$}}}\\
\end{center}
\begin{minipage}{0.5\columnwidth}
\centering
~~~~~~~~Phase Weighted ($a,b\rightarrow 0$)\\
\includegraphics[width=\columnwidth]{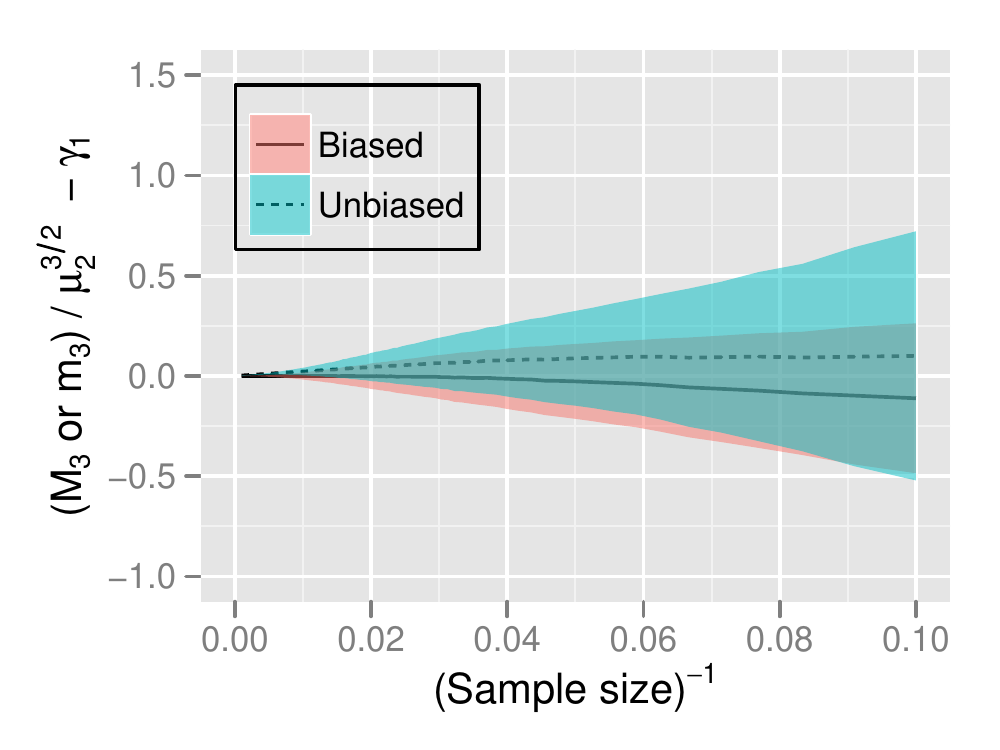}\\
~~~~~~~~Phase Weighted ($a=25,b=6$)\\
\includegraphics[width=\columnwidth]{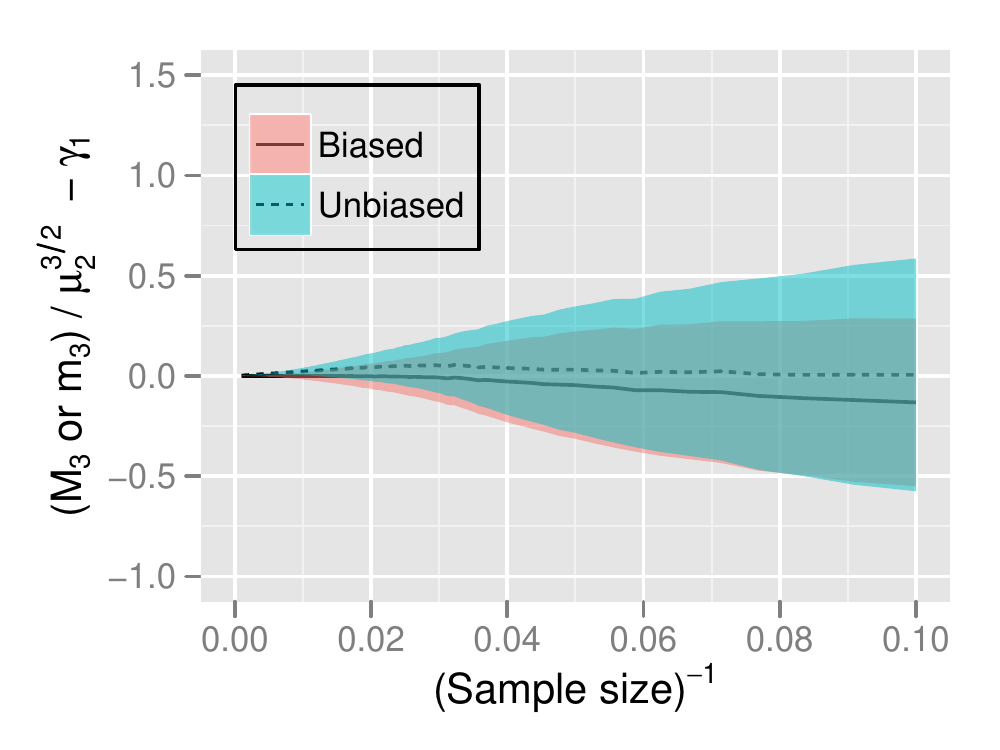}
\end{minipage}
\begin{minipage}{0.5\columnwidth}
\centering
~~~~~~~~Phase Weighted ($a,b\rightarrow 0$)\\
\includegraphics[width=\columnwidth]{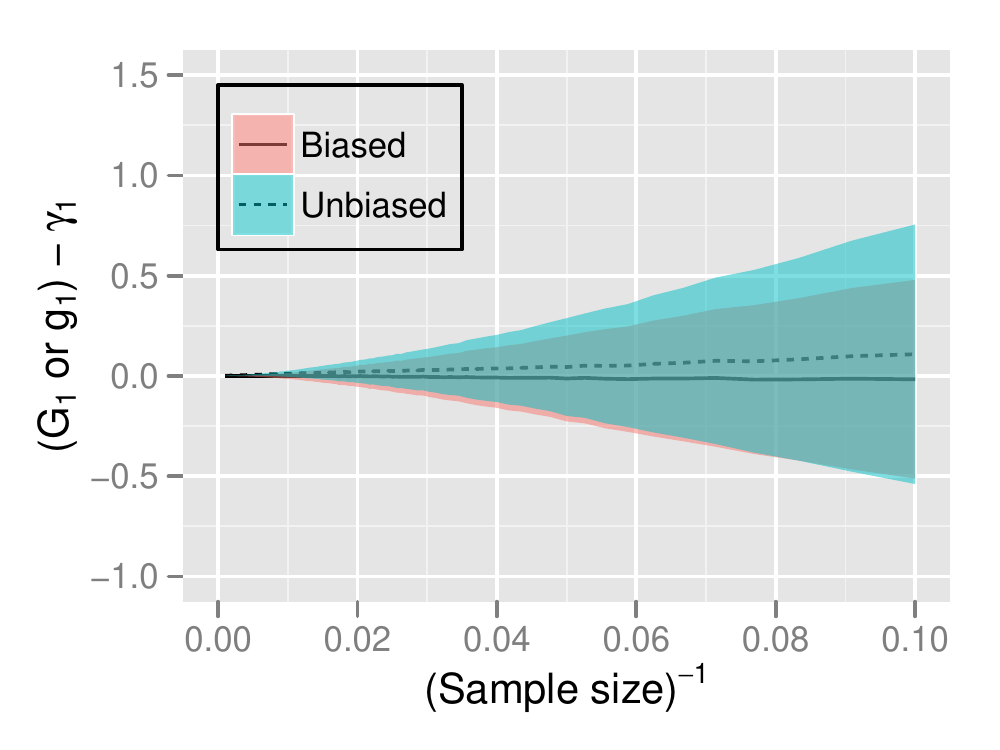}\\
~~~~~~~~Phase Weighted ($a=25,b=6$)\\
\includegraphics[width=\columnwidth]{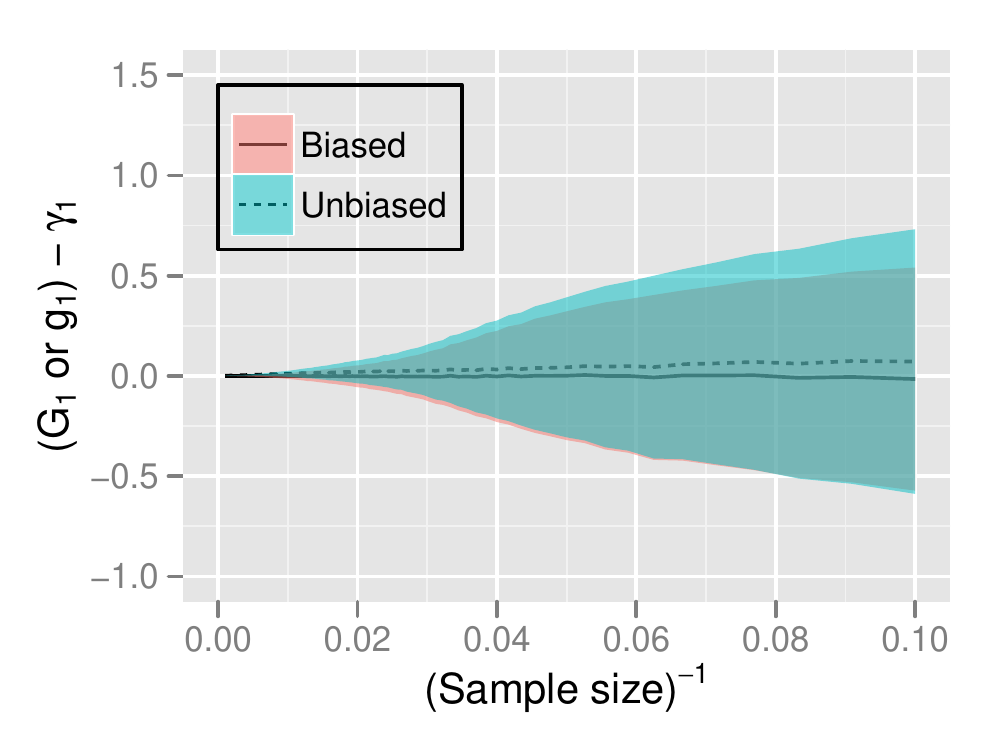}
\end{minipage}
\caption{Sample ({\it `biased'}\,) skewness $m_3$ of $\sin^4\phi$ versus its population ({\it `unbiased'}\,) estimate $M_3$ for $n\in(10,1000)$ and $S/N=100$, weighted by phase gaps, as defined by Eq.~(\ref{eq:phaseGap}), with different parameter values, as specified above each panel. 
The correlations introduced by weights are expected to bias the otherwise `unbiased' skewness. Also, estimators labeled as `unbiased' but involving ratios or powers of unbiased estimators are not expected to remain unbiased.
 Shaded areas encompass one standard deviation from the mean of the distribution of the skewness employing simulations defined by Eqs~(\ref{eq:simuStart})--(\ref{eq:simuCoreEnd}). }
\label{fig:M3_100wb}
\end{figure}

\begin{figure}
\begin{center}
~~~~~~~~{\bf\fbox{\parbox{0.15\textwidth}{\centering Kurtosis \\ $(\sin \phi)$}}}\\
\end{center}
\begin{minipage}{0.5\columnwidth}
\centering
~~~~~~Unweighted  \\
\includegraphics[width=\columnwidth]{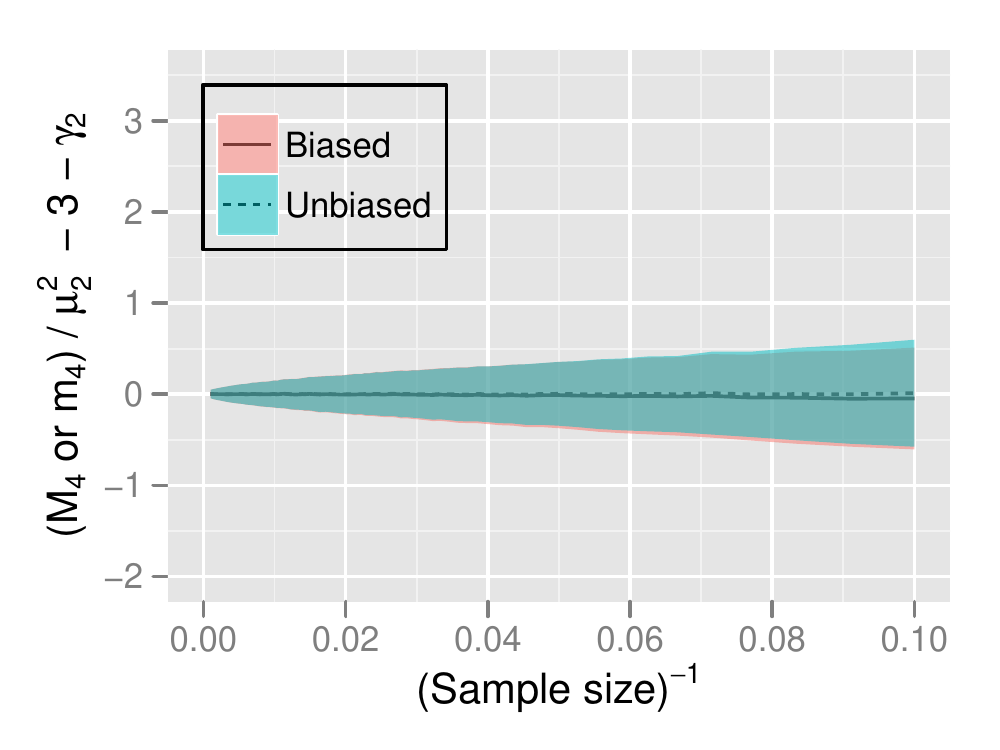}\\
~~~~~~~~Error Weighted \\
\includegraphics[width=\columnwidth]{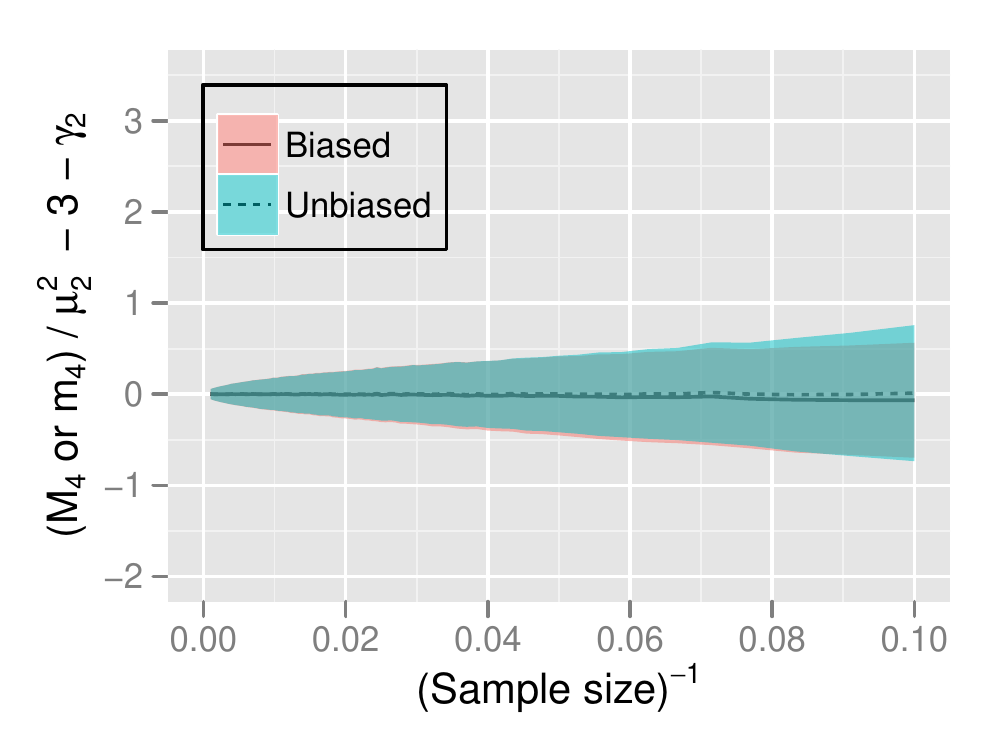}
\end{minipage}
\begin{minipage}{0.5\columnwidth}
\centering
~~~~~~Unweighted  \\
\includegraphics[width=\columnwidth]{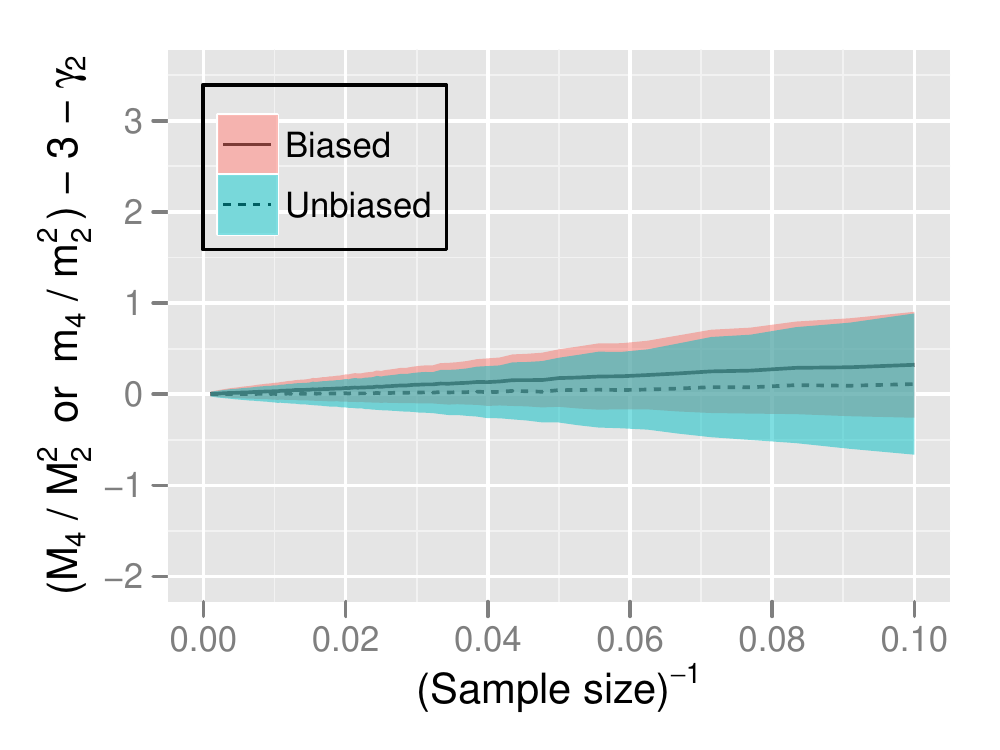}\\
~~~~~~~~Error Weighted\\
\includegraphics[width=\columnwidth]{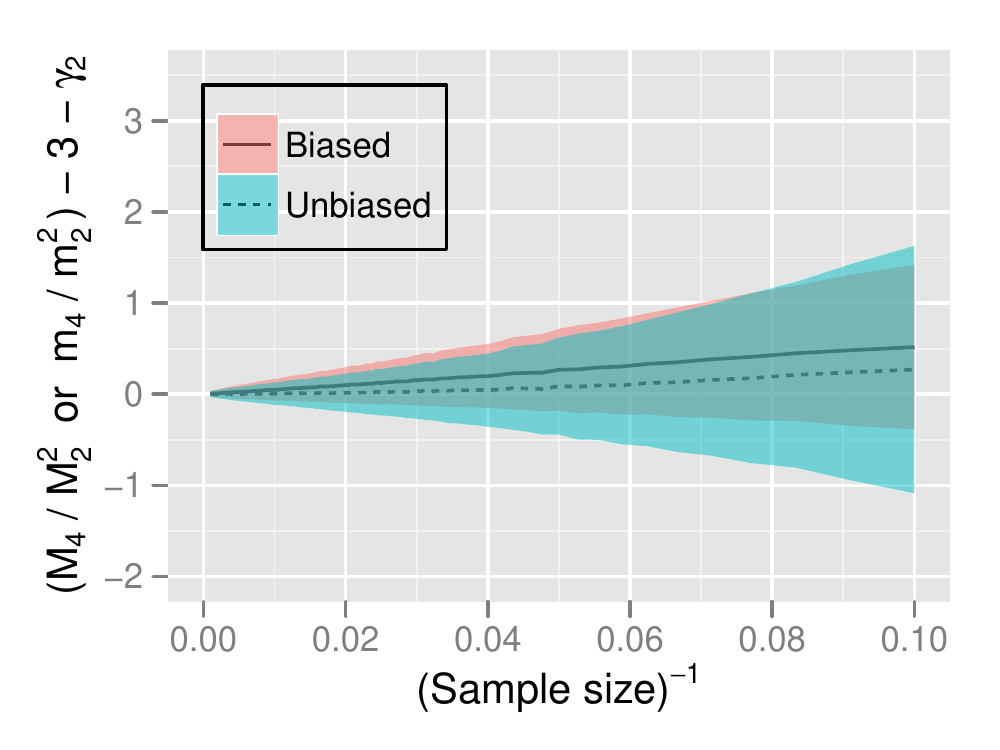}
\end{minipage}
\caption{Sample ({\it `biased'}\,) kurtosis $m_4$ of $\sin\phi$ versus its population ({\it `unbiased'}\,) estimate $M_4$ for $n\in(10,1000)$ and $S/N=100$: unweighted in the upper panels and weighted by the inverse of squared measurement errors in the lower panels.  
Estimators labeled as `unbiased' but involving ratios or powers of unbiased estimators are not expected to remain unbiased.
Shaded areas encompass one standard deviation from the mean of the distribution of the kurtosis employing simulations defined by Eqs~(\ref{eq:simuStart})--(\ref{eq:simuCoreEnd}). }
\label{fig:M4_100}
\end{figure}

\begin{figure}
\begin{center}
~~~~~~~~{\bf\fbox{\parbox{0.15\textwidth}{\centering Kurtosis \\ $(\sin \phi)$}}}\\
\end{center}
\begin{minipage}{0.5\columnwidth}
\centering
~~~~~~~~Phase Weighted ($a,b\rightarrow 0$)\\
\includegraphics[width=\columnwidth]{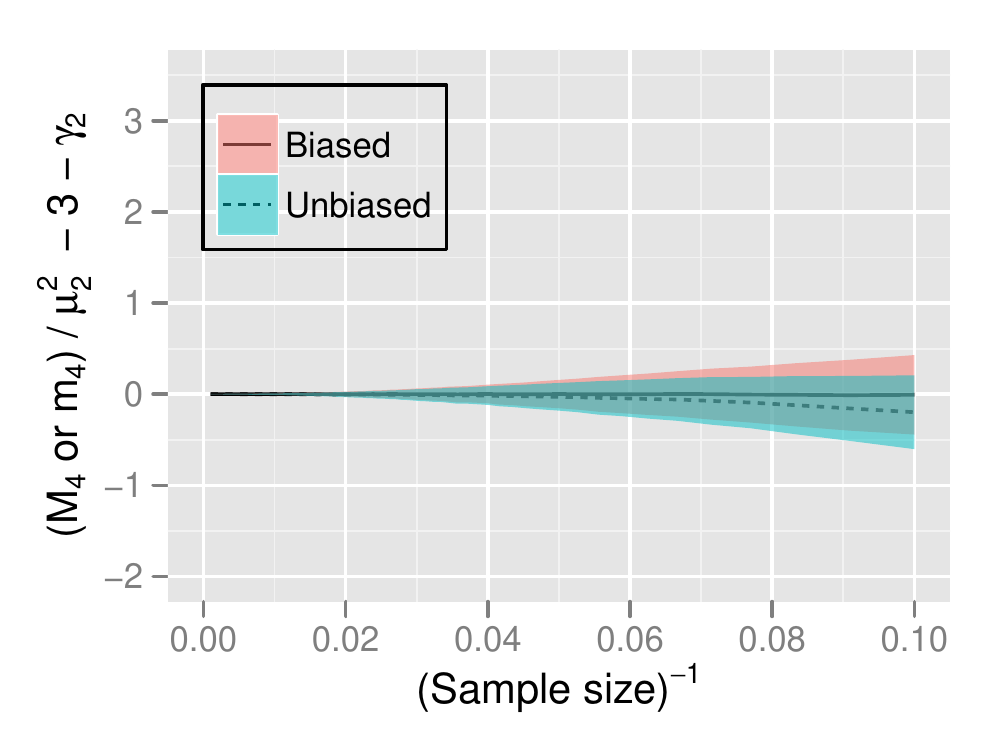}\\
~~~~~~~~Phase Weighted ($a=25,b=6$)\\
\includegraphics[width=\columnwidth]{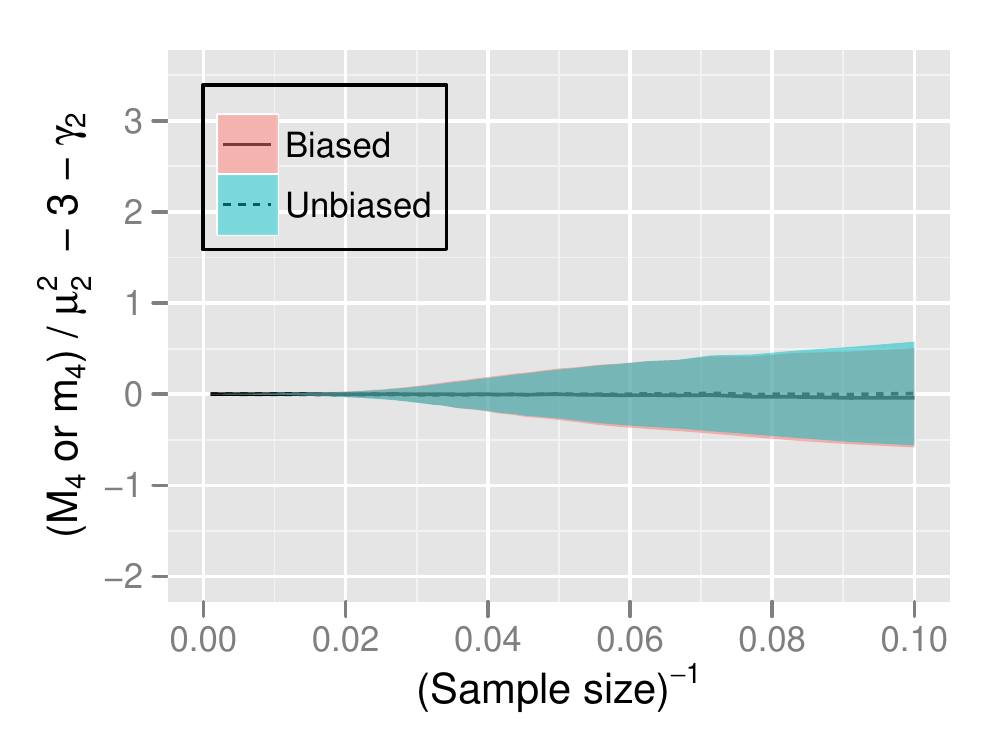}
\end{minipage}
\begin{minipage}{0.5\columnwidth}
\centering
~~~~~~~~Phase Weighted ($a,b\rightarrow 0$)\\
\includegraphics[width=\columnwidth]{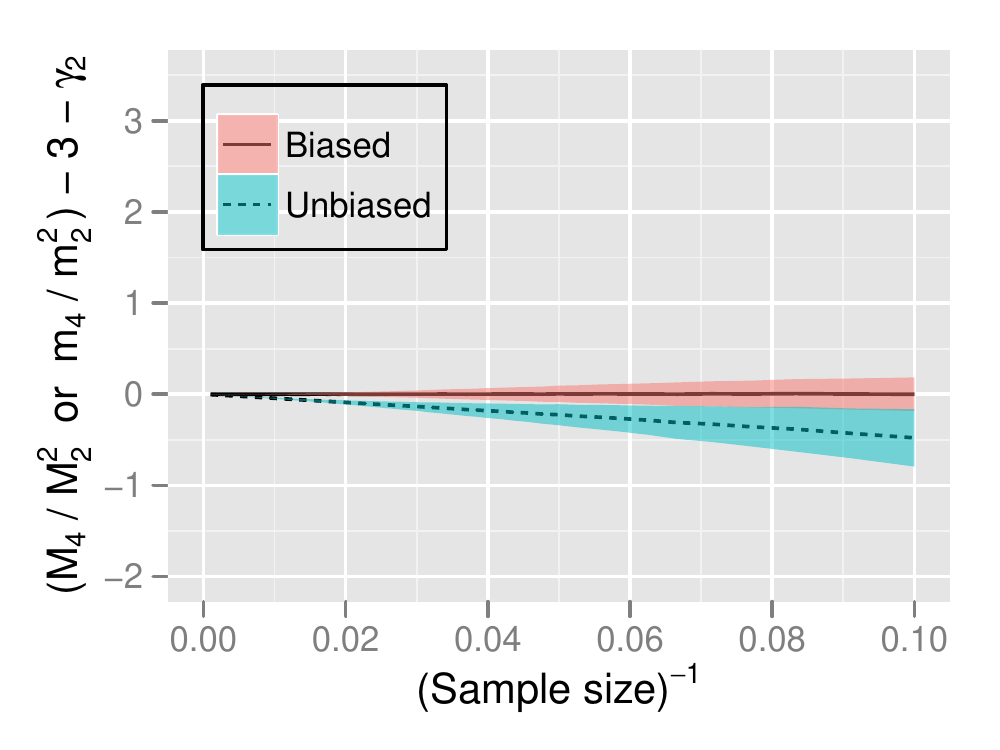}\\
~~~~~~~~Phase Weighted ($a=25,b=6$)\\
\includegraphics[width=\columnwidth]{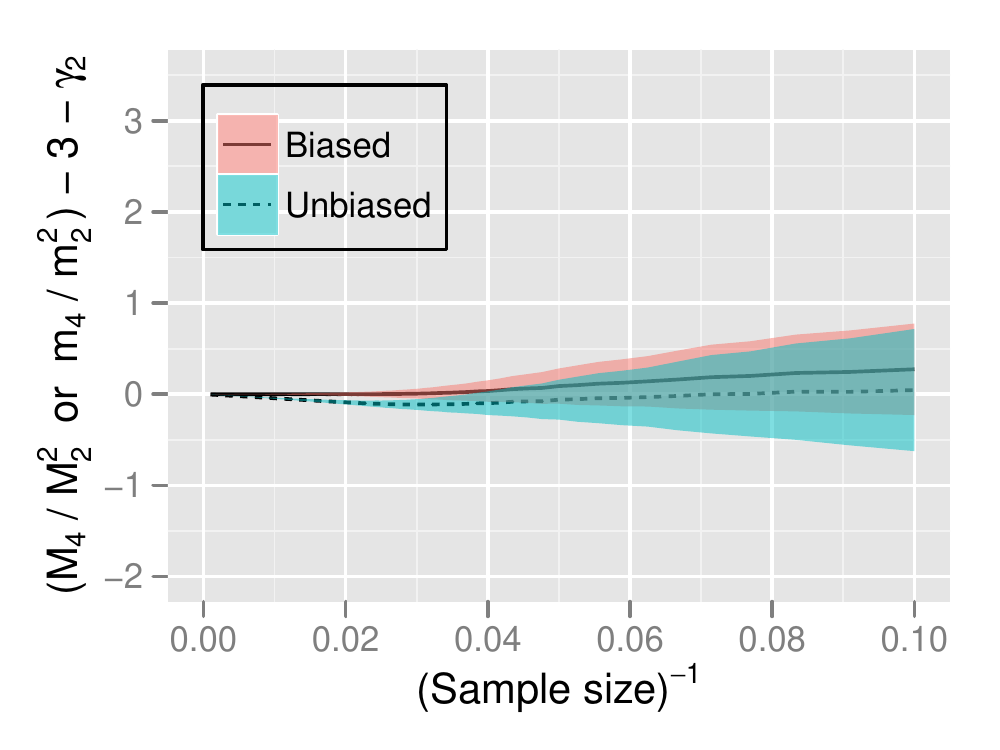}
\end{minipage}
\caption{Sample ({\it `biased'}\,) kurtosis $m_4$ of $\sin\phi$ versus its population ({\it `unbiased'}\,) estimate $M_4$ for $n\in(10,1000)$ and $S/N=100$, weighted by phase gaps, as defined by Eq.~(\ref{eq:phaseGap}), with different parameter values, as specified above each panel. 
The correlations introduced by weights are expected to bias the otherwise `unbiased' kurtosis. Also, estimators labeled as `unbiased' but involving ratios or powers of unbiased estimators are not expected to remain unbiased.
Shaded areas encompass one standard deviation from the mean of the distribution of the kurtosis employing simulations defined by Eqs~(\ref{eq:simuStart})--(\ref{eq:simuCoreEnd}).  }
\label{fig:M4_100w}
\end{figure}

\begin{figure}
\begin{center}
~~~~~~~~{\bf\fbox{\parbox{0.15\textwidth}{\centering Kurtosis \\ $(\sin^4 \phi)$}}}\\
\end{center}
\begin{minipage}{0.5\columnwidth}
\centering
~~~~~~Unweighted  \\
\includegraphics[width=\columnwidth]{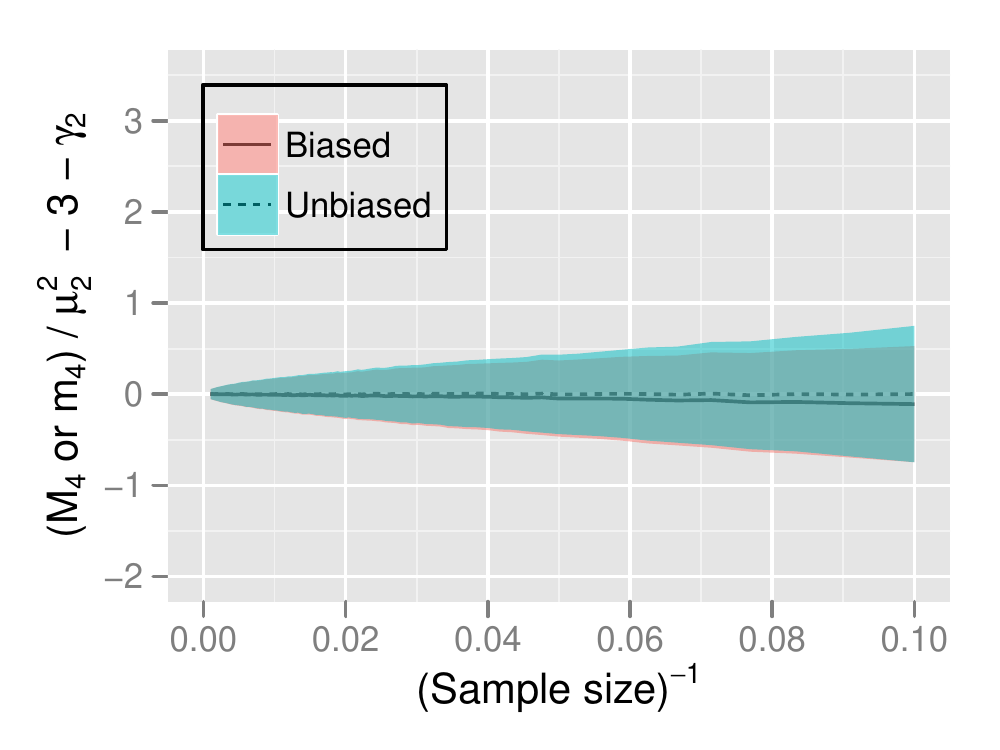}\\
~~~~~~~~Error Weighted \\
\includegraphics[width=\columnwidth]{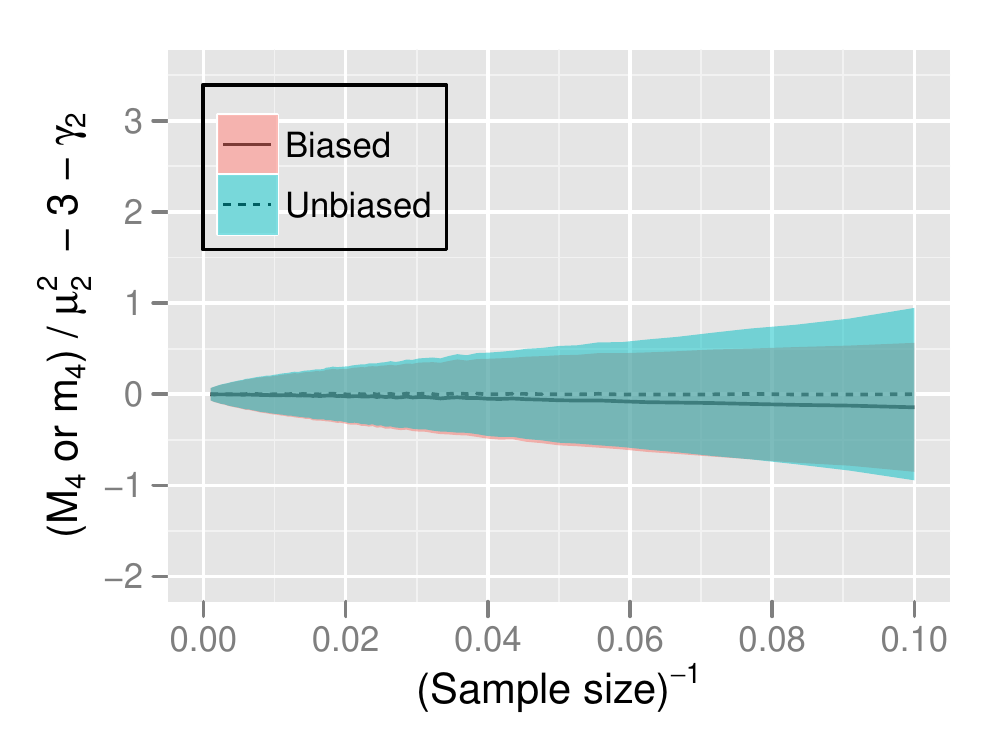}
\end{minipage}
\begin{minipage}{0.5\columnwidth}
\centering
~~~~~~Unweighted  \\
\includegraphics[width=\columnwidth]{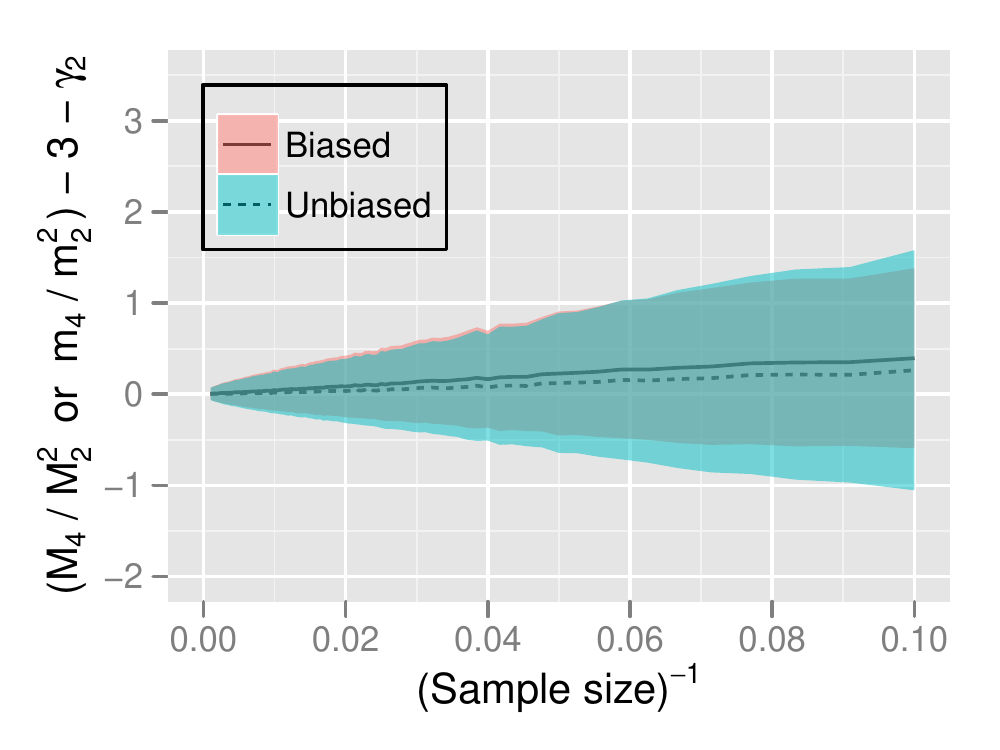}\\
~~~~~~~~Error Weighted\\
\includegraphics[width=\columnwidth]{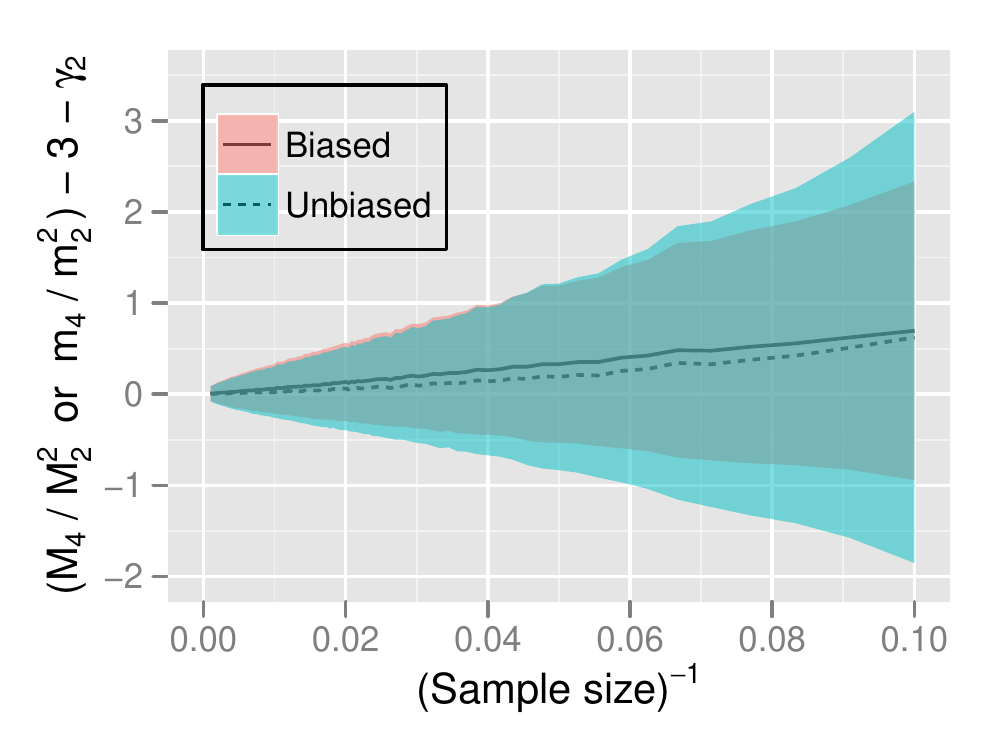}
\end{minipage}
\caption{Sample ({\it `biased'}\,) kurtosis $m_4$ of $\sin^4\phi$ versus its population ({\it `unbiased'}\,) estimate $M_4$ for $n\in(10,1000)$ and $S/N=100$: unweighted in the upper panels and weighted by the inverse of squared measurement errors in the lower panels.  
Estimators labeled as `unbiased' but involving ratios or powers of unbiased estimators are not expected to remain unbiased.
Shaded areas encompass one standard deviation from the mean of the distribution of the kurtosis employing simulations defined by Eqs~(\ref{eq:simuStart})--(\ref{eq:simuCoreEnd}). }
\label{fig:M4_100b}
\end{figure}

\begin{figure}
\begin{center}
~~~~~~~~{\bf\fbox{\parbox{0.15\textwidth}{\centering Kurtosis \\ $(\sin^4 \phi)$}}}\\
\end{center}
\begin{minipage}{0.5\columnwidth}
\centering
~~~~~~~~Phase Weighted ($a,b\rightarrow 0$)\\
\includegraphics[width=\columnwidth]{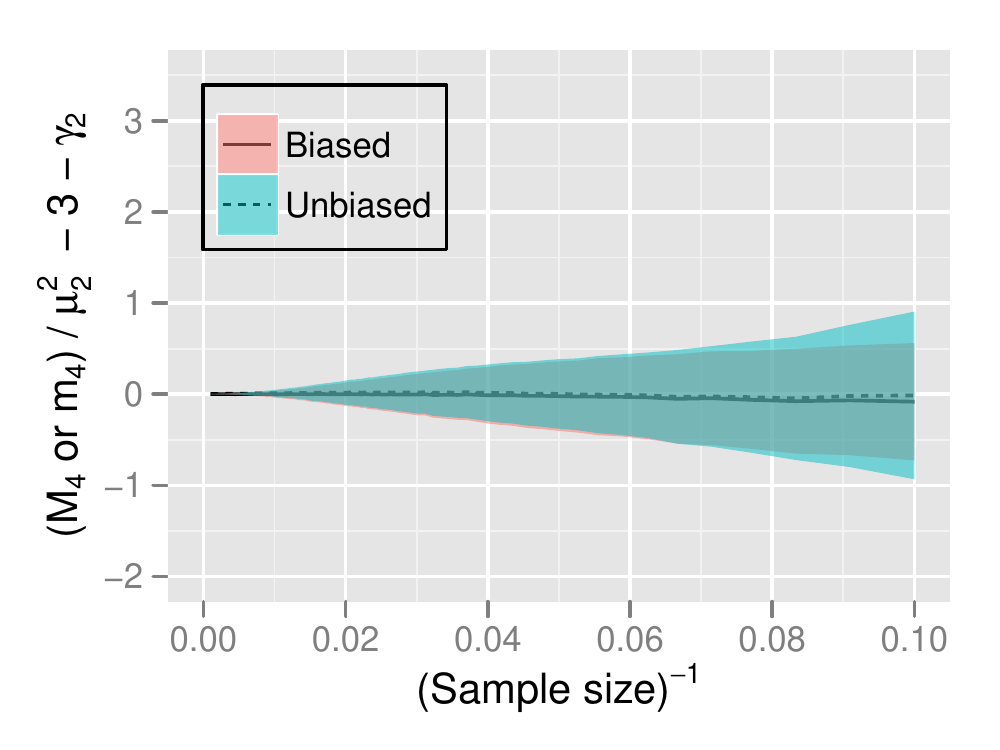}\\
~~~~~~~~Phase Weighted ($a=25,b=6$)\\
\includegraphics[width=\columnwidth]{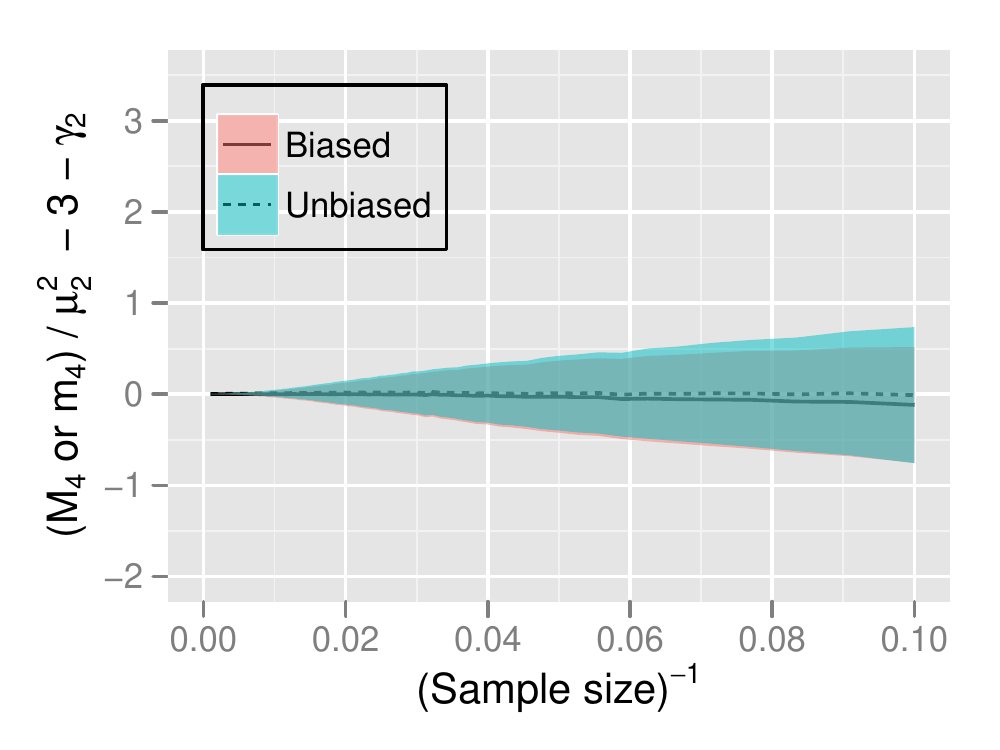}
\end{minipage}
\begin{minipage}{0.5\columnwidth}
\centering
~~~~~~~~Phase Weighted ($a,b\rightarrow 0$)\\
\includegraphics[width=\columnwidth]{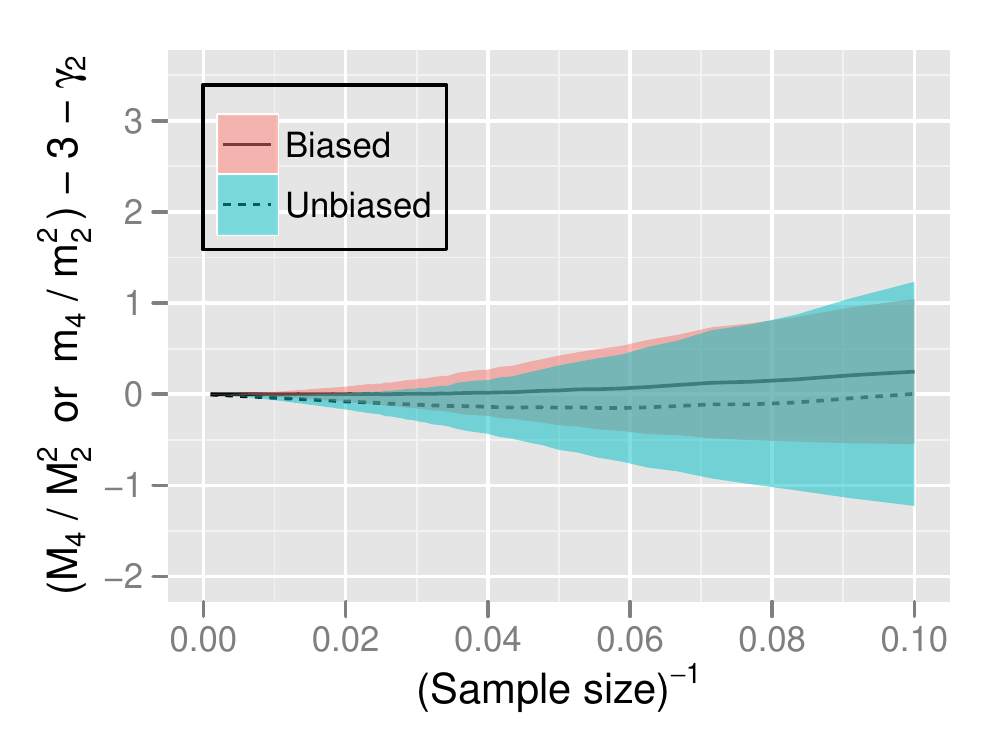}\\
~~~~~~~~Phase Weighted ($a=25,b=6$)\\
\includegraphics[width=\columnwidth]{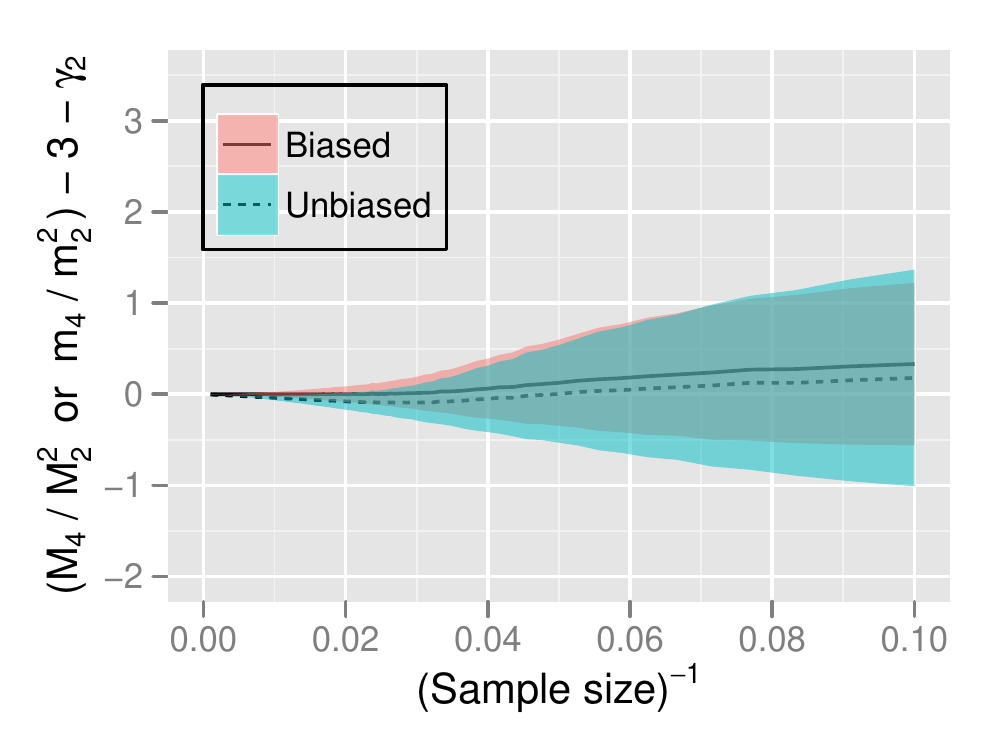}
\end{minipage}
\caption{Sample ({\it `biased'}\,) kurtosis $m_4$ of $\sin^4\phi$  versus its population ({\it `unbiased'}\,) estimate $M_4$ for $n\in(10,1000)$ and $S/N=100$, weighted by phase gaps, as defined by Eq.~(\ref{eq:phaseGap}), with different parameter values, as specified above each panel.
The correlations introduced by weights are expected to bias the otherwise `unbiased' kurtosis. Also, estimators labeled as `unbiased' but involving ratios or powers of unbiased estimators are not expected to remain unbiased.
 Shaded areas encompass one standard deviation from the mean of the distribution of the kurtosis employing simulations defined by Eqs~(\ref{eq:simuStart})--(\ref{eq:simuCoreEnd}).  }
\label{fig:M4_100wb}
\end{figure}

\begin{figure}
\begin{center}
~~~~~~~~{\bf\fbox{\parbox{0.15\textwidth}{\centering  {\em k-}Kurtosis \\ $(\sin \phi)$}}}\\
\end{center}
\begin{minipage}{0.5\columnwidth}
\centering
~~~~~~Unweighted  \\
\includegraphics[width=\columnwidth]{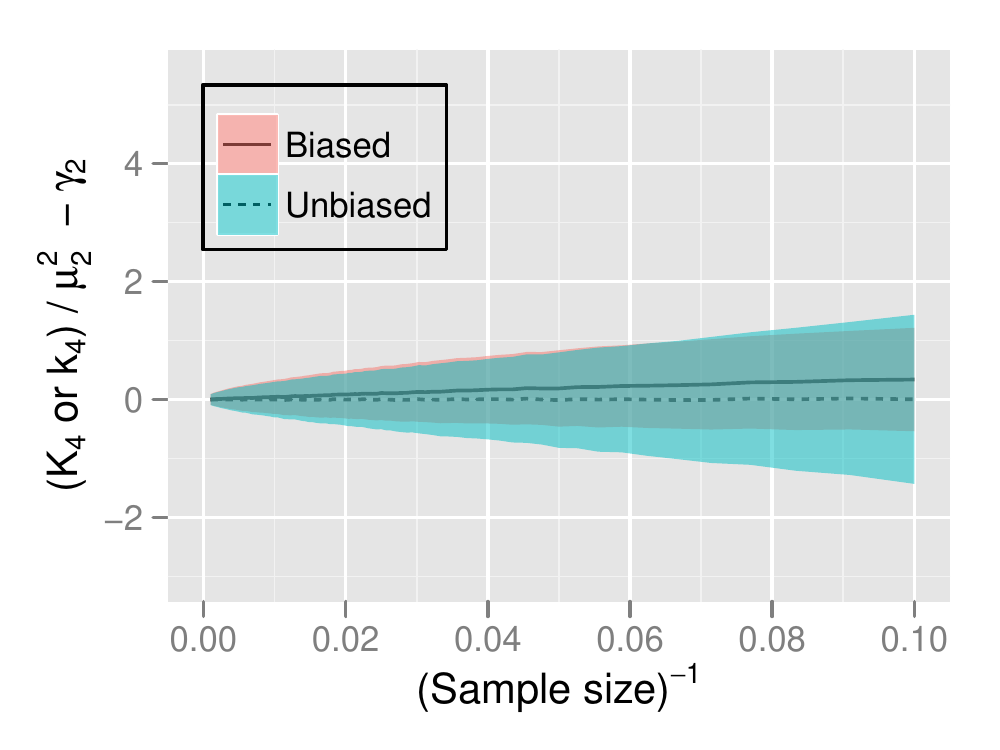}\\
~~~~~~~~Error Weighted \\
\includegraphics[width=\columnwidth]{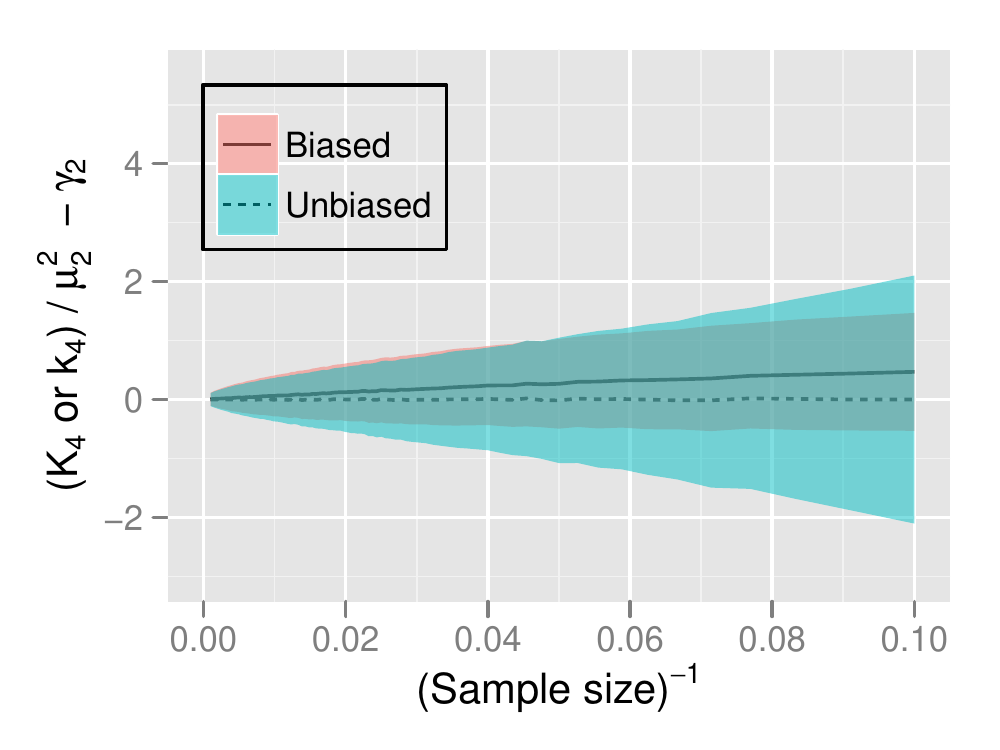}
\end{minipage}
\begin{minipage}{0.5\columnwidth}
\centering
~~~~~~Unweighted \\
\includegraphics[width=\columnwidth]{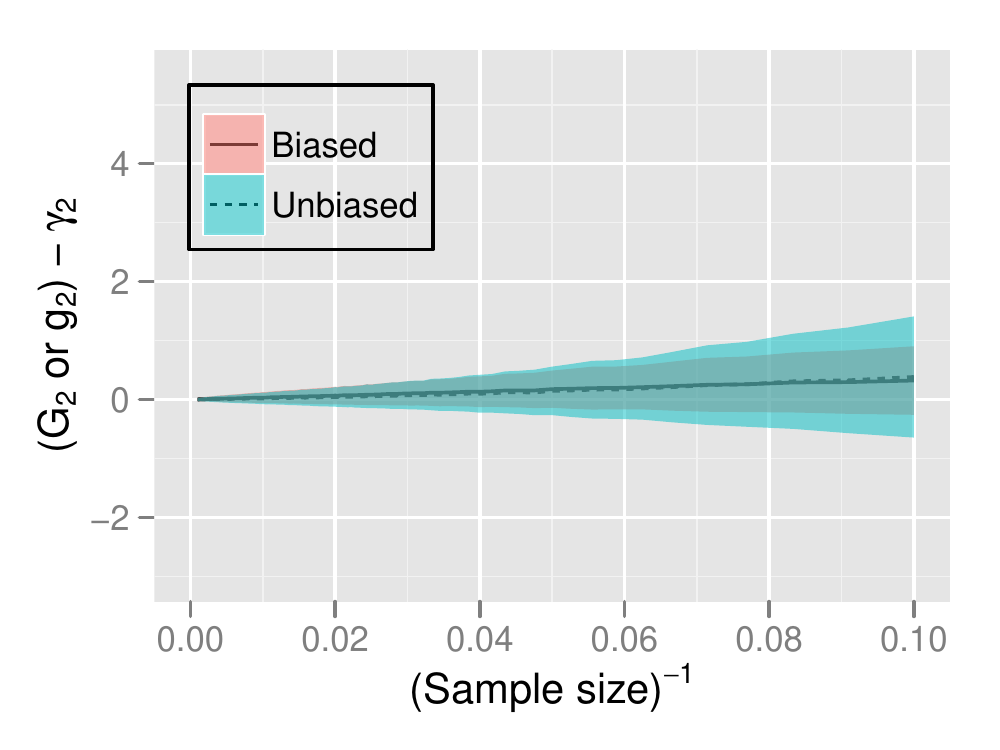}\\
~~~~~~~~Error Weighted \\
\includegraphics[width=\columnwidth]{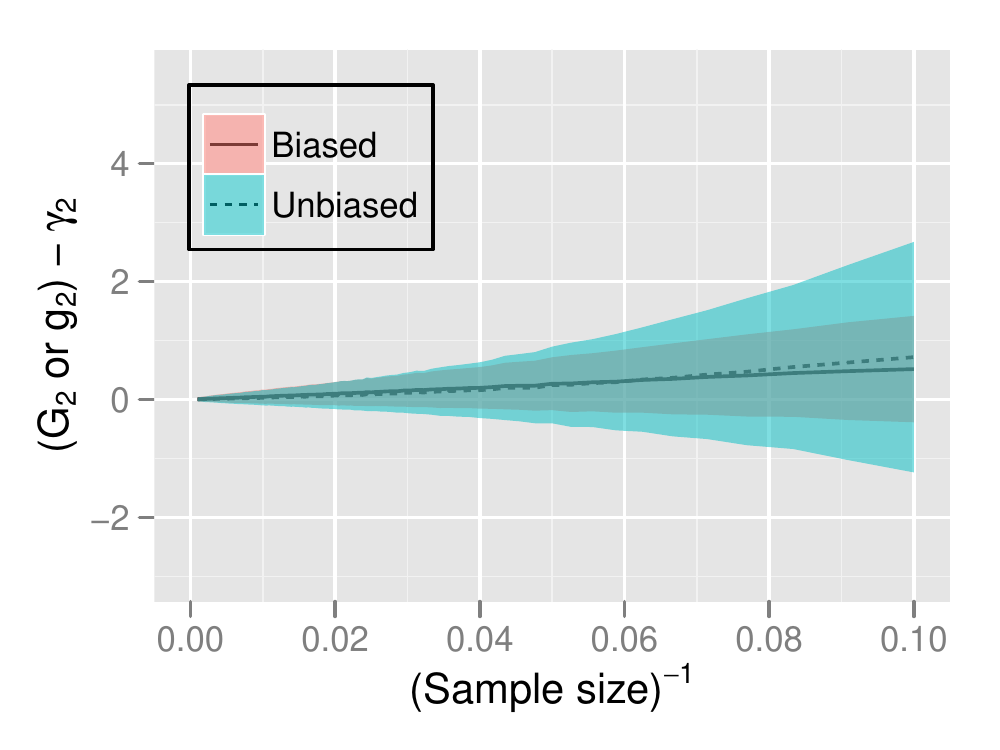}
\end{minipage}
\caption{Sample ({\it `biased'}\,) kurtosis $k_4$ of $\sin\phi$ versus its population ({\it `unbiased'}\,) estimate $K_4$ for $n\in(10,1000)$ and $S/N=100$: unweighted in the upper panels and weighted by the inverse of squared measurement errors in the lower panels.  
Estimators labeled as `unbiased' but involving ratios or powers of unbiased estimators are not expected to remain unbiased.
Shaded areas encompass one standard deviation from the mean of the distribution of the kurtosis employing simulations defined by Eqs~(\ref{eq:simuStart})--(\ref{eq:simuCoreEnd}). }
\label{fig:K4_100}
\end{figure}

\begin{figure}
\begin{center}
~~~~~~~~{\bf\fbox{\parbox{0.15\textwidth}{\centering  {\em k-}Kurtosis \\ $(\sin \phi)$}}}\\
\end{center}
\begin{minipage}{0.5\columnwidth}
\centering
~~~~~~~~Phase Weighted ($a,b\rightarrow 0$)\\
\includegraphics[width=\columnwidth]{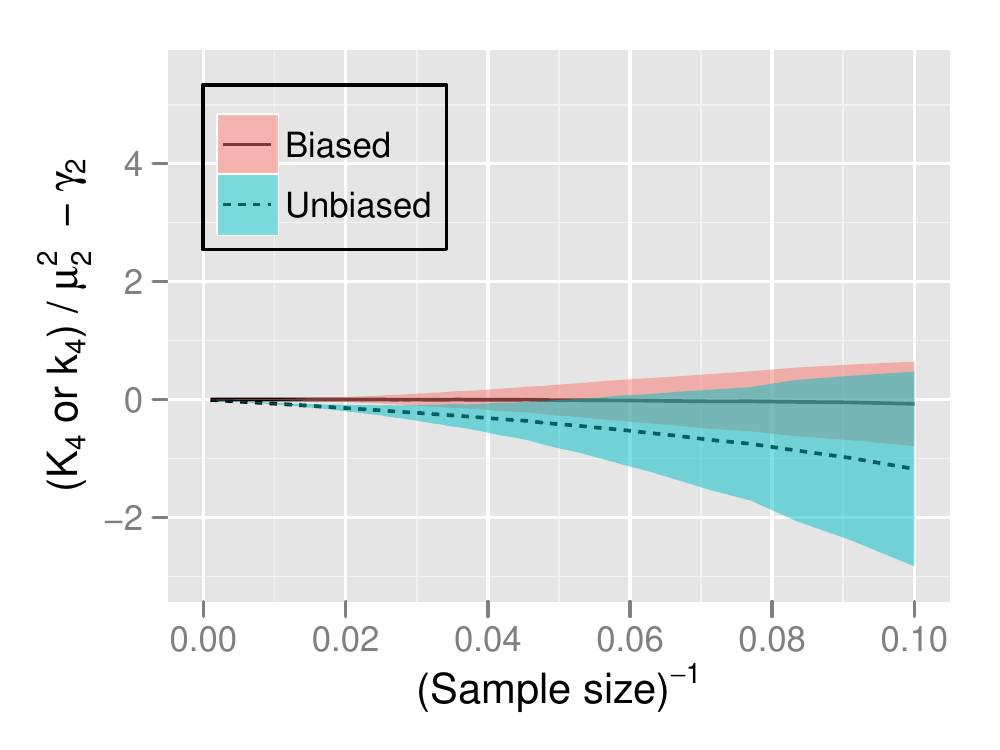}\\
~~~~~~~~Phase Weighted ($a=25,b=6$)\\
\includegraphics[width=\columnwidth]{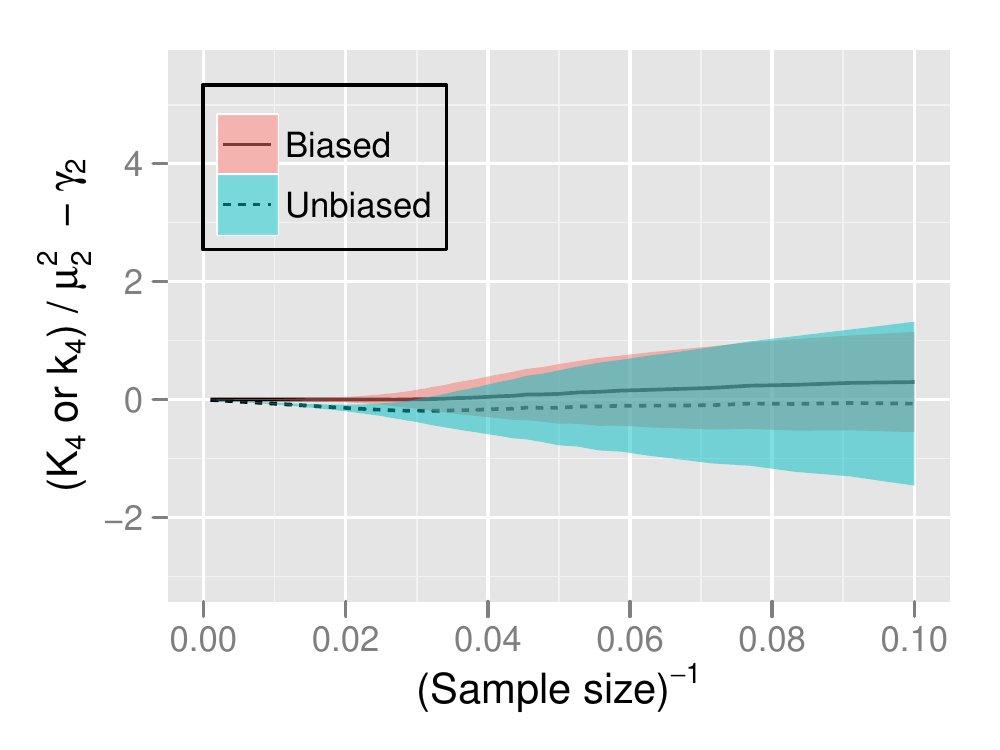}
\end{minipage}
\begin{minipage}{0.5\columnwidth}
\centering
~~~~~~~~Phase Weighted ($a,b\rightarrow 0$)\\
\includegraphics[width=\columnwidth]{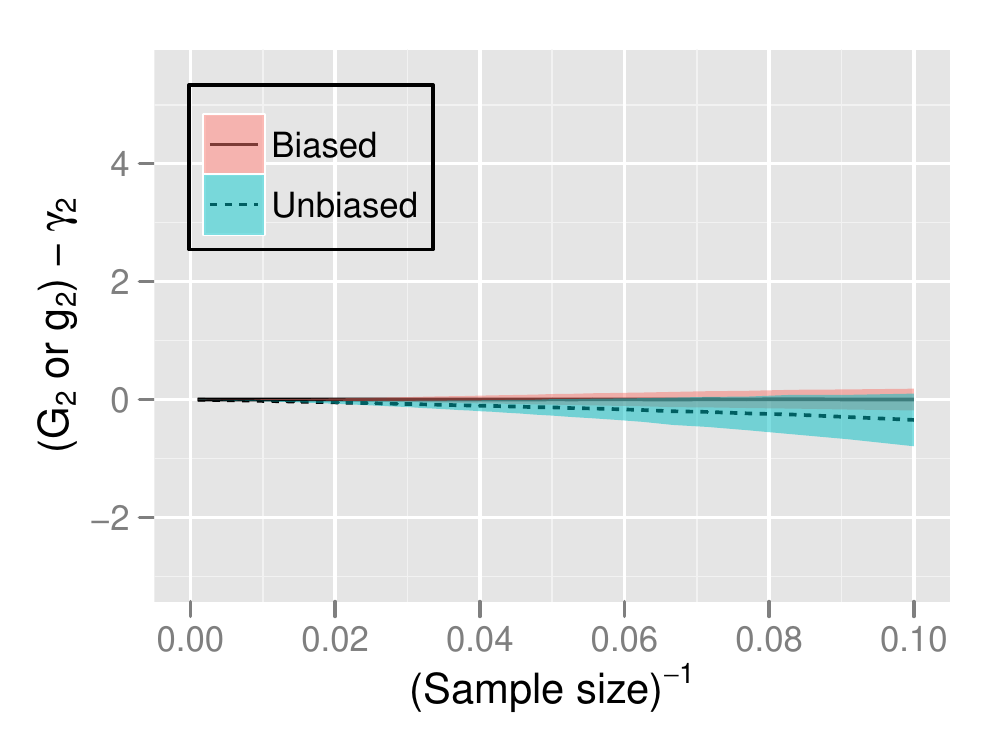}\\
~~~~~~~~Phase Weighted ($a=25,b=6$)\\
\includegraphics[width=\columnwidth]{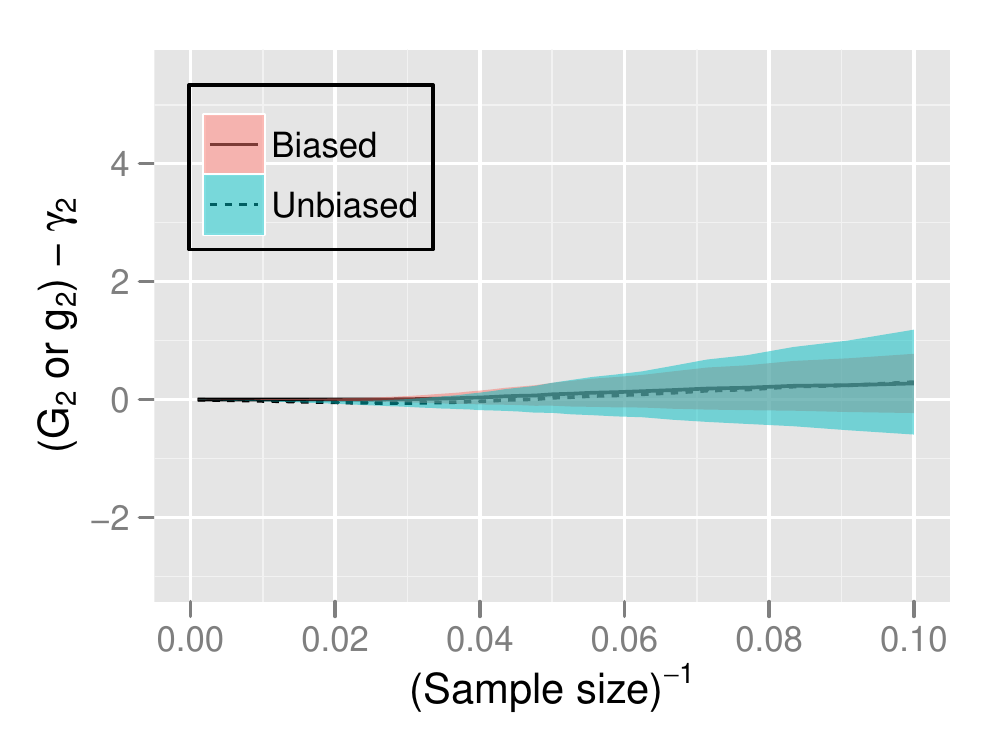}
\end{minipage}
\caption{Sample ({\it `biased'}\,) kurtosis $k_4$ of $\sin\phi$ versus its population ({\it `unbiased'}\,) estimate $K_4$ for $n\in(10,1000)$ and $S/N=100$, weighted by phase gaps, as defined by Eq.~(\ref{eq:phaseGap}), with different parameter values, as specified above each panel. 
The correlations introduced by weights are expected to bias the otherwise `unbiased' kurtosis. Also, estimators labeled as `unbiased' but involving ratios or powers of unbiased estimators are not expected to remain unbiased.
Shaded areas encompass one standard deviation from the mean of the distribution of the kurtosis employing simulations defined by Eqs~(\ref{eq:simuStart})--(\ref{eq:simuCoreEnd}). }
\label{fig:K4_100w}
\end{figure}

\begin{figure}
\begin{center}
~~~~~~~~{\bf\fbox{\parbox{0.15\textwidth}{\centering  {\em k-}Kurtosis \\ $(\sin^4 \phi)$}}}\\
\end{center}
\begin{minipage}{0.5\columnwidth}
\centering
~~~~~~Unweighted  \\
\includegraphics[width=\columnwidth]{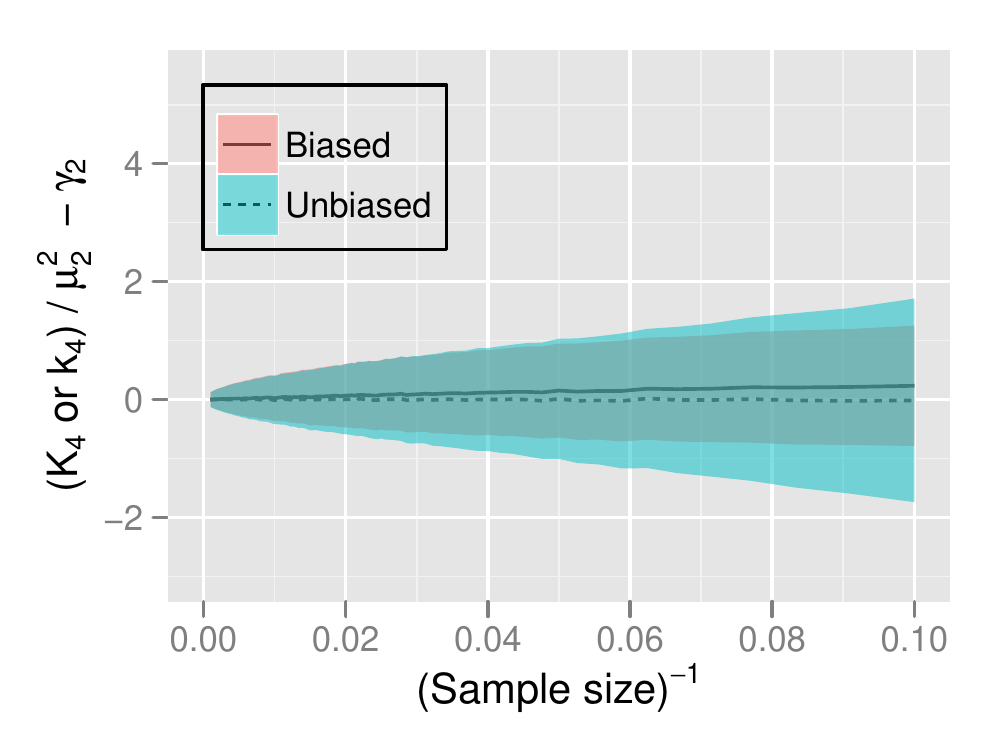}\\
~~~~~~~~Error Weighted \\
\includegraphics[width=\columnwidth]{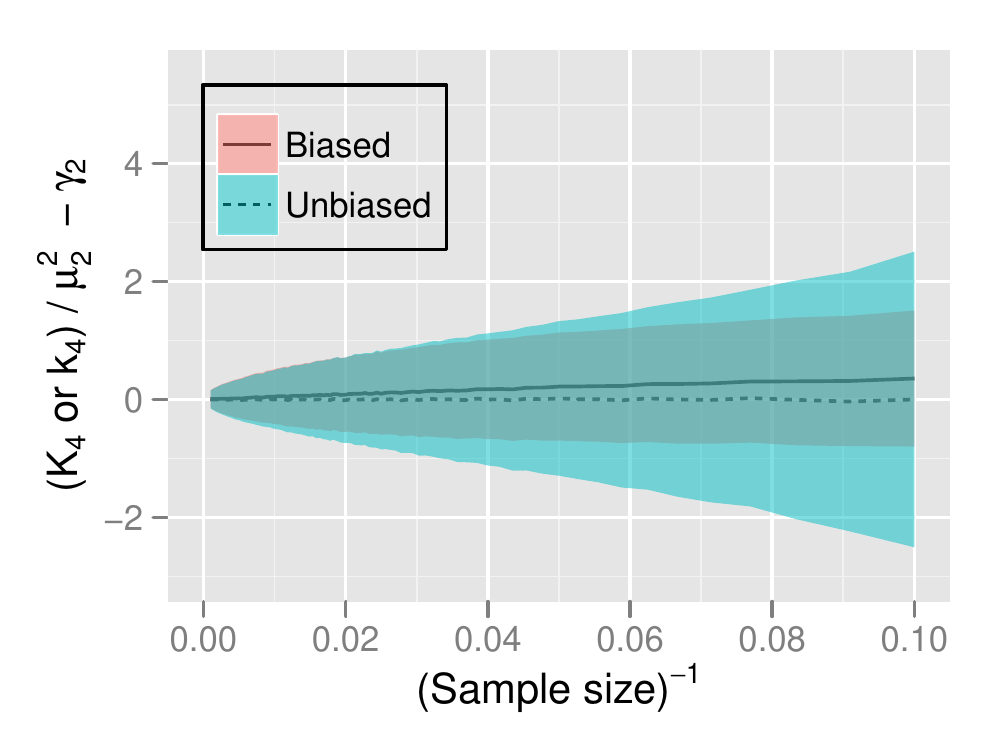}
\end{minipage}
\begin{minipage}{0.5\columnwidth}
\centering
~~~~~~Unweighted \\
\includegraphics[width=\columnwidth]{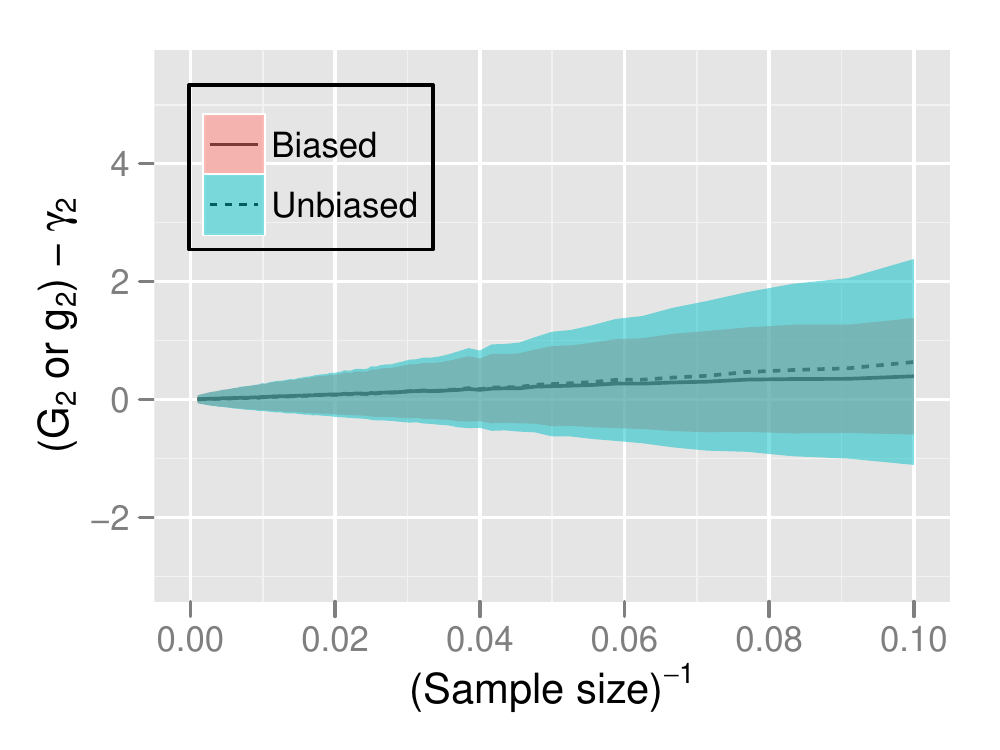}\\
~~~~~~~~Error Weighted \\
\includegraphics[width=\columnwidth]{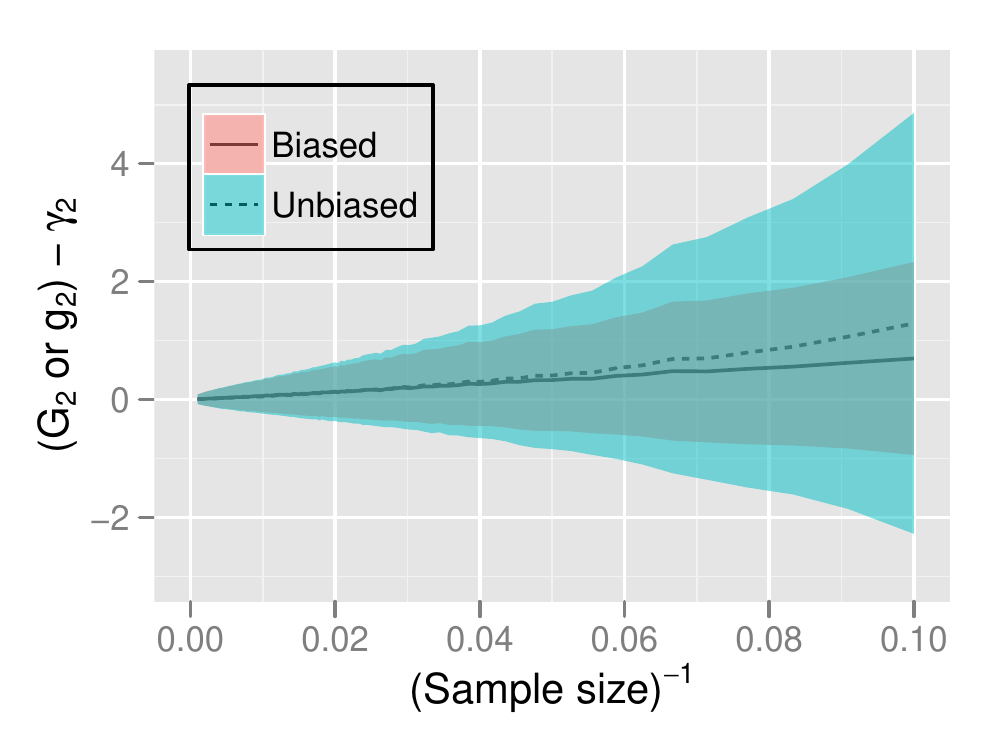}
\end{minipage}
\caption{Sample ({\it `biased'}\,) kurtosis $k_4$ of $\sin^4\phi$ versus its population ({\it `unbiased'}\,) estimate $K_4$ for $n\in(10,1000)$ and $S/N=100$: unweighted in the upper panels and weighted by the inverse of squared measurement errors in the lower panels.  
Estimators labeled as `unbiased' but involving ratios or powers of unbiased estimators are not expected to remain unbiased.
Shaded areas encompass one standard deviation from the mean of the distribution of the kurtosis employing simulations defined by Eqs~(\ref{eq:simuStart})--(\ref{eq:simuCoreEnd}). }
\label{fig:K4_100b}
\end{figure}

\begin{figure}
\begin{center}
~~~~~~~~{\bf\fbox{\parbox{0.15\textwidth}{\centering  {\em k-}Kurtosis \\ $(\sin^4 \phi)$}}}\\
\end{center}
\begin{minipage}{0.5\columnwidth}
\centering
~~~~~~~~Phase Weighted ($a,b\rightarrow 0$)\\
\includegraphics[width=\columnwidth]{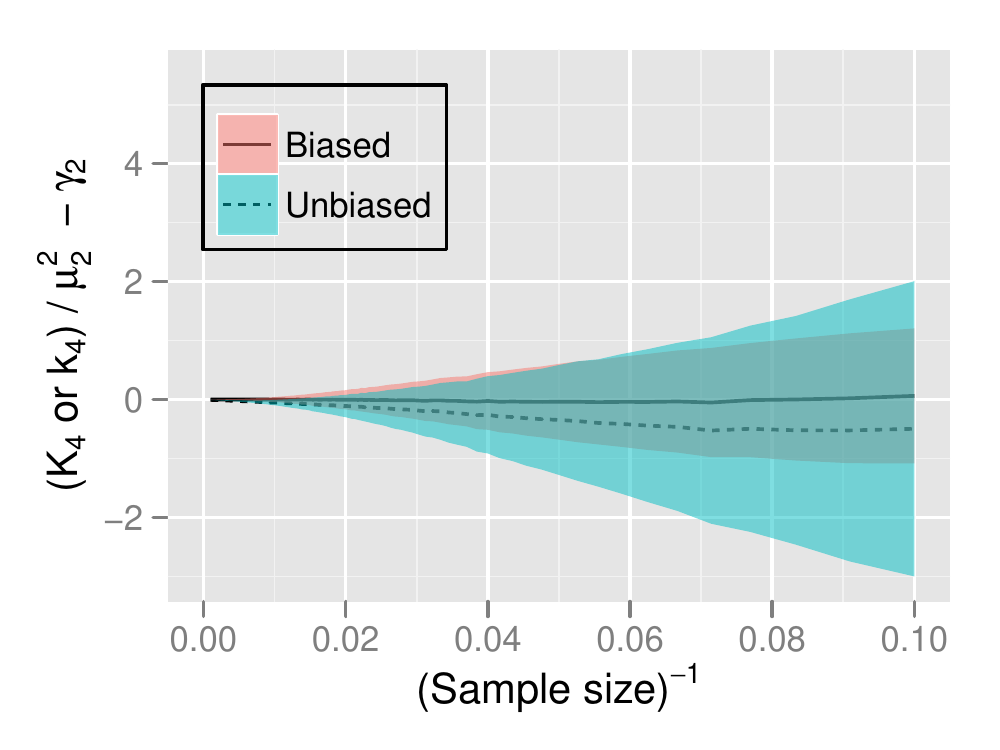}\\
~~~~~~~~Phase Weighted ($a=25,b=6$)\\
\includegraphics[width=\columnwidth]{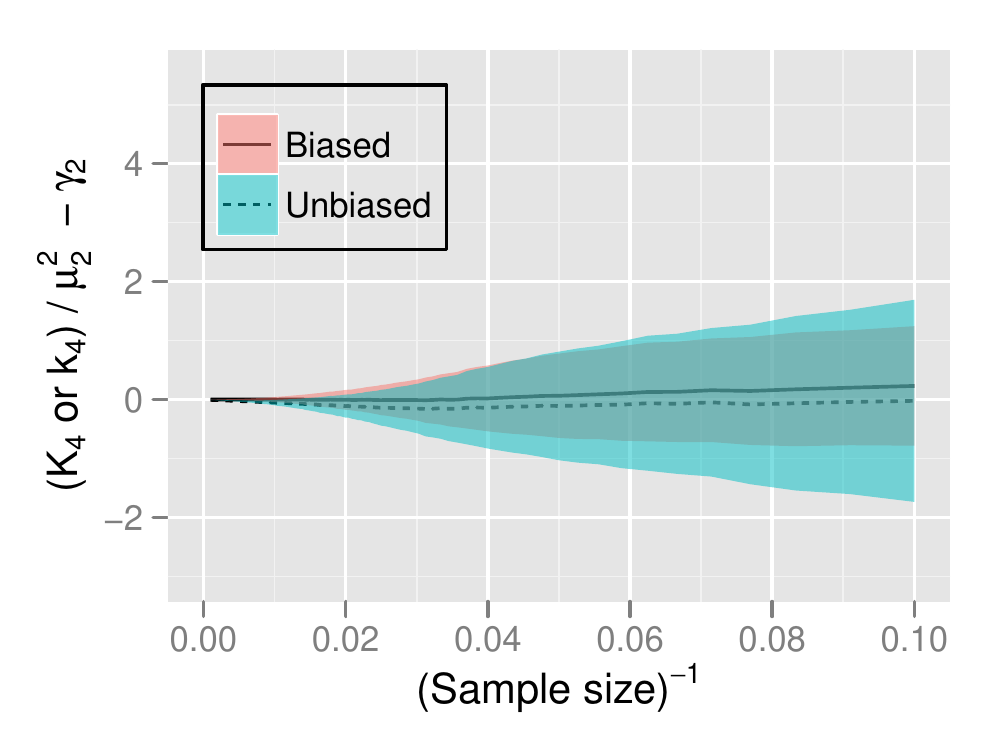}
\end{minipage}
\begin{minipage}{0.5\columnwidth}
\centering
~~~~~~~~Phase Weighted ($a,b\rightarrow 0$)\\
\includegraphics[width=\columnwidth]{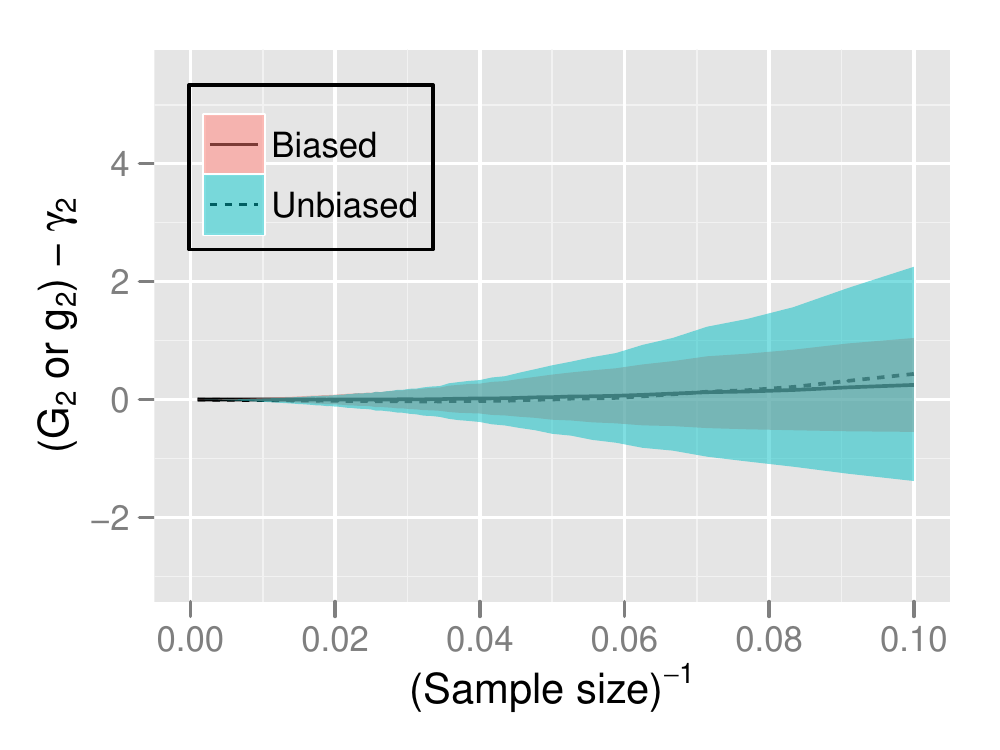}\\
~~~~~~~~Phase Weighted ($a=25,b=6$)\\
\includegraphics[width=\columnwidth]{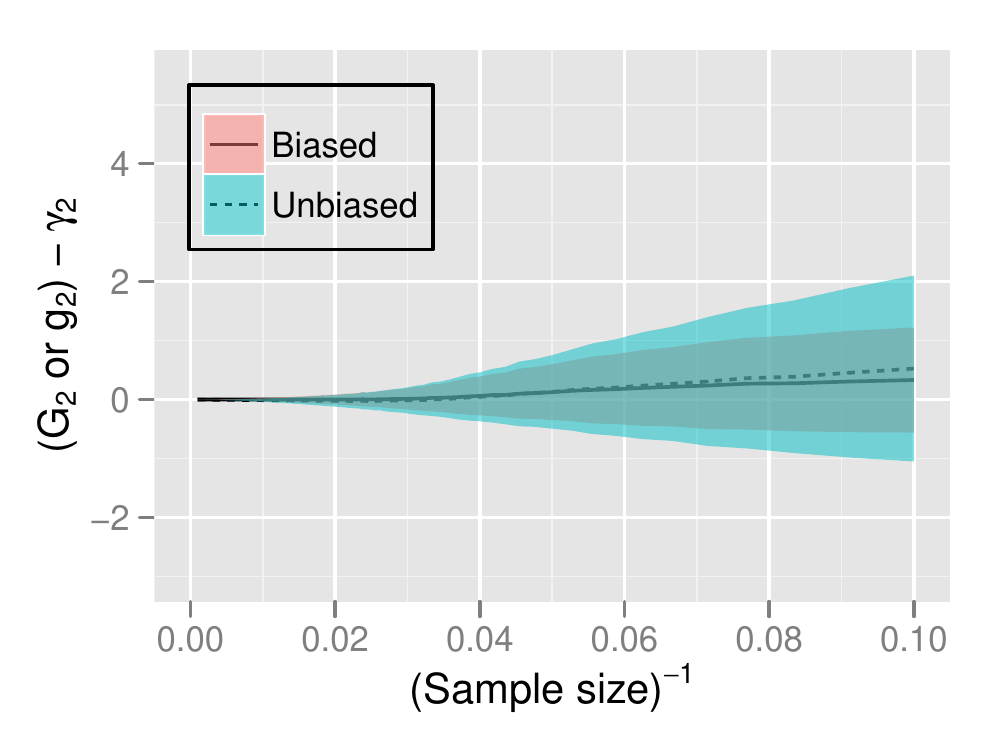}
\end{minipage}
\caption{Sample ({\it `biased'}\,) kurtosis $k_4$ of $\sin^4\phi$ versus its population ({\it `unbiased'}\,) estimate $K_4$ for $n\in(10,1000)$ and $S/N=100$, weighted by phase gaps, as defined by Eq.~(\ref{eq:phaseGap}), with different parameter values, as specified above each panel. 
The correlations introduced by weights are expected to bias the otherwise `unbiased' kurtosis. Also, estimators labeled as `unbiased' but involving ratios or powers of unbiased estimators are not expected to remain unbiased.
Shaded areas encompass one standard deviation from the mean of the distribution of the kurtosis employing simulations defined by Eqs~(\ref{eq:simuStart})--(\ref{eq:simuCoreEnd}). }
\label{fig:K4_100wb}
\end{figure}

Figures~\ref{fig:M2}--\ref{fig:K4_100wb} compare biased and unbiased estimators as a function of sample size, evaluating the following deviations from the population values:
\begin{align}
&M_2/\mu_2-1~~~~~~~~~\mbox{vs}~~m_2/\mu_2-1, \\
&M_3/\mu_2^{3/2}-\gamma_1~~~~~\mbox{vs}~~m_3/\mu_2^{3/2}-\gamma_1,~~~~~~G_1-\gamma_1~~~~~~~~~~~~~\mbox{vs}~~g_1-\gamma_1,\\
&M_4/\mu_2^2-3-\gamma_2~~\mbox{vs}~~m_4/\mu_2^2-3-\gamma_2,~~~M_4/M_2^2-3-\gamma_2~~\mbox{vs}~~m_4/m_2^2-3-\gamma_2,\\
&K_4/\mu_2^2-\gamma_2~~~~~~~\,\mbox{vs}~~k_4/\mu_2^2-\gamma_2,~~~~~~~~~\,G_2-\gamma_2~~~~~~~~~~~~~\,\mbox{vs}~~g_2-\gamma_2,
\end{align}
in both weighted and unweighted cases, for  signals of the forms $\sin\phi$ and $\sin^4\phi$, with $S/N=100$. 
Estimators standardized by both true and estimated variance are presented to help interpret the behaviour of the ratios from their components.

Biased and unbiased estimators are quite similar in the limit of large sample sizes (typically $n>100$). 
When weighting by  errors, or not weighting at all,  unbiased estimators are accurate  throughout the whole range of sample sizes, as expected, although their precision decreases at lower values of $n$. 
The biased counterparts are more precise but less accurate, with a degradation of both accuracy and precision at low $n$. 

Estimators which involve ratios (and powers) of unbiased estimators are not expected to be unbiased in general.
Figures ~\ref{fig:M3_100}--\ref{fig:M3_100wb} and \ref{fig:K4_100}--\ref{fig:K4_100wb} show that the  ratios $g_1,g_2$ of sample estimators (weighted or not) are more precise and accurate than the ratios $G_1,G_2$ of the respective unbiased counterparts.

Weighting by phase intervals leads to a significant improvement in precision of all estimators in the limit of large sample sizes $n$ and a reduction of  accuracy of many `unbiased' estimators at low $n$, because of the introduction of correlations through phase weights \citep[as the closer measurements are in phase, the more similar their values are likely to be;
see][]{RimoldiniWeighted}.
Tuning parameters such as $a=25$ and $b=6$ in Eq.~(\ref{eq:phaseGap}) have shown to be able to mitigate the inaccuracy of unbiased estimators at low $n$ and reduce to the unweighted results, which appear to be the most accurate and precise in the limit of small sample sizes (in these simulations). This  solution might provide a reasonable compromise between precision and accuracy of unbiased estimators, at least for sample sizes $n>10$. 

Estimators of sinusoidal signals appeared more precise than those arising from the fourth power of a sine, with the exception of the mean, for which the relative precision of the two signal shapes reversed in most cases.

From the comparison of sample-size biased and unbiased estimators with different weighting schemes for the two periodic signals considered, it appears that, {\em at high S/N ratios, sample-size biased phase-weighted estimators (with $a,b\rightarrow 0$) have the best precision and accuracy in general}, with the exception of $m_2$, $m_3$ and $m_4/m_2$, which become biased especially for $n<20$. 
Further improvements might be achieved by tuning parameters better fitted to  estimators and signals of interest, in view of specific requirements of precision and accuracy.

\section{Conclusions}
\label{sec:concl}
Exact expressions of weighted skewness and kurtosis corrected for sample-size biases are provided in Eqs~(\ref{eq:startResult})--(\ref{eq:endResult}) under the assumption of independent measurements and weights.
Such estimators are particularly useful when the adoption of a weighting scheme is important for the processing of the data and accuracy needs to be preserved at small sample sizes.

Simulations of irregularly sampled symmetric and skewed periodic signals were employed to compare sample-size biased and unbiased estimators as a function of sample size in the unweighted, inverse-squared error weighted and phase-interval weighted schemes.
While phase weighting introduced correlations  (not considered by the unbiased weighted expressions), a mixed phase and error weighting scheme was able to balance precision and accuracy on a wide range of sample sizes.

\section*{Acknowledgments}
The author thanks M. S\"uveges for many discussions and valuable comments on the original manuscript.

%%%%%%%%%%%%%%%%%%%%%%%%%%%%%%%%%%%%%%%%%%%%%%
\newpage

\appendix

\section{Derivation of sample-size unbiased weighted moments}
\label{app:derivations}

The derivations presented in this Appendix involve weighted estimators under the assumption of independent measurements and weights. %, and include the fourth cumulant too.
Definitions and some of the relations often employed herein are listed below.
\begin{itemize}
\item Averages are weighted as $\bar{\theta}=\sum_{i=1}^n w_i\theta_i/V_1$.
\item Ellipses indicate the existence of terms with null expected value.
\item $x_a$ and $x_b$ denote two different representatives from independent and identically distributed elements so that, for example, $E(x_a x_b)=E(x_a)E(x_b)=\mu^2$.
\item $\sum_i$ is implied to sum over all (from the 1-st to the $n$-th) terms, unless explicitly stated otherwise.
\item While $E(w_i^p)=w_i^p$ for the specific $i$-th value, for a generic weight $w_a$ it equals $E(w_a^p)=\sum_i w_i w_i^p/V_1=V_{p+1}/V_1$, so $E(w_a)=\bar{w}=V_2/V_1$, $E(w_a^2)=V_3/V_1$ and so on.
\item $V_1^2=\sum_{i}w_i\sum_{j}w_j\\
~~~~~=w_a\sum_{i} w_i + \sum_{i\neq a} w_i\sum_{j} w_j\\
~~~~~=w_a^2+w_a\sum_{i\neq a} w_i + w_a \sum_{i\neq a} w_i +\sum_{i\neq a} w_i\sum_{j\neq a} w_j\\
~~~~~=w_a^2+2 w_a\sum_{i\neq a} w_i +  \sum_{i\neq a} w_i^2+\sum_{i\neq a} w_i\sum_{j\neq i,a} w_j\\
~~~~~= \sum_{i} w_i^2+2 w_a\sum_{i\neq a} w_i +\sum_{i\neq a} w_i\sum_{j\neq i,a} w_j\\
~~~~~=V_2+2w_a\sum_{i\neq a} w_i+\sum_{i\neq a}w_i\sum_{j\neq i,a} w_j$.
\item $V_1^3=\sum_{i}w_i\sum_{j} w_j\sum_{k} w_k\\
~~~~=w_a\sum_{i} w_i\sum_{j} w_j + \sum_{i\neq a} w_i\sum_{j} w_j\sum_{k} w_k\\
~~~~=w_a^2 \sum_{i} w_i + w_a\sum_{i\neq a} w_i\sum_{j} w_j + w_a \sum_{i\neq a} w_i\sum_{j} w_j + \sum_{i\neq a} w_i\sum_{j\neq a} w_j\sum_{k} w_k\\
~~~~=w_a^3 + w_a^2 \sum_{i\neq a} w_i + w_a^2\sum_{i\neq a} w_i+ w_a\sum_{i\neq a} w_i\sum_{j\neq a} w_j +w_a^2 \sum_{i\neq a} w_i+ \\
~~~~~~~+w_a \sum_{i\neq a} w_i\sum_{j\neq a} w_j + w_a \sum_{i\neq a} w_i\sum_{j\neq a} w_j+  \sum_{i\neq a} w_i\sum_{j\neq a} w_j\sum_{k\neq a} w_k\\
~~~~=w_a^3 +3 w_a^2 \sum_{i\neq a} w_i + 3 w_a\sum_{i\neq a} w_i\sum_{j\neq a} w_j +  \sum_{i\neq a} w_i\sum_{j\neq a} w_j\sum_{k\neq a} w_k\\
~~~~=w_a^3 +3 w_a^2 \sum_{i\neq a} w_i + 3 w_a\sum_{i\neq a} w_i^2 +3 w_a\sum_{i\neq a} w_i\sum_{j\neq i,a} w_j + \sum_{i\neq a} w_i^3+ \\
~~~~~~~~~~~+\sum_{i\neq a} w_i^2\sum_{j\neq i,a} w_j+\sum_{i\neq a} w_i^2\sum_{j\neq i,a} w_j+\sum_{i\neq a} w_i\sum_{j\neq i,a} w_j^2+\sum_{i\neq a} w_i\sum_{j\neq i,a} w_j\sum_{k\neq i,ja} w_k\\
~~~~=V_3+3 w_a^2 \sum_{i\neq a} w_i + 3 w_a\sum_{i\neq a} w_i^2 +3 w_a\sum_{i\neq a} w_i\sum_{j\neq i,a} w_j +3\sum_{i\neq a} w_i^2\sum_{j\neq i,a} w_j+ \\
~~~~~~~+\sum_{i\neq a} w_i\sum_{j\neq i,a} w_j\sum_{k\neq i,ja} w_k$.
\item $V_1V_2=\sum_{i}w_i\sum_{j}w_j^2\\
~~~~~=w_a\sum_{i} w_i^2 + \sum_{i\neq a} w_i\sum_{j} w_j^2\\
~~~~~=w_a^3+w_a\sum_{i\neq a} w_i^2 + w_a^2 \sum_{i\neq a} w_i +\sum_{i\neq a} w_i\sum_{j\neq a} w_j^2\\
~~~~~=w_a^3+w_a\sum_{i\neq a} w_i^2 + w_a^2 \sum_{i\neq a} w_i +\sum_{i\neq a} w_i^3+\sum_{i\neq a} w_i\sum_{j\neq i,a} w_j^2\\
~~~~~=V_3+w_a\sum_{i\neq a} w_i^2 + w_a^2 \sum_{i\neq a} w_i +\sum_{i\neq a} w_i\sum_{j\neq i,a} w_j^2$.
\item $V_1^2-V_2=\sum_{i} \sum_{j\neq i} w_i w_j$.
\item $V_2^2-V_4=\sum_{i} \sum_{j\neq i} w_i^2 w_j^2$.
\item $w_a^q\sum_{i\neq a}^n w_i^p=V_pE(w_a^q)-E(w_a^{p+q})=(V_pV_{q+1}-V_{p+q+1})/V_1$.
\item $\sum_iw_i\sum_jw_j^2=w_a^3+w_a\sum_{i\neq a}w_i^2+w_a^2\sum_{i\neq a}w_i+\sum_{i\neq a}w_i^3+\sum_{i\neq a}w_i\sum_{j\neq i,a}w_j^2$, thus\\
$\sum_{i\neq a}w_i\sum_{j\neq i,a}w_j^2=V_1V_2-V_3-(V_2^2-V_{4})/V_1-(V_1V_{3}-V_{4})/V_1$.
\item $\sum_iw_i\sum_jw_j=w_a^2+2w_a\sum_{i\neq a}w_i+\sum_{i\neq a}w_i^2+\sum_{i\neq a}w_i\sum_{j\neq i,a}w_j$, thus\\
$w_a\sum_{i\neq a}w_i\sum_{j\neq i,a}w_j=V_1V_2-V_4/V_1-2(V_1V_{3}-V_{4})/V_1-(V_2^2-V_{4})/V_1$.
\end{itemize}

\subsection{Outline of results}
The expressions of the elements pursued along the derivation of sample-size unbiased estimators (detailed in Sec.~\ref{app:details}) are summarized below, following the notation introduced in Sec.~\ref{sec:notation}.
\begin{align}
&E\left[(\bar{x}-\mu)^2\right] =  V_2\mu_2/V_1^2\\
&E\left[(\bar{x}-\mu)^3\right] =  V_3\mu_3/V_1^3\\
&E\left[(\bar{x}-\mu)^4\right] = \left[V_4\mu_4 + 3 \left(V_2^2-V_4\right) \mu_2^2\right] / V_1^4\\
&\notag\\
&E(x_a^2) =  \mu_2+\mu^2\\
&E(x_a^3) =\mu_3+3\mu_2\mu+\mu^3\\
&E(x_a^4) =\mu_4+4\mu_3\mu+6\mu_2\mu^2+\mu^4\\
&\notag\\
&E(\bar{x}^2) =V_2\mu_2/V_1^2+\mu^2\\
&E(\bar{x}^3) =V_3\mu_3/V_1^3+3V_2\mu_2\mu/V_1^2+\mu^3\\
&E(\bar{x}^4) =V_4\mu_4/V_1^4+3\left(V_2^2-V_4\right)\mu^2_2/V_1^4+4V_3\mu_3\mu/V_1^3+6V_2\mu_2\mu^2/V_1^2+\mu^4\\
&\notag\\
&E(x_a\bar{x}) =V_2\mu_2/V_1^2+\mu^2=E(\bar{x}^2)\\
&E(x_a^2\bar{x}) =V_2 \mu_3/V_1^2+\left(1+2V_2/V_1^2\right)\mu_2\mu+\mu^3 \\
&E(x_a\bar{x}^2) = V_3\mu_3/V_1^3+3V_2\mu_2\mu/V_1^2+\mu^3=E(\bar{x}^3)\\
&E(x_a^3\bar{x}) = V_2\mu_4/V_1^2+\left(1+3V_2/V_1^2 \right)\mu_3\mu+3\left(1+V_2/V_1^2\right)\mu_2\mu^2+\mu^4\\
&E(x_a\bar{x}^3) = V_4\mu_4/V_1^4+3\left(V_2^2-V_4\right)\mu^2_2/V_1^4+4V_3\mu_3\mu/V_1^3+6V_2\mu_2\mu^2/V_1^2+\mu^4=E(\bar{x}^4)\\
&E(x_a^2\bar{x}^2) = V_3\mu_4/V_1^3+2\left(V_3/V_1^3+V_2/V_1^2\right)\mu_3\mu+\left(1+5V_2/V_1^2\right)\mu_2\mu^2+\left(V_2/V_1^2-V_3/V_1^3\right)\mu_2^2+\mu^4\\
&\notag\\
&E(m_2)=\left(1-V_2/V_1^2\right)\mu_2=\mu_2-E\left[(\bar{x}-\mu)^2\right] \\
&E(m_3)=\left(1-3V_2/V_1^2+2V_3/V_1^3\right)\mu_3=\mu_3-\left(3V_1V_2/V_3-2\right)E\left[(\bar{x}-\mu)^3\right] \\
&E(m_4)=\left(1-4V_2/V_1^2+6V_3/V_1^3-3V_4/V_1^4\right)\mu_4+\left[6(V_2/V_1^2-V_3/V_1^3)-9(V_2^2-V_4)/V_1^4\right]\mu_2^2\\
&~~~~~~~~~=\left(1-4V_2/V_1^2+6V_3/V_1^3\right)\mu_4+6\left(V_2/V_1^2-V_3/V_1^3\right)\mu_2^2-3E\left[(\bar{x}-\mu)^4\right] \\
&E(m_2^2)=\left(V_2/V_1^2-2V_3/V_1^3+V_4/V_1^4\right)\mu_4+\left[1-3V_2/V_1^2+2V_3/V_1^3+3(V_2^2-V_4)/V_1^4\right]\mu_2^2\\
&~~~~~~~~~ = \left(V_2/V_1^2-2V_3/V_1^3+V_4/V_1^4\right)\kappa_4 +\left[1-4V_3/V_1^3+3V_4/V_1^4+3(V_2^2-V_4)/V_1^4\right]\kappa_2^2\\
&E(k_4)=\left(1-7V_2/V_1^2+12V_3/V_1^3-6V_4/V_1^4\right)\kappa_4-6\left[V_2/V_1^2-4V_3/V_1^3+3V_4/V_1^4+3(V_2^2-V_4)/V_1^4\right]\kappa_2^2\\
%&\notag\\   % REMOVED OR TOO LONG AND BLANK PAGE
&M_2 = \frac{V_1^2}{V_1^2-V_2}\,m_2 = K_2 \\
&M_3 =  \frac{V_1^3}{V_1^3-3V_1V_2+2V_3}\,m_3 = K_3 \\
&M_4  =   \frac{V_1^2(V_1^4-3V_1^2V_2+2V_1V_3+3V_2^2-3V_4)\,m_4}{(V_1^2-V_2)(V_1^4-6V_1^2V_2+8V_1V_3+3V_2^2-6V_4)} 
-\frac{3V_1^2(2V_1^2V_2-2V_1V_3-3V_2^2+3V_4)\,m_2^2}{(V_1^2-V_2)(V_1^4-6V_1^2V_2+8V_1V_3+3V_2^2-6V_4)} \\
&K_4  =   \frac{V_1^2(V_1^4-4V_1V_3+3V_2^2)\,m_4}{(V_1^2-V_2)(V_1^4-6V_1^2V_2+8V_1V_3+3V_2^2-6V_4)} 
-\frac{3V_1^2(V_1^4-2V_1^2V_2+4V_1V_3-3V_2^2)\,m_2^2}{(V_1^2-V_2)(V_1^4-6V_1^2V_2+8V_1V_3+3V_2^2-6V_4)}
\end{align}

\subsection{Detailed computations}
\label{app:details}
\begin{align}
E\left[(\bar{x}-\mu)^2\right] &= E\left[\left(\frac{1}{V_1}\sum_i w_i x_i - \mu\right)^2 \right]\\
&=E\left[\left(\frac{1}{V_1}\sum_i w_i (x_i - \mu)\right)^2 \right]\\
&= \frac{1}{V_1^2}E\left[\sum_i w_i^2 (x_i - \mu)^2 + \sum_i w_i (x_i-\mu)\sum_{j\neq i} w_j(x_j-\mu)\right]\\
&= \frac{1}{V_1^2}\sum_i w_i^2 E\left[ (x_a - \mu)^2\right]\\
&=  \frac{V_2}{V_1^2}\,\mu_2\\
&\notag\\
E\left[(\bar{x}-\mu)^3\right] &= E\left[\left(\frac{1}{V_1}\sum_i w_i x_i - \mu\right)^3 \right]\\
&= E\left[\left(\frac{1}{V_1}\sum_i w_i (x_i - \mu)\right)^3 \right]\\
&= \frac{1}{V_1^3}E\left[\sum_i w_i^3 (x_i - \mu)^3 + 3\sum_i w_i^2 (x_i-\mu)^2\sum_{j\neq i} w_j(x_j-\mu)+...\right]\\
&= \frac{1}{V_1^3}\sum_i w_i^3 E\left[ (x_a - \mu)^3\right]\\
&=  \frac{V_3}{V_1^3}\,\mu_3\\
&\notag\\
E\left[(\bar{x}-\mu)^4\right] &= E\left[\left(\frac{1}{V_1}\sum_i w_i x_i - \mu\right)^4 \right]\\
&=E\left[\left(\frac{1}{V_1}\sum_i w_i (x_i - \mu)\right)^4 \right]\\
&= \frac{1}{V_1^4}E\left[\sum_i w_i^4 (x_i - \mu)^4 + 3\sum_i w_i^2 (x_i-\mu)^2\sum_{j\neq i} w_j^2(x_j-\mu)^2+...\right]\\
&= \frac{1}{V_1^4}\left[\sum_i w_i^4 E\left[ (x_a - \mu)^4\right]+3\sum_i\sum_{j\neq i}w_i^2 w_j^2 E\left[ (x_a-\mu)^2\right] E\left[(x_b-\mu)^2 \right]+...\right] \\
&= \frac{V_4}{V_1^4}\,\mu_4 + 3 \frac{V_2^2-V_4}{V_1^4}\, \mu_2^2
\end{align}

\begin{align}
E(x_a^2) &= E[(x_a-\mu+\mu)^2] \\
&= E[(x_a-\mu)^2]+2\mu E(x_a-\mu)+E(\mu^2)\\
&=  \mu_2+\mu^2\\
&\notag\\
E(x_a^3) &= E[(x_a-\mu+\mu)^3] \\
&= E[(x_a-\mu)^3]+3\mu E[(x_a-\mu)^2]+3\mu^2E(x_a-\mu)+E(\mu^3)\\
&=\mu_3+3\mu_2\mu+\mu^3\\
&\notag\\
E(x_a^4) &= E[(x_a-\mu+\mu)^4] \\
&= E[(x_a-\mu)^4]+4\mu E[(x_a-\mu)^3]+6\mu^2E[(x_a-\mu)^2]+4\mu^3E(x_a-\mu)+E(\mu^4)\\
&=\mu_4+4\mu_3\mu+6\mu_2\mu^2+\mu^4\\
&\notag\\
E(\bar{x}^2) &= E[(\bar{x}-\mu+\mu)^2]\\
&= E[(\bar{x}-\mu)^2]+2\mu E(\bar{x}-\mu)+E(\mu^2)\\
&=V_2\mu_2/V_1^2+\mu^2\\
&\notag\\
E(\bar{x}^3) &= E[(\bar{x}-\mu+\mu)^3]\\
&= E[(\bar{x}-\mu)^3]+3\mu E[(\bar{x}-\mu)^2]+3\mu^2E(\bar{x}-\mu)+E(\mu^3)\\
&=V_3\mu_3/V_1^3+3V_2\mu_2\mu/V_1^2+\mu^3\\
&\notag\\
E(\bar{x}^4) &= E[(\bar{x}-\mu+\mu)^4]\\
&=E[(\bar{x}-\mu)^4]+4\mu E[(\bar{x}-\mu)^3]+6\mu^2E[(\bar{x}-\mu)^2]+4\mu^3E(\bar{x}-\mu)+E(\mu^4) \\
&=V_4\mu_4/V_1^4+3(V_2^2-V_4)\mu^2_2/V_1^4+4V_3\mu_3\mu/V_1^3+6V_2\mu_2\mu^2/V_1^2+\mu^4\\
&\notag\\
E(x_a\bar{x}) &= E\left(x_a\frac{1}{V_1}\sum_i w_i x_i \right)\\
&= \frac{1}{V_1}E\left( w_a x_a^2+x_a\sum_{i\neq a}w_i x_i\right)\\
&=  \frac{1}{V_1}E(w_a x_a^2)+\frac{1}{V_1}E(x_a)\sum_{i\neq a}w_i E(x_b)\\
&= \frac{1}{V_1} E(w_a)\mu_2+\frac{1}{V_1}w_a \mu^2+\frac{1}{V_1}\,\mu^2\sum_{i\neq a}w_i\\
&= \frac{1}{V_1} \frac{V_2}{V_1}\,\mu_2+\frac{1}{V_1}\,\mu^2\sum_{i}w_i\\
&=\frac{V_2}{V_1^2}\,\mu_2+\mu^2=E(\bar{x}^2)
\end{align}

\begin{align}
E(x_a^2\bar{x}) &= E\left(x_a^2\frac{1}{V_1}\sum_i w_i x_i \right)\\
&= \frac{1}{V_1}E\left( w_a x_a^3+x_a^2\sum_{i\neq a}w_i x_i\right)\\
&=  \frac{1}{V_1}E(w_a x_a^3)+\frac{1}{V_1}E(x_a^2)\sum_{i\neq a}w_i E(x_b)\\
&= \frac{1}{V_1}E(w_a)\mu_3+ \frac{3}{V_1}w_a\mu_2\mu+ \frac{1}{V_1}w_a\mu^3+\frac{1}{V_1}(\mu_2+\mu^2)\mu\sum_{i\neq a}w_i\\
&= \frac{1}{V_1}\frac{V_2}{V_1}\,\mu_3+ \frac{2}{V_1}E(w_a)\mu_2\mu+\frac{1}{V_1}(\mu_2+\mu^2)\mu\sum_{i}w_i\\
&= \frac{V_2}{V_1^2}\,\mu_3+ 2\frac{V_2}{V_1^2}\,\mu_2\mu+(\mu_2+\mu^2)\mu\\
&=\frac{V_2}{V_1^2}\, \mu_3+\left(1+2\frac{V_2}{V_1^2}\right)\mu_2\mu+\mu^3 \\
&\notag\\
E(x_a\bar{x}^2) &=  E\left[x_a\left(\frac{1}{V_1}\sum_i w_i x_i \right)^2\right]\\
&= \frac{1}{V_1^2}E\left(x_a\sum_i w_i^2 x_i^2+x_a\sum_i w_i x_i\sum_{j\neq i}w_j x_j\right)\\
&= \frac{1}{V_1^2}E\left(w_a^2 x_a^3+x_a\sum_{i\neq a} w_i^2 x_i^2+w_a x_a^2\sum_{i \neq a} w_i x_i+x_a \sum_{i\neq a} w_i x_i\sum_{j\neq i}w_j x_j \right)\\
&= \frac{1}{V_1^2}E\left(w_a^2 x_a^3+x_a\sum_{i\neq a} w_i^2 x_i^2+2 w_a x_a^2\sum_{i \neq a} w_i x_i+ x_a \sum_{i\neq a} w_i x_i\sum_{j\neq i,a}w_j x_j  \right)\\
&= \frac{1}{V_1^2}E(w_a^2) \mu_3+\frac{3}{V_1^2}w_a^2\mu_2\mu+\frac{1}{V_1^2}w_a^2\mu^3+\frac{1}{V_1^2}(\mu_2\mu+\mu^3)\sum_{i\neq a} w_i^2+\notag\\
&~~~~ +\frac{2}{V_1^2} E(w_a) \mu_2\mu\sum_{i \neq a} w_i +\frac{2}{V_1^2} w_a\mu^2\mu\sum_{i \neq a} w_i + \frac{1}{V_1^2}\,\mu^3 \sum_{i\neq a} w_i \sum_{j\neq i,a}w_j \\
&= \frac{V_3}{V_1^3}\,\mu_3+\frac{1}{V_1^2}(\mu_2\mu+\mu^3)\sum_{i} w_i^2+2\frac{V_2}{V_1^3}\,\mu_2\mu \sum_{i} w_i +\frac{2}{V_1^2} w_a\mu^3\sum_{i \neq a} w_i + \notag\\
&~~~~ \frac{1}{V_1^2}\,\mu^3 \sum_{i\neq a} w_i \sum_{j\neq i,a}w_j \\
&= \frac{V_3}{V_1^3}\,\mu_3+\left(\frac{V_2}{V_1^2}+2\frac{V_2 V_1}{V_1^3}\right)\mu_2\mu +\mu^3 \\
&= \frac{V_3}{V_1^3}\,\mu_3+3\frac{V_2}{V_1^2}\,\mu_2\mu+\mu^3=E(\bar{x}^3)
\end{align}

\begin{align}
E(x_a^3\bar{x}) &= E\left(x_a^3\frac{1}{V_1}\sum_i w_i x_i \right)\\
&= \frac{1}{V_1}E\left( w_a x_a^4+x_a^3\sum_{i\neq a}w_i x_i\right)\\
&=  \frac{1}{V_1}E(w_a x_a^4)+\frac{1}{V_1}E(x_a^3)\sum_{i\neq a}w_i E(x_b)\\
&= \frac{1}{V_1} E(w_a)\mu_4+\frac{3}{V_1}E(w_a) \mu_3\mu+\frac{1}{V_1}w_a \mu_3\mu+\frac{3}{V_1}E(w_a)\mu_2\mu^2+\frac{3}{V_1}w_a\mu_2\mu^2+\frac{1}{V_1}w_a\mu^4+\notag\\
&~~~~ +\frac{1}{V_1}\left(\mu_3+3\mu_2\mu+\mu^3\right)\mu\sum_{i\neq a}w_i\\
&=  \frac{V_2}{V_1^2}\,\mu_4+3\frac{V_2}{V_1^2}\, \mu_3\mu+3\frac{V_2}{V_1^2}\,\mu_2\mu^2+\frac{1}{V_1}\left(\mu_3+3\mu_2\mu+\mu^3\right)\mu\sum_{i}w_i\\
&= \frac{V_2}{V_1^2}\,\mu_4+\left(1+3\frac{V_2}{V_1^2} \right)\mu_3\mu+3\left(1+\frac{V_2}{V_1^2}\right)\mu_2\mu^2+\mu^4\\
&\notag\\
E(x_a\bar{x}^3) &=  E\left[x_a\left(\frac{1}{V_1}\sum_i w_i x_i \right)^3\right]\\
&= \frac{1}{V_1^3}E\left(x_a\sum_i w_i x_i\sum_j w_j x_j\sum_k w_k x_k\right)\\
&= \frac{1}{V_1^3}E\left[x_a\sum_i w_i x_i\left(\sum_j w_j^2 x_j^2+\sum_j w_j x_j\sum_{k\neq j} w_k x_k\right)\right]\\
&= \frac{1}{V_1^3}E\left(x_a\sum_i w_i^3 x_i^3+3x_a\sum_i w_i x_i\sum_{j\neq i}w_j^2 x_j^2+x_a\sum_i w_i x_i\sum_{j\neq i}w_j x_j \sum_{k\neq i,j}w_k x_k\right)\\
&= \frac{1}{V_1^3}E\left(w_a^3x_a^4+x_a\sum_{i\neq a} w_i^3 x_i^3+3w_ax_a^2\sum_{i\neq a}w_i^2 x_i^2+3w_a^2x_a^3\sum_{i\neq a} w_i x_i+3x_a\sum_{i\neq a} w_i x_i\sum_{j\neq i,a}w_j^2 x_j^2+\right.\notag\\
&~~~~ \left.  +3 w_a x_a^2\sum_{i\neq a}w_i x_i \sum_{j\neq i,a}w_j x_j +x_a\sum_{i\neq a} w_i x_i\sum_{j\neq i,a}w_j x_j \sum_{k\neq i,j,a}w_k x_k \right)\\
&= \frac{1}{V_1^3}E(w_a^3)\mu_4+\frac{4}{V_1^3}w_a^3\mu_3\mu+\frac{6}{V_1^3}w_a^3\mu_2\mu^2+\frac{1}{V_1^3}w_a^3\mu^4
+\frac{1}{V_1^3}(\mu_3+3\mu_2\mu+\mu^3)\mu\sum_{i\neq a} w_i^3 +\notag\\
&~~~~ +\frac{3}{V_1^3}w_a(\mu_2+\mu^2)^2\sum_{i\neq a}w_i^2
+\frac{3}{V_1^3}w_a^2 (\mu_3+3\mu_2\mu+\mu^3) \mu \sum_{i\neq a} w_i +\notag\\
&~~~~ +\frac{3}{V_1^3}(\mu_2+\mu^2)\mu^2\sum_{i\neq a} w_i \sum_{j\neq i,a}w_j^2 
+\frac{3}{V_3} w_a (\mu_2+\mu^2)\mu^2\sum_{i\neq a}w_i \sum_{j\neq i,a}w_j +\notag\\
&~~~~ +\frac{1}{V_1^3}\,\mu^4\sum_{i\neq a} w_i \sum_{j\neq i,a}w_j \sum_{k\neq i,j,a}w_k
\end{align}
\begin{align}
&= \frac{1}{V_1^3}\frac{V_4}{V_1}\,\mu_4+\frac{1}{V_1^3}\,\mu_3\mu\sum_{i} w_i^3+ \frac{3}{V_1^3}E(w_a^2) \mu_3 \mu \sum_{i} w_i +\frac{6}{V_1^3}E(w_a^2) \mu_2\mu^2 \sum_{i} w_i
+\frac{1}{V_1^3}\,\mu^4\sum_{i} w_i^3+\notag\\
&~~~~ +\frac{3}{V_1^3}\,\mu_2\mu^2\sum_{i\neq a} w_i^3  
+\frac{3}{V_1^3}w_a(\mu_2^2+\mu^4+2\mu_2\mu^2)\sum_{i\neq a}w_i^2
+\frac{3}{V_1^3}w_a^2 (\mu_2\mu^2+\mu^4) \sum_{i\neq a} w_i +\notag\\
&~~~~ +\frac{3}{V_1^3}(\mu_2\mu^2+\mu^4)\sum_{i\neq a} w_i \sum_{j\neq i,a}w_j^2 
+\frac{3}{V_3} w_a (\mu_2\mu^2+\mu^4)\sum_{i\neq a}w_i \sum_{j\neq i,a}w_j +\notag\\
&~~~~ +\frac{1}{V_1^3}\,\mu^4\sum_{i\neq a} w_i \sum_{j\neq i,a}w_j \sum_{k\neq i,j,a}w_k \\
&= \frac{V_4}{V_1^4}\,\mu_4+\left(\frac{V_3}{V_1^3}+3\frac{V_3}{V_1^3}\right)\mu_3\mu 
+6\frac{V_3}{V_1^3}\,\mu_2\mu^2
+\mu^4+\frac{3}{V_1^3}\,\mu_2\mu^2\sum_{i\neq a} w_i^3  +\notag\\
&~~~~ +\frac{3}{V_1^3}w_a\mu_2^2\sum_{i\neq a}w_i^2
+\frac{6}{V_1^3}w_a\mu_2\mu^2\sum_{i\neq a}w_i^2
+\frac{3}{V_1^3}w_a^2 \mu_2\mu^2 \sum_{i\neq a} w_i +\notag\\
&~~~~ +\frac{3}{V_1^3}\,\mu_2\mu^2\sum_{i\neq a} w_i \sum_{j\neq i,a}w_j^2 
+\frac{3}{V_3} w_a \mu_2\mu^2\sum_{i\neq a}w_i \sum_{j\neq i,a}w_j \\
&= \frac{V_4}{V_1^4}\,\mu_4+4\frac{V_3}{V_1^3}\,\mu_3\mu +\mu^4+\frac{3}{V_1^3}\frac{V_2^2-V_4}{V_1}\,\mu^2_2
+\frac{3}{V_1^3}\left(2V_3+\frac{V_1V_3-V_4}{V_1}+2\frac{V_2^2-V_4}{V_1}+\frac{V_1V_3-V_4}{V_1}+\right.\notag\\
&~~~~ \left.+V_1V_2-V_3-\frac{V_2^2-V_{4}}{V_1}-\frac{V_1V_{3}-V_{4}}{V_1}+V_1V_2-\frac{V_4}{V_1}-2\frac{V_1V_{3}-V_{4}}{V_1}-\frac{V_2^2-V_{4}}{V_1} \right)\mu_2\mu^2\\
&= \frac{V_4}{V_1^4}\,\mu_4+3\frac{V_2^2-V_4}{V_1^4}\,\mu^2_2+4\frac{V_3}{V_1^3}\,\mu_3\mu+6\frac{V_2}{V_1^2}\,\mu_2\mu^2+\mu^4=E(\bar{x}^4)\\
&\notag\\
E(x_a^2\bar{x}^2) &=  E\left[x_a^2\left(\frac{1}{V_1}\sum_i w_i x_i \right)^2\right]\\
&= \frac{1}{V_1^2}E\left(x_a^2\sum_i w_i^2 x_i^2+x_a^2\sum_i w_i x_i\sum_{j\neq i}w_j x_j\right)\\
&= \frac{1}{V_1^2}E\left(w_a^2 x_a^4+x_a^2\sum_{i\neq a} w_i^2 x_i^2+w_a x_a^3\sum_{i \neq a} w_i x_i+x_a^2 \sum_{i\neq a} w_i x_i\sum_{j\neq i}w_j x_j \right)\\
&= \frac{1}{V_1^2}E\left(w_a^2 x_a^4+x_a^2\sum_{i\neq a} w_i^2 x_i^2+2 w_a x_a^3\sum_{i \neq a} w_i x_i+ x_a^2 \sum_{i\neq a} w_i x_i\sum_{j\neq i,a}w_j x_j  \right)\\
&= \frac{1}{V_1^2}E(w_a^2) \mu_4+\frac{2}{V_1^2}E(w_a^2)\mu_3\mu+\frac{2}{V_1^2}w_a^2\mu_3\mu+\frac{6}{V_1^2}w_a^2\mu_2\mu^2+\frac{1}{V_1^2}w_a^2\mu^4+\frac{1}{V_1^2}(\mu_2+\mu^2)^2\sum_{i\neq a} w_i^2+\notag\\
&~~~~ +\frac{2}{V_1^2}  w_a\mu_3 \mu\sum_{i \neq a} w_i +\frac{6}{V_1^2}w_a\mu_2\mu^2\sum_{i \neq a} w_i +\frac{2}{V_1^2}w_a \mu^4\sum_{i \neq a} w_i + \frac{1}{V_1^2}(\mu_2+\mu^2)\mu^2 \sum_{i\neq a} w_i \sum_{j\neq i,a}w_j
\end{align}
\begin{align}
&= \frac{V_3}{V_1^3}\, \mu_4+2\frac{V_3}{V_1^3}\,\mu_3\mu+\frac{2}{V_1^2} E(w_a)\mu_3 \mu\sum_i w_i+\frac{1}{V_1^2}\,\mu_2^2\sum_{i\neq a} w_i^2 + \frac{2}{V_1^2}\,\mu_2\mu^2\sum_{i} w_i^2 +\frac{1}{V_1^2}\,\mu^4\sum_i w_i^2+\notag\\
&~~~~ + \frac{4}{V_1^2} E(w_a)\mu_2\mu^2\sum_{i} w_i + \frac{2}{V_1^2} w_a\mu_2\mu^2\sum_{i\neq a} w_i +\frac{2}{V_1^2}w_a\mu^4 \sum_{i\neq a}w_i+ \frac{1}{V_1^2}\,\mu_2\mu^2 \sum_{i\neq a} w_i \sum_{j\neq i,a}w_j +\notag\\
&~~~~  +\frac{1}{V_1^2}\,\mu^4 \sum_{i\neq a} w_i \sum_{j\neq i,a}w_j \\
&= \frac{V_3}{V_1^3}\, \mu_4+2\left(\frac{V_3}{V_1^3}+\frac{V_2}{V_1^2}\right)\mu_3\mu+\frac{1}{V_1^2}\,\mu_2^2\sum_{i\neq a} w_i^2 
 +2 \frac{V_2}{V_1^2}\,\mu_2\mu^2+4 \frac{V_2}{V_1^2}\,\mu_2\mu^2 + \frac{2}{V_1^2} w_a\mu_2\mu^2\sum_{i\neq a} w_i+\notag \\
 &~~~~ + \frac{1}{V_1^2}\,\mu_2\mu^2 \sum_{i\neq a} w_i \sum_{j\neq i,a}w_j +\mu^4 \\
&= \frac{V_3}{V_1^3}\,\mu_4+2\left(\frac{V_3}{V_1^3}+\frac{V_2}{V_1^2}\right)\mu_3\mu+\left(1+5\frac{V_2}{V_1^2}\right)\mu_2\mu^2+\left(\frac{V_2}{V_1^2}-\frac{V_3}{V_1^3}\right)\mu_2^2+\mu^4\\
&\notag\\
E(m_2)
&= E\left[ \frac{1}{V_1}\sum_i w_i(x_i-\bar{x})^2\right]\\
&= E\left[(x_a-\bar{x})^2 \right]\\
&= E\left(x_a^2-2x_a\bar{x}+\bar{x}^2 \right)\\
&= E(x_a^2)-E(\bar{x}^2)\\
&= \mu_2+\mu^2-V_2\mu_2/V_1^2-\mu^2\\
&=\left(1-\frac{V_2}{V_1^2}\right)\mu_2\\
&=\mu_2-E\left[(\bar{x}-\mu)^2\right] \\
&\notag\\
E(m_3)
&= E\left[ \frac{1}{V_1}\sum_i w_i(x_i-\bar{x})^3\right]\\
&= E\left[(x_a-\bar{x})^3 \right]\\
&= E\left(x_a^3-3x_a^2\bar{x}+3x_a\bar{x}^2-\bar{x}^3 \right)\\
&= E(x_a^3)-3E(x_a^2\bar{x})+2E(\bar{x}^3)\\
&= \mu_3+3\mu_2\mu+\mu^3 - 3\left[\frac{V_2}{V_1^2}\, \mu_3+\left(1+2\frac{V_2}{V_1^2}\right)\mu_2\mu+\mu^3  \right]
+2\left(\frac{V_3}{V_1^3}\,\mu_3+3\frac{V_2}{V_1^2}\,\mu_2\mu+\mu^3\right)\\
&=\left(1-3\frac{V_2}{V_1^2}+2\frac{V_3}{V_1^3}\right)\mu_3\\
&=\mu_3-\left(3\frac{V_1V_2}{V_3}-2\right)E\left[(\bar{x}-\mu)^3\right] \\
&\notag\\
E(m_4)
&=  E\left[ \frac{1}{V_1}\sum_i w_i(x_i-\bar{x})^4\right]\\
&= E\left[(x_a-\bar{x})^4 \right]\\
&= E(x_a^4)-4E(x_a^3\bar{x})+6E(x_a^2\bar{x}^2)-4E(x_a\bar{x}^3) +E(\bar{x}^4)
\end{align}
\begin{align}
&= \mu_4+4\mu_3\mu+6\mu_2\mu^2+\mu^4
-4\left[\frac{V_2}{V_1^2}\,\mu_4+\left(1+3\frac{V_2}{V_1^2} \right)\mu_3\mu+3\left(1+\frac{V_2}{V_1^2}\right)\mu_2\mu^2+\mu^4 \right]+\notag\\
&~~~~ +6\left[\frac{V_3}{V_1^3}\,\mu_4+2\left(\frac{V_3}{V_1^3}+\frac{V_2}{V_1^2}\right)\mu_3\mu+\left(1+5\frac{V_2}{V_1^2}\right)\mu_2\mu^2+\left(\frac{V_2}{V_1^2}-\frac{V_3}{V_1^3}\right)\mu_2^2+\mu^4 \right] + \notag\\
&~~~~ -4\left[ \frac{V_4}{V_1^4}\,\mu_4+3\frac{V_2^2-V_4}{V_1^4}\,\mu^2_2+4\frac{V_3}{V_1^3}\,\mu_3\mu+6\frac{V_2}{V_1^2}\,\mu_2\mu^2+\mu^4\right] + \notag \\
&~~~~ +\frac{V_4}{V_1^4}\,\mu_4+3\frac{V_2^2-V_4}{V_1^4}\,\mu^2_2+4\frac{V_3}{V_1^3}\,\mu_3\mu+6\frac{V_2}{V_1^2}\,\mu_2\mu^2+\mu^4\\
&=\left(1-4\frac{V_2}{V_1^2}+6\frac{V_3}{V_1^3}-3\frac{V_4}{V_1^4}\right)\mu_4+\left[6\left(\frac{V_2}{V_1^2}-\frac{V_3}{V_1^3}\right)-9\frac{V_2^2-V_4}{V_1^4}\right]\mu_2^2\\
&=\left(1-4\frac{V_2}{V_1^2}+6\frac{V_3}{V_1^3}\right)\mu_4+6\left(\frac{V_2}{V_1^2}-\frac{V_3}{V_1^3}\right)\mu_2^2-3E\left[(\bar{x}-\mu)^4\right] \\
&\notag\\
E(m_2^2)&=  E\left[ \left(\frac{1}{V_1}\sum_i w_i(x_i-\bar{x})^2\right)^2\right]\\
&= E\left[ \left(\frac{1}{V_1}\sum_i w_i(x_i^2-\bar{x}^2)\right)^2\right]\\
&=\frac{1}{V_1^2}E\left[ \left(\sum_i w_i x_i^2\right)^2\right]-\frac{2}{V_1^2}E\left(\bar{x}^2\sum_i w_ix_i^2\right)+E(\bar{x}^4)\\
&=\frac{1}{V_1^2}E\left(\sum_i w_i^2 x_i^4+\sum_iw_i x_i^2\sum_{j\neq i}w_j x_j^2\right)
-2E(x_a^2\bar{x}^2) + E(\bar{x}^4)\\
&=\frac{V_2}{V_1^2}\left(\mu_4+4\mu_3\mu+6\mu_2\mu^2+\mu^4\right)+\frac{V_1^2-V_2}{V_1^2}\left(\mu_2+\mu^2 \right)^2+\notag\\
&~~~~ -2\left[\frac{V_3}{V_1^3}\,\mu_4+2\left(\frac{V_3}{V_1^3}+\frac{V_2}{V_1^2}\right)\mu_3\mu+\left(1+5\frac{V_2}{V_1^2}\right)\mu_2\mu^2+\left(\frac{V_2}{V_1^2}-\frac{V_3}{V_1^3}\right)\mu_2^2+\mu^4\right] +\notag\\
&~~~~ +\frac{V_4}{V_1^4}\,\mu_4+3\frac{V_2^2-V_4}{V_1^4}\,\mu^2_2+4\frac{V_3}{V_1^3}\,\mu_3\mu+6\frac{V_2}{V_1^2}\,\mu_2\mu^2+\mu^4\\
&=\left(\frac{V_2}{V_1^2}-2\frac{V_3}{V_1^3}+\frac{V_4}{V_1^4} \right)\mu_4
+4\left(\frac{V_2}{V_1^2}-\frac{V_3}{V_1^3}-\frac{V_2}{V_1^2}+\frac{V_3}{V_1^3} \right)\mu_3\mu+\notag\\
&~~~~ +\left(6\frac{V_2}{V_1^2}+2\frac{V_1^2-V_2}{V_1^2}-2-10\frac{V_2}{V_1^2}+6\frac{V_2}{V_1^2} \right)\mu_2\mu^2
+\left(\frac{V_2}{V_1^2}+\frac{V_1^2-V_2}{V_1^2}-2+1 \right)\mu^4+\notag\\
&~~~~ +\left(\frac{V_1^2-V_2}{V_1^2}-2\frac{V_2}{V_1^2}+2\frac{V_3}{V_1^3}+3\frac{V_2^2-V_4}{V_1^4} \right)\mu_2^2\\
&=\left(\frac{V_2}{V_1^2}-2\frac{V_3}{V_1^3}+\frac{V_4}{V_1^4} \right)\mu_4+\left(1-3\frac{V_2}{V_1^2}+2\frac{V_3}{V_1^3}+3\frac{V_2^2-V_4}{V_1^4} \right)\mu_2^2\\
E(m_2^2)&=\frac{V_2}{V_1^2}\left(\mu_4-3\mu_2^2\right)-2\frac{V_3}{V_1^3}\left(\mu_4-3\mu_2^2\right)+\frac{V_4}{V_1^4}\left(\mu_4-3\mu_2^2\right)+\notag\\
&~~~~ +\left(1-4\frac{V_3}{V_1^3}+3\frac{V_4}{V_1^4}+3\frac{V_2^2-V_4}{V_1^4}\right)\mu_2^2\\
&=\left(\frac{V_2}{V_1^2}-2\frac{V_3}{V_1^3}+\frac{V_4}{V_1^4}\right)\kappa_4 +\left(1-4\frac{V_3}{V_1^3}+3\frac{V_4}{V_1^4}+3\frac{V_2^2-V_4}{V_1^4}\right)\kappa_2^2
\end{align}

\begin{align}
E(k_4)&= E(m_4)-3E(m_2^2)\\
&=\left(1-4\frac{V_2}{V_1^2}+6\frac{V_3}{V_1^3}-3\frac{V_4}{V_1^4}\right)\mu_4+\left[6\left(\frac{V_2}{V_1^2}-\frac{V_3}{V_1^3}\right)-9\frac{V_2^2-V_4}{V_1^4}\right]\mu_2^2 +\notag\\
&~~~~ -3\left[\left(\frac{V_2}{V_1^2}-2\frac{V_3}{V_1^3}+\frac{V_4}{V_1^4} \right)\mu_4+\left(1-3\frac{V_2}{V_1^2}+2\frac{V_3}{V_1^3}+3\frac{V_2^2-V_4}{V_1^4} \right)\mu_2^2 \right]\\
&=\left(1-7\frac{V_2}{V_1^2}+12\frac{V_3}{V_1^3}-6\frac{V_4}{V_1^4} \right)\mu_4 - 3\left(1-5\frac{V_2}{V_1^2}+4\frac{V_3}{V_1^3}+6\frac{V_2^2-V_4}{V_1^4} \right)\mu_2^2\\
&=\left(1-7\frac{V_2}{V_1^2}+12\frac{V_3}{V_1^3}-6\frac{V_4}{V_1^4} \right)\left(\mu_4-3\mu_2^2\right) - 6\left( \frac{V_2}{V_1^2}-4\frac{V_3}{V_1^3}+3\frac{V_4}{V_1^4}+3\frac{V_2^2-V_4}{V_1^4}\right)\mu_2^2\\
&=\left(1-7\frac{V_2}{V_1^2}+12\frac{V_3}{V_1^3}-6\frac{V_4}{V_1^4}\right)\kappa_4-6\left(\frac{V_2}{V_1^2}-4\frac{V_3}{V_1^3}+3\frac{V_4}{V_1^4}+3\frac{V_2^2-V_4}{V_1^4}\right)\kappa_2^2 \\
&\notag\\
E(M_2) &=\mu_2= E(m_2)\left(1-\frac{V_2}{V_1^2}\right)^{-1}\\
M_2&= \frac{V_1^2}{V_1^2-V_2}\,m_2 = K_2 \\
&\notag\\
E(M_3) &= \mu_3=E(m_3)\left(1-3\frac{V_2}{V_1^2}+2\frac{V_3}{V_1^3}\right)^{-1}\\
M_3&=  \frac{V_1^3}{V_1^3-3V_1V_2+2V_3}\,m_3 = K_3 \\
&\notag\\
E(M_4) &= \mu_4=\left\{E(m_4)-\left[6\left(\frac{V_2}{V_1^2}-\frac{V_3}{V_1^3}\right)-9\frac{V_2^2-V_4}{V_1^4}\right]\mu_2^2 \right\}\left( 1-4\frac{V_2}{V_1^2}+6\frac{V_3}{V_1^3}-3\frac{V_4}{V_1^4}\right)^{-1}\\
&= \left\{E(m_4)-\left[6\left(\frac{V_2}{V_1^2}-\frac{V_3}{V_1^3}\right)-9\frac{V_2^2-V_4}{V_1^4}\right]
\left[E(m_2^2)-\left(\frac{V_2}{V_1^2}-2\frac{V_3}{V_1^3}+\frac{V_4}{V_1^4} \right)E(M_4) \right]\times\right.\notag\\
&~~~~ \left.\times\left( 1-3\frac{V_2}{V_1^2}+2\frac{V_3}{V_1^3}+3\frac{V_2^2-V_4}{V_1^4} \right)^{-1}
 \right\}\left( 1-4\frac{V_2}{V_1^2}+6\frac{V_3}{V_1^3}-3\frac{V_4}{V_1^4}\right)^{-1}\\
 &= \left[\frac{E(m_4)}{1-4\frac{V_2}{V_1^2}+6\frac{V_3}{V_1^3}-3\frac{V_4}{V_1^4}}-\frac{\left( 6\frac{V_2}{V_1^2}-6\frac{V_3}{V_1^3}-9\frac{V_2^2-V_4}{V_1^4}\right)E(m_2^2)}{\left( 1-4\frac{V_2}{V_1^2}+6\frac{V_3}{V_1^3}-3\frac{V_4}{V_1^4}\right)\left( 1-3\frac{V_2}{V_1^2}+2\frac{V_3}{V_1^3}+3\frac{V_2^2-V_4}{V_1^4} \right)} \right] \times \notag\\
&~~~~ \times \left[1-\frac{\left( 6\frac{V_2}{V_1^2}-6\frac{V_3}{V_1^3}-9\frac{V_2^2-V_4}{V_1^4}\right)\left(\frac{V_2}{V_1^2}-2\frac{V_3}{V_1^3}+\frac{V_4}{V_1^4} \right)}{\left( 1-4\frac{V_2}{V_1^2}+6\frac{V_3}{V_1^3}-3\frac{V_4}{V_1^4}\right)\left( 1-3\frac{V_2}{V_1^2}+2\frac{V_3}{V_1^3}+3\frac{V_2^2-V_4}{V_1^4} \right)} \right]^{-1}\\
&= \left[\frac{V_1^4 \,E(m_4)}{V_1^4-4V_1^2V_2+6V_1V_3-3V_4}-\frac{V_1^4(6V_1^2V_2-6V_1V_3-9V_2^2+9V_4)\,E(m_2^2)}{(V_1^4-4V_1^2V_2+6V_1V_3-3V_4)(V_1^4-3V_1^2V_2+2V_1V_3+3V_2^2-3V_4)} \right] \times \notag\\
&~~~~ \times \left[1-\frac{\left(6V_1^2V_2-6V_1V_3-9V_2^2+9V_4\right)\left(V_1^2V_2-2V_1V_3+V_4\right)}{\left(V_1^4-4V_1^2V_2+6V_1V_3-3V_4\right)\left(V_1^4-3V_1^2V_2+2V_1V_3+3V_2^2-3V_4 \right)} \right]^{-1}\\
M_4&=   \frac{V_1^2(V_1^4-3V_1^2V_2+2V_1V_3+3V_2^2-3V_4)\,m_4}{(V_1^2-V_2)(V_1^4-6V_1^2V_2+8V_1V_3+3V_2^2-6V_4)} 
-\frac{3V_1^2(2V_1^2V_2-2V_1V_3-3V_2^2+3V_4)\,m_2^2}{(V_1^2-V_2)(V_1^4-6V_1^2V_2+8V_1V_3+3V_2^2-6V_4)} 
\end{align}

\begin{align}
E(K_4) &=\kappa_4=\left[E(k_4)+6\left(\frac{V_2}{V_1^2}-4\frac{V_3}{V_1^3}+3\frac{V_4}{V_1^4}+3\frac{V_2^2-V_4}{V_1^4}\right)\kappa_2^2 \right]\left( 1-7\frac{V_2}{V_1^2}+12\frac{V_3}{V_1^3}-6\frac{V_4}{V_1^4}\right)^{-1}\\
&= \left\{E(k_4)+6\left(\frac{V_2}{V_1^2}-4\frac{V_3}{V_1^3}+3\frac{V_4}{V_1^4}+3\frac{V_2^2-V_4}{V_1^4}\right)
\left[E(m_2^2)-\left(\frac{V_2}{V_1^2}-2\frac{V_3}{V_1^3}+\frac{V_4}{V_1^4} \right)E(K_4) \right]\times\right.\notag\\
&~~~~ \left.\times\left( 1-4\frac{V_3}{V_1^3}+3\frac{V_4}{V_1^4}+3\frac{V_2^2-V_4}{V_1^4} \right)^{-1}
 \right\}\left(1-7\frac{V_2}{V_1^2}+12\frac{V_3}{V_1^3}-6\frac{V_4}{V_1^4}\right)^{-1}\\
&= \left[\frac{E(m_4-3m_2^2)}{1-7\frac{V_2}{V_1^2}+12\frac{V_3}{V_1^3}-6\frac{V_4}{V_1^4}}+\frac{6\left(\frac{V_2}{V_1^2}-4\frac{V_3}{V_1^3}+3\frac{V_4}{V_1^4}+3\frac{V_2^2-V_4}{V_1^4}\right)E(m_2^2) }{\left(1-7\frac{V_2}{V_1^2}+12\frac{V_3}{V_1^3}-6\frac{V_4}{V_1^4}\right)\left( 1-4\frac{V_3}{V_1^3}+3\frac{V_4}{V_1^4}+3\frac{V_2^2-V_4}{V_1^4} \right)}\right] \times \notag\\
&~~~~ \times \left[1+\frac{6\left(\frac{V_2}{V_1^2}-4\frac{V_3}{V_1^3}+3\frac{V_4}{V_1^4}+3\frac{V_2^2-V_4}{V_1^4}\right)\left(\frac{V_2}{V_1^2}-2\frac{V_3}{V_1^3}+\frac{V_4}{V_1^4} \right)}{\left(1-7\frac{V_2}{V_1^2}+12\frac{V_3}{V_1^3}-6\frac{V_4}{V_1^4}\right)\left( 1-4\frac{V_3}{V_1^3}+3\frac{V_4}{V_1^4}+3\frac{V_2^2-V_4}{V_1^4} \right)} \right]^{-1}\\
&= \left[\frac{V_1^4\,E(m_4-3m_2^2)}{V_1^4-7V_1^2V_2+12V_1V_3-6V_4}+\frac{6V_1^4\left(V_1^2V_2-4V_1V_3+3V_4+3V_2^2-3V_4\right)E(m_2^2)}{\left(V_1^4-7V_1^2V_2+12V_1V_3-6V_4\right)\left(V_1^4-4V_1V_3+3V_4+3V_2^2-3V_4 \right)}\right] \times \notag\\
&~~~~ \times \left[1+\frac{6\left(V_1^2V_2-4V_1V_3+3V_4+3V_2^2-3V_4\right)\left(V_1^2V_2-2V_1V_3+V_4 \right)}{\left(V_1^4-7V_1^2V_2+12V_1V_3-6V_4\right)\left(V_1^4-4V_1V_3+3V_4+3V_2^2-3V_4 \right)} \right]^{-1}\\
K_4&=   \frac{V_1^2(V_1^4-4V_1V_3+3V_2^2)\,m_4}{(V_1^2-V_2)(V_1^4-6V_1^2V_2+8V_1V_3+3V_2^2-6V_4)} 
-\frac{3V_1^2(V_1^4-2V_1^2V_2+4V_1V_3-3V_2^2)\,m_2^2}{(V_1^2-V_2)(V_1^4-6V_1^2V_2+8V_1V_3+3V_2^2-6V_4)}
\end{align}


\begin{thebibliography}{99}
\bibitem[\protect\citeauthoryear{Bowley}{1920}]{Bowley}Bowley A.L., 1920, Elements of Statistics, Charles Scribner's Sons, New York
\bibitem[\protect\citeauthoryear{Cram\'er}{1961}]{Cramer}Cram\'er H., 1961, Mathematical Methods of Statistics, Princeton University Press
\bibitem[\protect\citeauthoryear{D'Agostino}{1986}]{Dagostino}D'Agostino R.B., 1986, Goodness-of-fit techniques, D'Agostino \& Stephens eds., Marcel Dekker, New York, p.~367
\bibitem[\protect\citeauthoryear{Dressel}{1940}]{Dressel}Dressel P.L., 1940, Annals of Mathematical Statistics, 11, 33
\bibitem[\protect\citeauthoryear{Fisher}{1929}]{Fisher}Fisher R.A., 1929, Proceedings of the London Mathematical Society, Series 2, 30, 199
\bibitem[\protect\citeauthoryear{Groeneveld \& Meeden}{1984}]{Groeneveld}Groeneveld R.A., Meeden G., 1984, The Statistician, 33, 391
\bibitem[\protect\citeauthoryear{Heijmans}{1999}]{Heijmans}Heijmans R., 1999, Statistical Papers, 40, 107
\bibitem[\protect\citeauthoryear{Hosking}{1990}]{Hosking}Hosking J.R.M., 1990, J. R. Statist. Soc. B, 52, 105
\bibitem[\protect\citeauthoryear{Moors et al.}{1996}]{Moors}Moors J.J.A., Wagemakers R.Th.A., Coenen V.M.J., Heuts R.M.J., Janssens M.J.B.T., 1996, Statistica Neerlandica, 50, 417
\bibitem[\protect\citeauthoryear{Rimoldini}{2013a}]{RimoldiniIntrinsic}Rimoldini L., 2013a, preprint (\href{http://xxx.lanl.gov/abs/1304.6715}{arXiv:1304.6715})
\bibitem[\protect\citeauthoryear{Rimoldini}{2013b}]{RimoldiniWeighted}Rimoldini L., 2013b, preprint (\href{http://xxx.lanl.gov/abs/1304.6616}{arXiv:1304.6616})
\bibitem[\protect\citeauthoryear{Stuart \& Ord}{1969}]{Kendall}Stuart A., Ord J., 1969, Kendall's Advanced Theory of Statistics, Charles Griffin \& Co.~Ltd, London
\bibitem[\protect\citeauthoryear{Thiele}{1889}]{Thiele}Thiele T.N., 1889, Forlaesinger over almindelig iagttagelseslaere: sandsynlighedsregning og mindste kvadraters methode, C.A. Reitzel, Copenhagen
\end{thebibliography}
\end{document}